\documentclass[fleqn,usenatbib]{mnras}

\usepackage{txfonts}

\usepackage[T1]{fontenc}
\usepackage{ae,aecompl}

\usepackage[normalem]{ulem}

\usepackage{graphicx}

\usepackage{amssymb}	
\usepackage{epsfig}
\usepackage{upgreek}
\usepackage{mathptmx}
\usepackage{txfonts}
\usepackage[utf8]{inputenc}
\usepackage{dcolumn}
\usepackage{natbib}
\usepackage{enumerate}
\usepackage{longtable,lscape}
\usepackage{color}
\usepackage{grffile}
\usepackage{array,float}
\usepackage{multirow}
\usepackage{lscape}
\usepackage{hyperref}
\usepackage{pdflscape}
\usepackage{afterpage}
\usepackage{capt-of}
\usepackage{xargs}
\usepackage{adjustbox}
\usepackage{booktabs}

\newcolumntype{R}[2]{%
    >{\adjustbox{angle=#1,lap=\width-(#2)}\bgroup}%
    l%
    <{\egroup}%
}
\newcommand*\rot{\multicolumn{1}{R{45}{1em}}}

\newcommand{\arcs}{\mbox{\ensuremath{^{\prime\prime}}}}

\def\NTHP {$\mathrm{N_2H^+}$} 
\def\solmass {$\hbox{M}_\odot$}

\def\HII{H{\sc ii}}

\def\NTHP {$\mathrm{N_2H^+}$} 

\newcolumntype{d}[1]{D{.}{\cdot}{#1}}

\newcolumntype{.}{D{.}{.}{-1}}

\newcommand{\malt}{MALT90}

\newcommand{\lsun}{L$_\odot$}
\newcommand{\msun}{M$_\odot$}

\newcommand{\vlsr}{V$_{\rm{LSR}}$}

\newcommand{\mum}{$\mu$m}

\newcommand{\kms}{km\,s$^{-1}$}

\newcommand{\hii}{H{\sc ii}}

\newcommand{\KS}{Kolmogorov-Smirnov}

\newcommand{\mcweeds}{\texttt{MCWeeds}}
\newcommand{\weeds}{\texttt{Weeds}}

\title[ATLASGAL --- Fingerprints of dense clumps]{ATLASGAL --- Molecular fingerprints of a sample
           of massive star forming clumps\thanks{The full version of Tables\,\ref{tab:source_list}, \ref{tab:fitted_parameters} and \ref{table:integrated_intensities} are only available in electronic form at the CDS via anonymous ftp to cdsarc.u-strasbg.fr (130.79.125.5) or via http://cdsweb.u-strasbg.fr/cgi-bin/qcat?J/MNRAS/.}}
\author[J.\,S.\,Urquhart et al.]{J.\,S.\,Urquhart$^{1,2}$\thanks{E-mail: j.s.urquhart@gmail.com}, C.\,Figura$^{3}$, F. Wyrowski$^{2}$, A.\,Giannetti$^{4,2}$, W.-J.\,Kim$^{2}$, M.\,Wienen$^{2}$, \newauthor  S.\,Leurini$^{5, 2}$, T.\,Pillai$^{2}$, T.\,Csengeri$^{2}$, S.\,J.\,Gibson$^{1}$, K.\,Menten$^{2}$, T.\,J.\,T.\,Moore$^{6}$,\newauthor M.\,A.\,Thompson$^{7}$ \\
\\
$^{1}$ Centre for Astrophysics and Planetary Science, University of Kent, Canterbury, CT2\,7NH, UK \\
$^{2}$ Max-Planck-Institut f\"ur Radioastronomie, Auf dem H\"ugel 69, D-53121 Bonn, Germany \\
{\color{black}$^{3}$ Wartburg College, Waverly, IA, 50677, USA}\\
{$^{4}$ INAF - Istituto di Radioastronomia \& Italian ALMA Regional Centre, Via P. Gobetti 101, I-40129 Bologna, Italy  }\\
{\color{black}$^{5}$ INAF - Osservatorio Astronomico di Cagliari, Via della Scienza 5, I-09047, Selargius (CA), Italy} \\
{\color{black}$^{6}$ Astrophysics Research Institute, Liverpool John Moores University, Liverpool Science Park, 146 Brownlow Hill, Liverpool, L3\,5RF, UK}\\
{\color{black}$^{7}$ Science and Technology Research Institute, University of Hertfordshire, College Lane, Hatfield, AL10 9AB, UK}\\
}

\date{Accepted XXX. Received YYY; in original form ZZZ}

\pubyear{2017}

\begin{document}
\label{firstpage}
\pagerange{\pageref{firstpage}--\pageref{lastpage}}

\maketitle

\begin{abstract}

{We have conducted a 3-mm molecular-line survey towards 570 high-mass star-forming clumps, using the Mopra telescope. The sample is selected from the 10,000 clumps identified by the ATLASGAL survey and includes all of the most important embedded evolutionary stages associated with massive star formation, classified into five distinct categories (quiescent, protostellar, young stellar objects, \hii\ regions and photo-dominated regions). The observations were performed in broadband mode with frequency coverage of 85.2 to 93.4\,GHz and a velocity resolution of $\sim$0.9\,\kms, detecting emission from 26 different transitions. We find significant evolutionary trends in the detection rates, integrated line intensities, and abundances of many of the transitions and also identify a couple of molecules that appear to be invariant to changes in the dust temperature and evolutionary stage (N$_2$H$^+$\,(1-0) and HN$^{13}$C\,(1-0)). We use the K-ladders for CH$_3$C$_2$H\,(5-4) and CH$_3$CH\,(5-4) to calculate the rotation temperatures and find $\sim$1/3 of the quiescent clumps have rotation temperatures that suggest the presence of an internal heating source. These sources may constitute a population of very young protostellar objects that are still dark at 70\,\mum\ and suggest that the fraction of truly quiescent clumps may only be a few per cent. We also identify a number of line ratios that show a strong correlation with the evolutionary stage of the embedded objects and discuss their utility as diagnostic probes of evolution.}

\end{abstract}




\section{\label{intro}Introduction}

Massive stars play a dominant role in the evolution of galaxies through the release of vast amounts of radiative and mechanical energy (e.g., jets, molecular outflows and powerful stellar winds) into the interstellar medium (ISM). They are responsible for the production of nearly all of the heavy elements, which are returned to the ISM in the late stages of their evolution, or at the time of their death via supernova explosions. The enrichment of the ISM and the energy pumped into their local environments change the chemical composition and structure of the surrounding gas, which can have a significant impact on future star formation activity: this can lead to an enhancement of the local star formation rate  (such as triggering; \citealt{urquhart2007d}) or can result in the disruption of molecular material before star formation has a chance to begin. Massive stars can, therefore, play an important role in shaping their local environments and regulating star formation.

Due to their short lifetime, massive stars do not escape significantly from their birth places. Observations of nearby spiral galaxies have revealed a tight correlation between spiral arms and the locations of massive star formation, and so the spatial distribution of massive stars in our own Galaxy can help us derive the structure and dynamics of the Milky Way.  These points taken together indicate that understanding the formation and evolution of massive stars is therefore of crucial importance to many areas of astrophysics.  At present, there is no well-established evolutionary scheme for high mass star formation, in contrast to the detailed framework of classes that exists for the early evolution of low-mass stars. In the 1990s, targeted surveys found many ultracompact \HII\ regions  \citep{wood1989a}, and hot molecular cores were found to be associated with them in subsequent follow-up observations \citep[e.g.][]{cesaroni+1992}. More recently, so called high-mass protostellar objects (HMPOs) or massive young stellar objects (MYSOs) were recognized as likely to represent an even earlier stage of massive star formation \citep[e.g.][]{beuther+2002a,lumsden2013}. Infra-red dark clumps have also been found to be promising candidates for even earlier stages in the formation of massive stars \citep[see][and references therein]{yuan2017}. 

These targeted surveys are, however, only partial, as they usually only trace one of the stages of massive star formation (MSF); UC\,\HII\ regions, massive young stellar objects (MYSOs) or CH$_3$OH masers and infrared darks clouds (e.g. \citealt{egan+1998,simon+2006b,peretto2009}) for instance.  As a result, they may miss entire classes of objects, or have very incomplete statistics. In contrast, an unbiased sample using complete dust continuum imaging at the scale of the molecular complex Cygnus-X \citep{motte+2007} was able to derive some more systematic results on the existence of a cold phase for high-mass star formation. To remedy this situation, APEX/LABOCA was used to produce the first unbiased  submillimetre continuum survey of the inner Galactic plane. 

In this paper we will use a molecular line survey of $\sim$600 massive clumps that include examples of all evolutionary stages associated with high-mass star formation in an effort to identify trends that might be utilised to develop a chemical clock for massive star formation. There have already been some efforts to address this recently with chemical evolution studies towards samples of IRDCs (e.g., \citealt{sanhueza2012, sakai2010}) and massive clump selected from their dust emission (e.g., \citealt{hoq2013,rathborne2016}). These have identified trends in specific line ratios and abundances, however, these have either been limited in the number of sources or in sensitivity and so we aim to build on these findings of these previous studies.

\subsection{ATLASGAL -- The APEX/LABOCA Galactic plane survey}

The APEX Telescope Large Area Survey of the Galaxy (ATLASGAL) is an unbiased 870\,$\mu$m\ submillimetre survey covering 420\,sq.\,degrees of the inner Galactic plane (\citealt{schuller2009}). The regions covered by this survey are $|\ell| < 60\degr$ with $|b|< 1.5\degr$ and $280\degr < \ell < 300\degr$ with $b$ between $-$2\degr\ and 1\degr; the change in latitude was necessary to account for the warp in the Galactic disc in the outer Galaxy extension. 

 Only a large unbiased survey can provide the statistical base to study the scarce and short-living protoclusters as the origin of the massive stars and the richest clusters in the Galaxy. The dust continuum emission in the (sub)millimetre range is the best tracer of the earliest phases of (high-mass) star formation since it is directly probing the material from which the stars form.  Cross-correlation with other complementary Galactic plane continuum surveys (e.g., GLIMPSE \citep{benjamin2003_ori}; MSX \citep{price2001}; WISE \citep{Wright2010}; MIPSGAL \citep{carey2009}; HiGAL \citep{Molinari2010}; and CORNISH \citep{hoare2012}) and molecular line surveys (e.g., CHIMPS \citep{rigby2016}; SEDIGISM \citep{schuller2017}; HOPS \citep{walsh2011}; and the MMB survey \citep{caswell2010_mmb}) will help considerably to answer a wide range of questions including: (1) What are the properties of the cold phase of massive star-formation?  (2) What is the evolutionary sequence for high-mass stars? (3) How important is triggering to form new generations of high-mass stars?  (4) What are the earliest phases of the richest clusters of the Galaxy?

Source extraction has identified $\sim$10,000 distinct massive clumps (\citealt{contreras2013,urquhart2014a,csengeri2014}). Analyses by \citet{wienen2015} of the physical properties of these clumps revealed that $\sim$92\,per\,cent satisfy the \citet{kauffmann2010a} size-mass criterion for massive star formation. Furthermore, the majority ($\sim$90\,per\,cent) of these clumps also are already associated with star formation (i.e., hosting an embedded 70\,\mum\ point source; \citealt{urquhart2018_csc}). Further analysis has confirmed that $\sim$15\% of these infrared bright sources are associated with massive young stellar objects and compact \hii\ regions (e.g., \citealt{urquhart2013a, urquhart2013b,urquhart2014_atlas}). Only a small fraction appear to be in a quiescent phase ($\sim $15\,per\,cent; \citealt{urquhart2018_csc}) and are therefore likely to be cold precursors to massive protostars.  This is probably an upper limit as some of these have been found to be associated with outflows (\citealt{Feng2016ApJ100F, Tan2016ApJ3T} and \citealt{Yang2018}) and, as we will find from rotation temperature calculations, many more appear to host an internal heating source. 

\subsection{Molecular fingerprints with Mopra}

Large scale continuum surveys are crucial for the study of massive star formation but there are a couple of shortcomings: perhaps the most significant is that, a priori, the distances to the newly found sources, which are needed to determine important parameters such as mass and luminosity of the clumps,  are unknown. Furthermore, their dust temperatures and densities are only loosely constrained by spectral energy distributions (in conjunction with infrared surveys). However, all of these issues can be more tightly constrained by molecular line observations, such as NH$_3$, CH$_3$C$_2$H and CH$_3$CN, of the massive star forming regions (e.g., \citealt{urquhart2011_nh3,wienen2015,tang2017,giannetti2017}).

We have embarked on a project to use the high sensitivity and large bandwidth of the Mopra telescope to study the physical and chemical properties of a large and unbiased sample of Galactic massive star forming clumps.  The results from this Mopra survey are vital for the analysis of the Galactic plane survey, since they allow us to investigate the kinematic, thermal and chemical properties of the clumps.  These can be thought of as the ``molecular fingerprints'' of the dust clumps and allow us to determine their kinematic distances, virial masses, infall and outflow activity, as well as probing their temperature, density and chemistry.

In this paper we present an overview of the Mopra spectral line survey and discuss the statistical properties of the sample. Detailed analysis of individual transitions is ongoing and will be published separately, and some of the transitions have already been published (e.g., mm-radio recombination lines --- \citealt{kim2017}; temperature determination --- \citealt{giannetti2017}). Furthermore, the line velocities have been used to derive distances, and aperture photometry has been performed on MSX/WISE and HiGAL and ATLASGAL emission maps to fit their spectral energy distributions to estimate clump bolometric luminosities, dust temperatures and masses (see \citealt{konig2017} and \citealt{urquhart2018_csc} for more details).  The observational data have been used to construct an evolutionary sequence for the formation of massive stars. Here, we examine the molecular line detection rates and fitted properties to see if they are consistent with this sequence and whether they can provide a deeper insight into the processes involved. 

The structure of the paper is as follows: in Sect.\,\ref{sect:classification} we outline the source selection criteria, in Sect.\,\ref{sect:obs} we describe the observational set up and the data reduction methodology.  Sect.\,\ref{sect:results} presents the results and discusses the overall statistical properties of the sample.  We discuss the viability of using these molecular transitions to trace the evolution of the ongoing embedded star formation in Sect.\,\ref{sect:chemical_clock}. In Sect.\,\ref{sect:summary} we present a summary of our results and highlight our main findings.

\section{Source selection}
\label{sect:classification}

\begin{figure}
\centering
\includegraphics[width=0.45\textwidth, trim= 40 0 40 0]{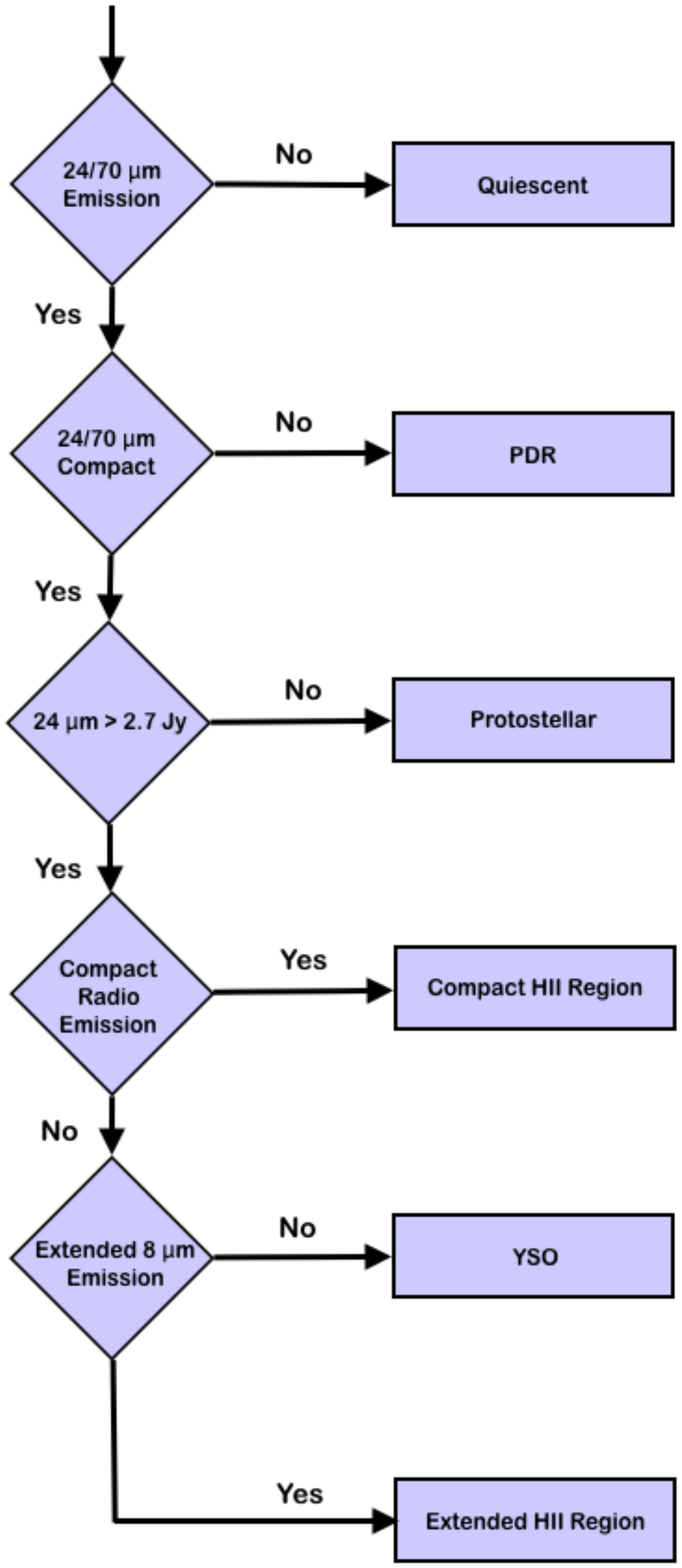}

\caption{Flow chart showing the criteria used to classify the current level of star formation taking place within the dense clumps. \label{fig:classification_flowchart} } 

\end{figure}

We have selected a flux-limited sample of clumps ($>$2\,Jy\,beam$^{-1}$ for clumps associated with MSX 21\,$\mu$m point sources  ($F_{21} > 2.7$\,Jy) and $>$\,1\,Jy\,beam$^{-1}$ for mid-infrared weak sources ($F_{21} < 2.7$\,Jy) due to their lower temperatures)\footnote{Clumps associated with mid-infrared sources have dust temperatures between 20 and 30\, while less evolved clumps have temperature between 10 and 20\,K; the difference in temperature results in the clumps associated with embedded objects being a factor of 2 brighter at submillimetre wavelengths than clumps where the star formation is less evolved. }, ensuring sufficiently high column densities for the Mopra line detections and  coverage of all stages of massive star formation. As shown in Fig.\,\ref{fig:classification_flowchart}, we consider all sources without a clear MSX 21-$\mu$m  point source coincident within 10\arcsec\ of the peak submillimetre emission to be quiescent or in the very earliest stages of star formation. We are able to probe all massive clumps down to a mass limit of 100\,\solmass\ within 5\,kpc, which is the distance of the Galactic molecular ring.

A total 600 sources that satisfied the selection criteria and are located in the Galactic longitude range $\ell=300-358\degr$\ and latitude $|b|<\,1\degr$ were observed as part of this line survey. These were drawn from a preliminary version of the catalogue produced using a Gaussian source-finder from an early reduction of the maps. A consequence of this is that a small number of the observed sources were spurious detections (9 sources corresponding to $\sim$2\,per\,cent of the sample) and do not appear in the final ATLASGAL Compact Source Catalogue (CSC; \citealt{contreras2013,urquhart2014_csc}). 

\begin{figure}
\centering
\includegraphics[width=0.45\textwidth]{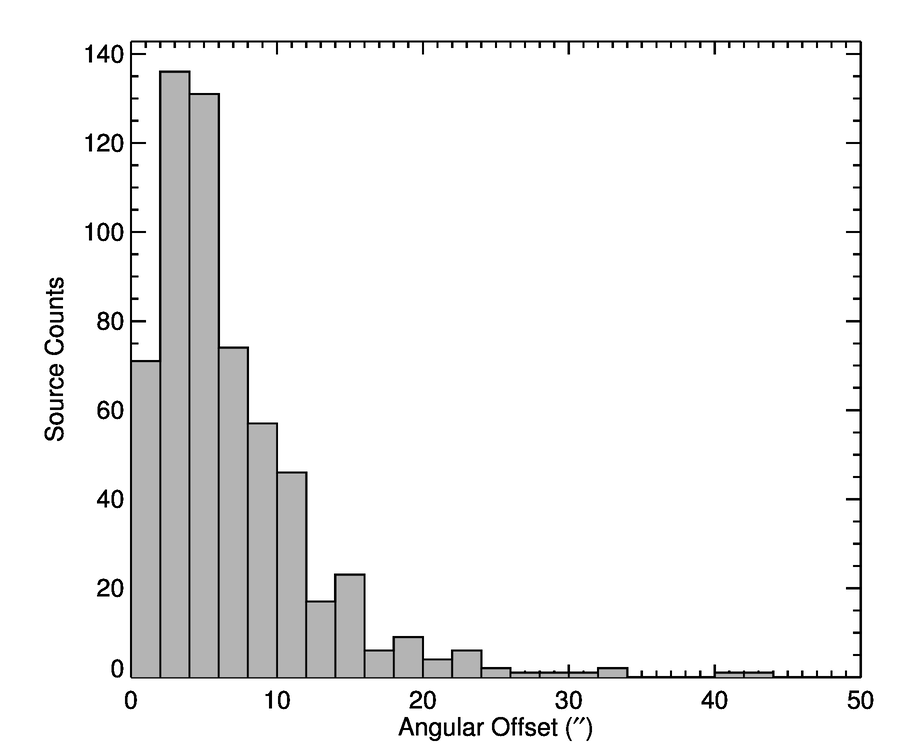}

  \caption{Offset between the observed position and the peak of the submillimetre emission that defines the centre of the clump. The bin size used is 2\arcsec.}
  \label{fig:offset}  
\end{figure}

Another consequence was that the peak positions were not always well-constrained, and this has led to offsets between the peak of the submillimetre emission and the telescope point position; these offsets are shown in Fig.\,\ref{fig:offset}. The resolution  and pointing accuracy of the Mopra telescope is 36\arcsec\ and $\sim$6\arcsec, respectively. We use a value of 20\arcsec\ (approximately half of the Mopra beam) to match the spectral emission features with the ATLASGAL clumps: this reduces the sample to 570 clumps, which corresponds to 95\,per\,cent of the observed sources. 

\begin{figure}
\centering
\includegraphics[width=0.45\textwidth]{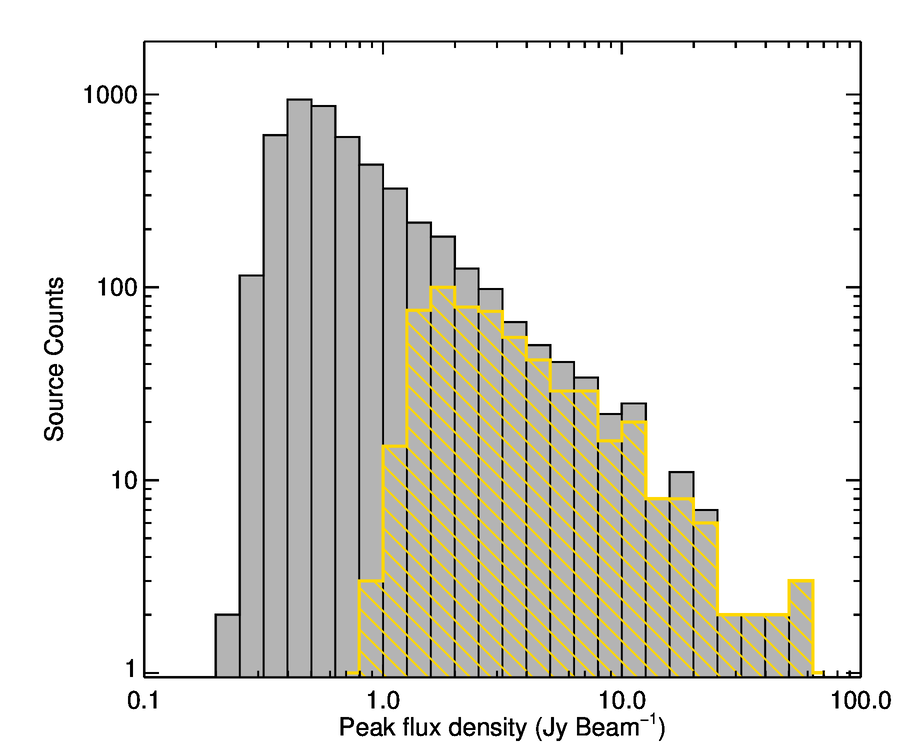}

  \caption{Peak flux distribution of all ATLASGAL sources located in the Fourth quadrant of the Galaxy (i.e., $300\degr < \ell < 358.5\degr$, shown in grey) and the 570 clumps observed as part of this project where the angular offset between peak submillimetre and target position is $<$ 20\arcsec\, (shown in yellow).}
  \label{fig:peak_flux}  
\end{figure}

In Fig.\,\ref{fig:peak_flux}, we show the peak flux distribution of the reduced sample (yellow hatching) and all of the ATLASGAL CSC sources in the same region of the sky. This plot shows that this sample is almost complete for all bright clumps located in the fourth quadrant above $\sim$2\,Jy\,beam$^{-1}$, and is therefore representative of the Galactic population of bright clumps. It is this reduced sample of 570 clumps that will be the focus of this paper.

These sources have been classified using the method and criteria used to classify the ATLASGAL Top 100 sample (\citealt{konig2017} and \citealt{urquhart2018_csc}) into four evolutionary types: quiescent, protostellar, YSO and \hii\ regions. The \hii\ regions and YSOs have similar mid-infrared colours and appear similar in mid-infrared images, and so we use the presence of compact radio continuum emission to classify compact \hii\ regions while extended 8-\mum\ emission identifies more evolved \hii\ emission regions. Following \citet{jackson2013},  we also include an additional classification to identify clumps located towards the edges of more evolved \hii\ regions, where the chemistry of these clumps and their structure are likely to be driven by the photo-dominated region (PDR). While the   quiescent, protostellar, YSO and \hii\ regions are an observationally defined evolutionary sequence for embedded stars, the PDRs are driven by much more evolved stars that have emerged from their natal environment. We include the PDRs in the plots and note any interesting trends for completeness and to facilitate comparison with other studies (e.g., \citealt{rathborne2016}), however,  we refrain from including these sources in any of the detailed analyses presented in this study.

\begin{table*}
\begin{center}

\caption{Summary of source, their infrared classification, the observed position, offset from the peak of the submillimetre emission and their physical properties. Clumps with Galactic latitudes $> |b| =1\degr$ are not classified or had their spectral energy distribution fitted as these are not covered by the HiGAL survey (\citealt{molinari2010a}). }

\begin{tabular}{lc...cc...}
\hline
\multirow{ 2}{*}{ATLASGAL name} & \multirow{ 2}{*}{Classification} & \multicolumn{1}{c}{$\ell_{\rm{obs}}$}  &  \multicolumn{1}{c}{$b_{\rm{obs}}$} & \multicolumn{1}{c}{Offset} & \multicolumn{1}{c}{Distance} & \multicolumn{1}{c}{$T_{\rm{dust}}$}& \multicolumn{1}{c}{$N_{\rm{H_2}}$} & \multicolumn{1}{c}{Log(M)} & \multicolumn{1}{c}{Log(L)} \\
 &   & \multicolumn{1}{c}{(\degr)} & \multicolumn{1}{c}{(\degr)} & \multicolumn{1}{c}{(\arcsec)} & \multicolumn{1}{c}{(kpc)} & \multicolumn{1}{c}{(K)} & \multicolumn{1}{c}{(cm$^{-2}$)} &\multicolumn{1}{c}{(\msun)} & \multicolumn{1}{c}{(\lsun)}  \\
\hline\hline
AGAL300.164$-$00.087	&	YSO	&	300.1635	&	-0.0889	&	5.37	&	4.30$\pm$0.95	&	18.89$\pm$1.39	&	22.82	&	2.42	&	3.19	\\
AGAL300.504$-$00.176	&	HII	&	300.5038	&	-0.1762	&	2.2	&	9.20$\pm$0.62	&	26.10$\pm$4.48	&	23.03	&	3.30	&	4.97	\\
AGAL300.721+01.201	&	YSO	&	300.7212	&	1.2006	&	1.17	&	3.40$\pm$0.91	&	\multicolumn{1}{c}{$\cdots$}	&	\multicolumn{1}{c}{$\cdots$}	&	\multicolumn{1}{c}{$\cdots$}	&	\multicolumn{1}{c}{$\cdots$}	\\
AGAL300.826+01.152	&	Uncertain	&	300.8248	&	1.1511	&	5.69	&	3.40$\pm$0.91	&	13.10$\pm$2.36	&	23.24	&	2.64	&	2.79	\\
AGAL300.911+00.881	&	Protostellar	&	300.9094	&	0.8812	&	5.51	&	3.40$\pm$0.91	&	23.59$\pm$4.38	&	22.87	&	2.28	&	3.63	\\
AGAL301.014+01.114	&	Uncertain	&	301.0140	&	1.1129	&	3.97	&	3.40$\pm$0.91	&	39.06$\pm$4.46	&	22.45	&	1.86	&	4.31	\\
AGAL301.116+00.959	&	PDR	&	301.1162	&	0.9596	&	2.35	&	3.40$\pm$0.91	&	20.87$\pm$3.34	&	23.11	&	2.52	&	4.19	\\
AGAL301.116+00.977	&	PDR	&	301.1178	&	0.9770	&	7.18	&	3.40$\pm$0.91	&	20.63$\pm$3.29	&	23.11	&	2.51	&	4.15	\\
AGAL301.136$-$00.226	&	HII	&	301.1364	&	-0.2256	&	2.18	&	4.60$\pm$1.14	&	34.17$\pm$1.20	&	23.64	&	3.30	&	5.39	\\
AGAL301.139+01.009	&	Quiescent	&	301.1373	&	1.0077	&	8.3	&	3.40$\pm$0.91	&	17.27$\pm$2.41	&	23.06	&	2.46	&	3.59	\\
\hline
\end{tabular}
\label{tab:source_list}
\end{center}

Notes: Only a small portion of the data is provided here, the full table is available in electronic form at the CDS via anonymous ftp to cdsarc.u-strasbg.fr (130.79.125.5) or via http://cdsweb.u-strasbg.fr/cgi-bin/qcat?J/MNRAS/.\\

\end{table*}

\begin{figure*}
\centering
\includegraphics[width=0.95\textwidth]{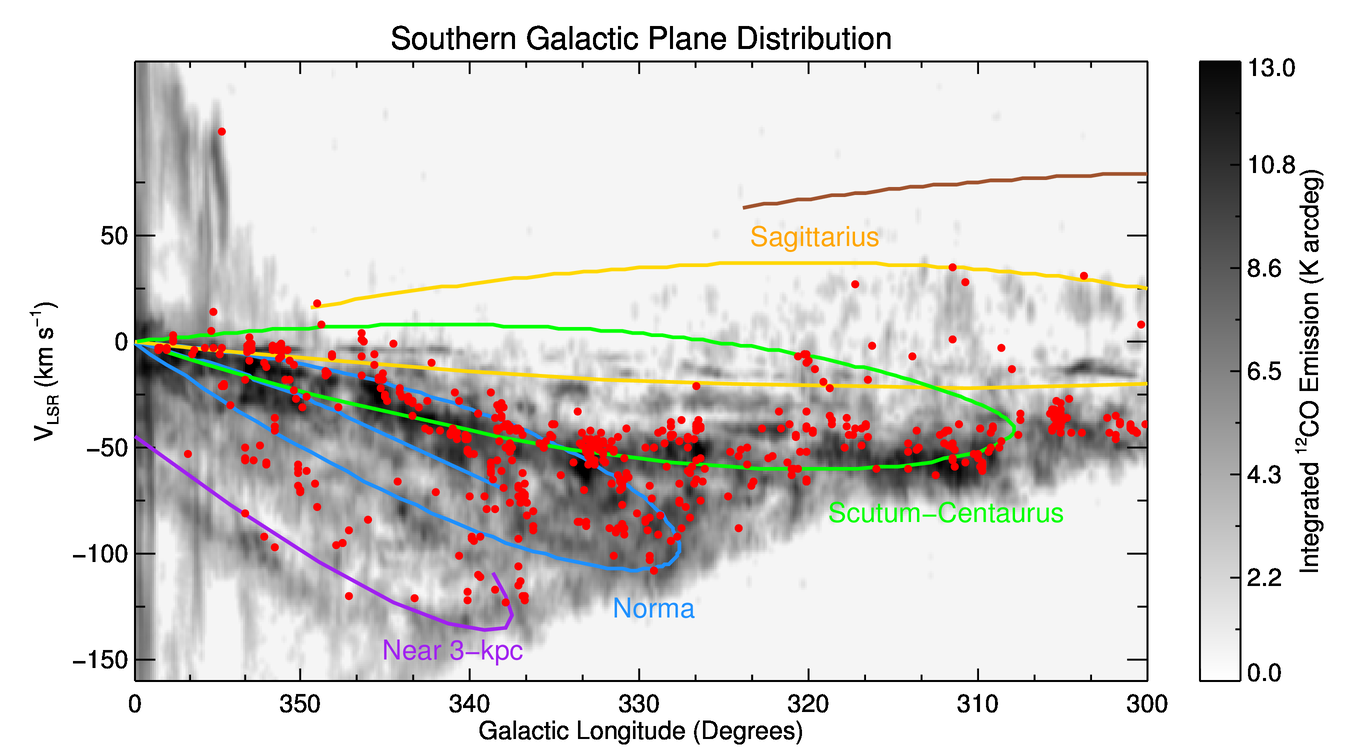}

  \caption{Longitude-velocity distribution of all ATLASGAL sources observed by Mopra toward which molecular emission has been detected. The background image shows the distribution of all molecular gas as traced by the $^{12}$CO (1-0) transition (\citealt{dame2001}), and the loci of the spiral arms are taken from \citet{taylor1993} and \citet{bronfman2000}. The measured velocities of the sources have been determined using \NTHP. The sources with unusual high velocities at $\ell = 354$\degr\ belong to Bania clump 1 \citep{bania+1986}. }
  \label{fig:showvel}  
\end{figure*}

We present a flowchart that outlines the classification scheme in Fig.\,\ref{fig:classification_flowchart}.  Individual source classifications and physical properties are summarised in Table\,\ref{tab:source_list}. The distances, dust temperatures and their uncertainties are taken from \citet{urquhart2018_csc} while the column densities and masses have been recalculated using the 870-\mum\ flux measured using a 36\arcsec\ aperture centred on the Mopra position. The measurement uncertainties (i.e., distance, flux and temperature) in the masses and column densities are between 20-40\,per\,cent, however, these are dominated by uncertainties in the dust to gas ratio and dust opacity. We therefore do not explicitly give the uncertainties for these parameters in Table\,\ref{tab:source_list} but estimate the mass and column densities to be correct within a factor of two. The uncertainty in the luminosity is a combination of the distance and flux measurements used for the SED fits are also estimated to be correct within a factor of two. The longitude-velocity ($\ell v$) distribution of the sample with respect to the molecular gas and the loci of the spiral arms is shown in Fig\,\ref{fig:showvel}. The systemic velocities of the sources, determined using \NTHP\ since its lines are strong, have a high detection rate ($\sim$96\,per\,cent) and do not show self-absorption.

\setlength{\tabcolsep}{6pt}

\begin{table*}




\begin{center}
\caption{Summary of the physical parameters for the 570 clumps matched with the observations located within 20\arcsec\ of the peak submillimetre emission.  The uncertainties given for the column density, mass and luminosities are the standard deviations of the logs of the parameters and therefore provide an indication of the spread in the values.}
\begin{tabular}{lrrcccc}
\hline
 \multirow{ 2}{*}{Classification} & \multicolumn{2}{c}{Number} &  \multicolumn{1}{c}{$T_{\rm{dust}}$}& \multicolumn{1}{c}{Log($N_{\rm{H_2}}$)} & \multicolumn{1}{c}{Log($M$)} & \multicolumn{1}{c}{Log($L$)} \\
 &  \multicolumn{2}{c}{of clumps}  & \multicolumn{1}{c}{(K)} & \multicolumn{1}{c}{(cm$^{-2}$)} &\multicolumn{1}{c}{(\msun)} & \multicolumn{1}{c}{(\lsun)}  \\
\hline\hline
Quiescent	&	29	&	(5.1\%)	&	18.2$\pm$2.9	&	23.11$\pm$0.18	&	2.39$\pm$0.46	&	3.35$\pm$0.73	\\
Protostellar	&	153	&	(26.8\%)	&	18.7$\pm$3.3	&	23.10$\pm$0.25	&	2.58$\pm$0.57	&	3.40$\pm$0.75	\\
YSO	&	128	&	(22.5\%)	&	22.9$\pm$4.1	&	23.08$\pm$0.28	&	2.55$\pm$0.53	&	3.91$\pm$0.73	\\
\hii\ region	&	166	&	(29.1\%)	&	28.2$\pm$4.8	&	23.11$\pm$0.32	&	2.82$\pm$0.58	&	4.73$\pm$0.68	\\
PDR	&	48	&	(8.4\%)	&	27.4$\pm$4.7	&	23.03$\pm$0.21	&	2.37$\pm$0.43	&	4.42$\pm$0.69	\\

\hline
Uncertain	&	46	&	(8.1\%)	&	25.0$\pm$6.3	&	23.09$\pm$0.22	&	2.47$\pm$0.63	&	4.23$\pm$0.82	\\
\hline
\end{tabular}

\label{tab:classification_summary}
\end{center}

\end{table*}

\setlength{\tabcolsep}{6pt}

We give a summary of the classifications and their average properties in Table\,\ref{tab:classification_summary}. The classification process has resulted in 29 quiescent, 153 protostellar, 128 YSO and 166 \hii\ region-associated clumps. The latter three evolutionary stages are relatively well-represented. It is interesting to note that less than 5\,per\,cent of the observed sources are classified as quiescent.  The infrared surveys we have used to identify the embedded protostellar sources are flux-limited, and therefore may not be sensitive to lower-mass protostars embedded in more distant clumps.  As a result, this quiescent fraction is likely to be an  upper limit. This would suggest that the dense quiescent phase of clumps is statistically very short-lived (cf. \citealt{urquhart2018_csc,csengeri2014}). We have classified 48 sources as likely to be associated with PDRs.

We have been unable to classify 46 clumps: for 42 of these the mid-infrared images where too badly saturated or too  complicated to make a reliable classification, while the remaining 4 clumps are located outside the HiGAL latitude range (i.e., $|b| > 1\degr$) and consequently there are no 70-\mum\ images available to allow a classification to be made.  The line parameters and physical properties are given for these sources, but these will not be included in any of the statistical analysis of the evolutionary types that follows; this reduces the number of sources to 524.

It is clear from a comparison of the average physical properties for the different source types that the dust temperature and luminosity increases with evolution (quiescent to \hii\ region stages) but that the column densities and masses remain similar.  This is confirmed by a \KS\ test for the column densities and masses, although clump masses for the \hii\ regions are found to be significantly larger. These findings are in line with those reported by \citet{konig2017} for a similar but smaller sample.\citet{contreras2017} calculate the masses for a larger sample of ATLASGAL sources ($\sim$1200) as part of the MALT90  project and also found the clump masses were similar for all their evolutionary stages. Their masses were estimated from an independent set of SED fits (\citealt{guzman2015}).  The similarity in the column density for the different evolutionary samples allows us to compare properties and detection rates for the subsamples independent of any significant sensitivity bias.

\section{Observations}
\label{sect:obs}

\subsection{Observational setup}

\begin{table}
\begin{center}
\caption{Summary of the observational parameters}
\begin{tabular}{ll}
\hline
Parameter & Value \\
\hline\hline
Galactic longitude ($\ell$) range      &      300\degr\ to 358.5\degr\ \\
Galactic latitude ($|b|$) )range      &      $-$1.15\degr\ to 1.58\degr\ \\
Number of observations & 601 \\
Number of CSCs observed & 570\\
Frequency range & 85.2\,GHz to 93.4\,GHz \\
Angular resolution (FWHM)$^a$ & 36$\pm$2\arcsec\ \\
Spectral resolution & 0.9\,\kms\ \\
Typical noise (T$^{*}_{\rm{A}}$) & $\sim$20-25\,mK\,channel$^{-1}$ \\
Typical system temperatures (T$_{\rm{sys}})$ & $\sim$200\,K\\
Typical pointing offset & $\sim$6\arcsec\ \\
Integration time (on-source) & 15\,mins\\
Main beam efficiency ($\eta_{\rm{mb}}$)$^a$ & 0.49 \\
Mean observation offset & 7.8\arcsec\ \\
\hline
\end{tabular}

\label{tab:obs_parameters}
\end{center}
Notes: $^a$ These values have been determined by \citet{ladd2005} for the Mopra telescope for a frequency of 86\,GHz.
\end{table}

The observations were carried out with the ATNF Mopra 22-m telescope during the Australian winter in 2008 and 2009 (Project IDs:M327-2008 and M327-2009; \citealt{wyrowski2008,wyrowski2009}).\footnote{The Mopra radio telescope is part of the Australia Telescope National Facility which is funded by the Australian Government for operation as a National Facility managed by CSIRO.}  The clumps were observed as pointed observations in position switching mode using the MOPS spectrometer.\footnote{The University of New South Wales Digital Filter Bank used for the observations with the Mopra Telescope was provided with support from the Australian Research Council.} The HEMT receiver was tuned to 89.3\,GHz and the MOPS spectrometer was used in broadband mode to cover the frequency range from 85.2 -- 93.4\,GHz with a velocity resolution of $\sim$0.9\,\kms. The beam size at these frequencies is $\sim$36\arcs. 

The typical observing time was 15\,min per source. The weather conditions were stable with typical system temperature of  $\sim$200\,K. The telescope pointing was checked approximately every hour with line pointings on SiO masers (\citealt{indermuehle2013}) and the absolute flux scale was checked each day by observing the standard line calibrators G327 and M17. We estimate the flux uncertainty is of order 20\,per\,cent.

\begin{figure}

\includegraphics[width=0.45\textwidth]{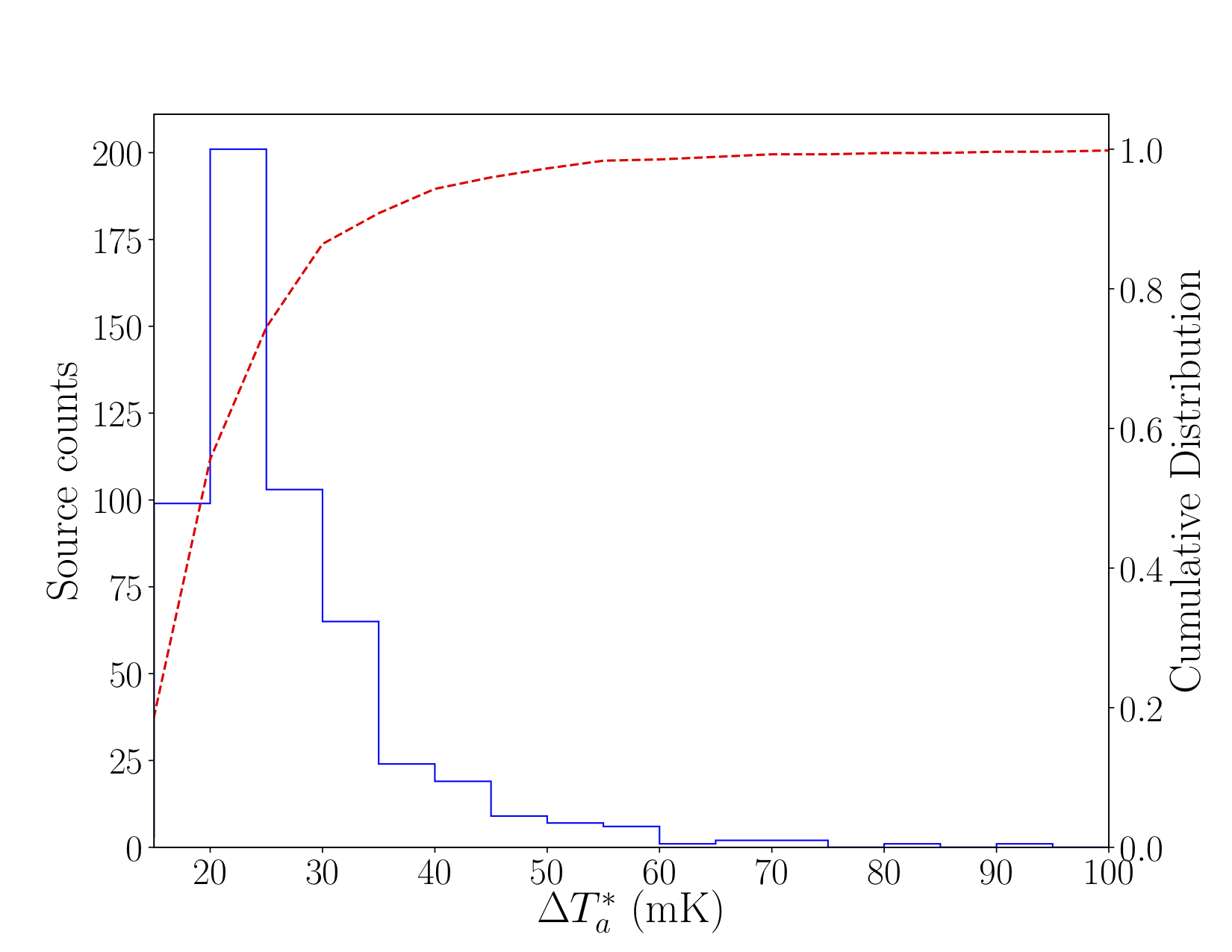}

  \caption{Histogram of the noise level reached for N$_2$H$^+$ lines for the observed sources.  Bin size is 5\,mK, median and mean noise levels are 29\,mK and 34\,mK, respectively.}
 \label{fig:show_noise}  
 
\end{figure}

\begin{figure*}

	\includegraphics[width=0.9\textwidth, trim=30 10 40 30, clip]{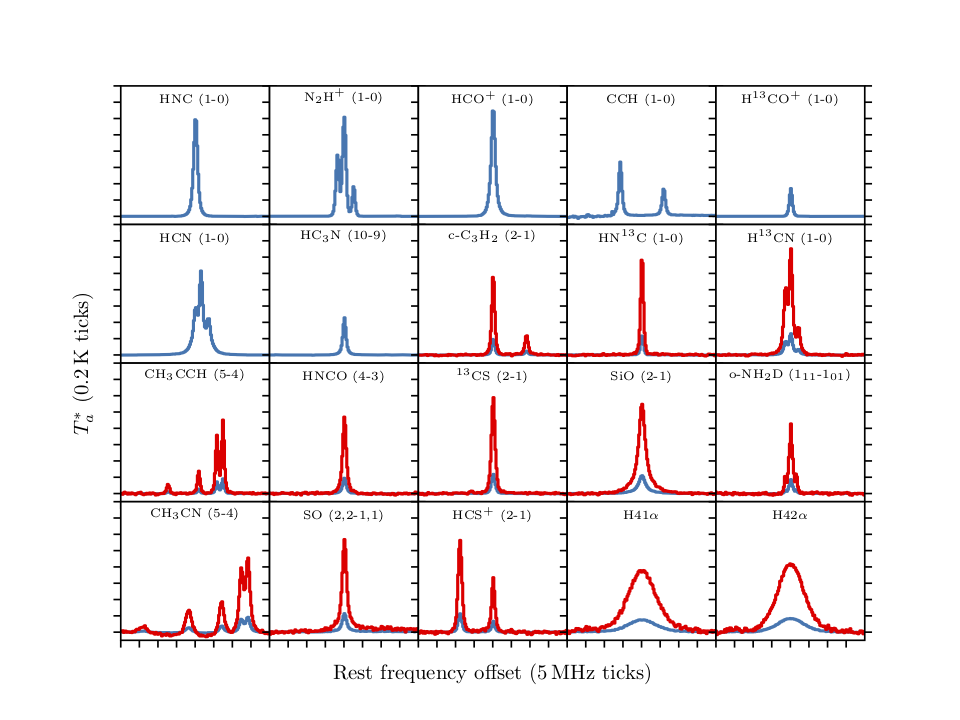}
    \caption{ Averaged emission for the 20 most commonly detected transitions. The horizontal axis is measured in 5\,MHz ticks from the line rest frequency.  The antenna temperature scale ranges from 0 to 1.6\,K in 0.2\,K ticks.  Weak lines are shown in red magnified by a factor of five. \label{fig:showaverage} }

\end{figure*} 

\subsection{Data reduction}

Initial processing of the data, which included processing of the on-off observing mode, the time and polarization averaging and baseline subtraction, were done with the ATNF Spectral line Analysis (ASAP) package.\footnote{\texttt{http://svn.atnf.csiro.au/trac/asap.}} Data were calibrated to the $T_{\rm A}^*$ temperature scale in ASAP. The spectra were then analysed using PySpecKit (\citealt{pyspeckit2011})\footnote{\texttt{http://http://pyspeckit.bitbucket.org/}}. No further smoothing was performed given the relatively large channel widths (0.95\,km\,s$^{-1}$\,channel$^{-1}$).  100\,km\,s$^{-1}$ wide spectral windows were set centred at the CSC source velocities (\citealt{urquhart2018_csc}), and each emission peak within these windows was fitted with a Gaussian profile and removed from the spectrum until no significant emission remained. Transitions that are associated with hyperfine structure were fitted with multiple Gaussian components assuming common source conditions, and utilised fixed velocity offsets and equal line-widths for each component.

Once all of the emission components had been identified, the regions fitted were excluded and the non-emission channels were baselined with a second-order polynomial. Emission regions were then re-fitted using prior values as seeds to optimise the resulting fit parameters. These final emission-line fits were determined to be significant if  $T_{\rm {A}}^* >  4 \sigma_{\rm{rms}}$ and if line integrated intensity $W > 5 \Delta W$, where $\Delta W = \sigma_{\rm{rms}}\times {\rm channel~width} \times \sqrt{{\rm number~of~channels}}$.  The median value of $\Delta W$ across all detected lines was 0.051\,K\,\kms, with a sample standard deviation of 0.045\,K\,\kms. In addition to extracting the bulk emission parameters, we have performed  $K$-ladder fitting for the CH$_3$CN and CH$_3$CCH transitions; these are described in detail in Sect.\,\ref{sect:k_ladder}. 

In Fig.\,\ref{fig:show_noise}, we show a histogram of the spectral noise for the N$_2$H$^+$ transition. The typical noise is $\sim$20\,mK\,channel$^{-1}$, which means we are approximately 3-4 times more sensitive than the \malt\ survey (\citealt{jackson2013,rathborne2016}) when the noise is estimated using the same channel width. We show the average significant spectra for all transitions detected towards more than 50 clumps in Fig.\,\ref{fig:showaverage}. Multiple components are detected towards a significant number of sources for some transitions ($\sim$50\,per\,cent for HCN and $\sim$20-25\,per\,cent for C$_2$H, HCO$^+$ and HNC). The number of source detections for each transition, as well as the number of multiple detections, are given in Column\,5 of Table\,\ref{tab:mollines}.


\subsection{Comparison with the MALT90 survey detections}

The overlap between the MALT90 survey and sources observed in this study is approximately 100\,per\,cent. Furthermore, both  the MALT90 survey and our observations  have included many of the same transitions; however, our use of the broad band allowed complete coverage of the 8\,GHz bandpass between $\sim$85-93\,GHz rather than the $16\times137$\,MHz zoom windows used by the \malt\ survey.  This has the advantage of covering $\sim$50\,per\,cent more spectral transitions including c-C$_3$H$_2$, CH$_3$C$_2$H, o-NH$_2$D and H$^{13}$CN, all of which are well-detected. The combination of longer integration times and specific targeting the brighter sources in the ATLASGAL catalogue (i.e., $>$1\,Jy\,beam$^{-1}$) has resulted in significantly higher detection rates for the four most detected lines by MALT90 (HCO$^+$, HNC, N$_2$H$^+$ and HCN), as well as many more transitions overall. We have detected at least 12-13 transitions towards more than 50\,per\,cent of the sample: this is approximately three times more than detected by the MALT90 survey. Although MALT90 has a poorer sensitivity it has mapped small regions around each clump ($\sim 3\arcmin \times 3\arcmin$) and so allows the spatial distribution of molecules to be investigated. 

The higher bandwidth comes at the expense of spectral resolution, which is one-eighth that of the \malt\ survey. Since most spectral lines have a FWHM line-width of $\sim$3\,\kms\, this lower spectral resolution does not represent a significant problem when analysing bulk line properties (e.g., radial velocities, line-widths, peak and integrated intensities) but does limit detailed analyses of line profiles that are often used to identify infall signatures. We are sensitive to outflow signatures (e.g., traced by SiO; \citealt{csengeri2016_sio}) but these single-pointing observations provide no information of their direction or extent.

\begin{figure*}
\centering
	\includegraphics[width=0.99\textwidth]{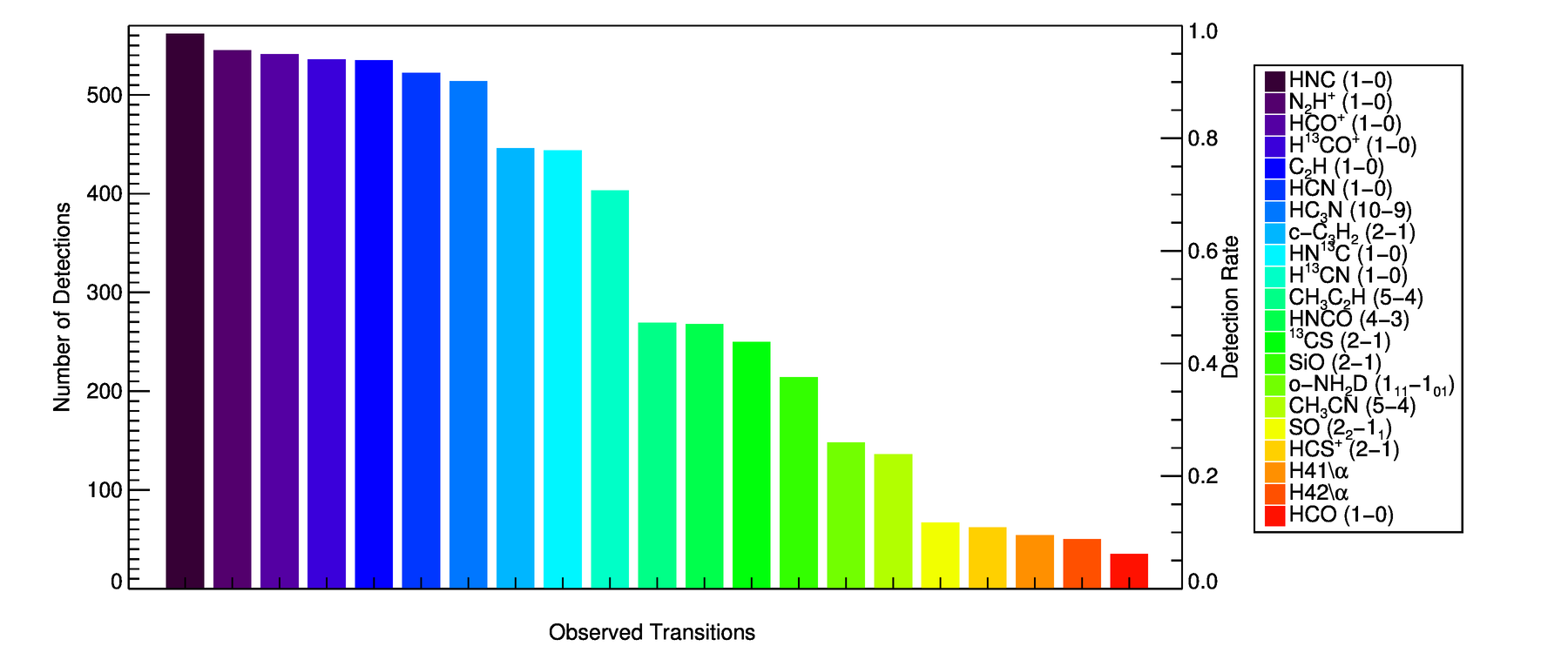}
  	\caption{Detection statistics for the molecular and radio recombination lines found within the 8\,GHz band observed with Mopra.}
 \label{fig:transitions_bargraph}  	 

\end{figure*}

\begin{figure*}

	\centering
	\includegraphics[width=0.9\textwidth]{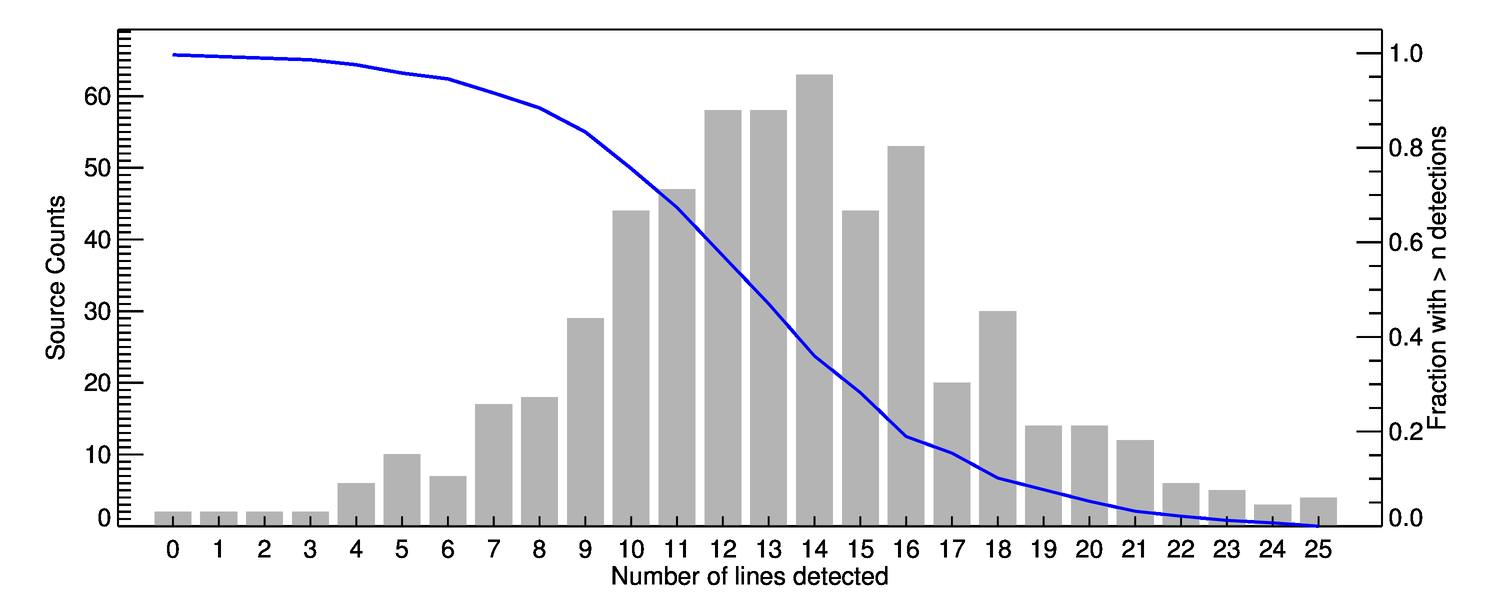}

  	\caption{The distribution of detections per clump.  We have detected 12-13 transitions towards $\sim$50\,per\,cent of the sample. The blue curved line shows the fraction of sources with $>$ n detections; these can be read off from the right y-axis.}
 	\label{fig:detection_stats}  
  
\end{figure*}

\section{\label{sect:results}Results and analysis}
\label{sect:results}

\subsection{Detection statistics}

\begin{figure*}
\includegraphics[width=0.45\textwidth, trim= 45 15 10 0, clip]{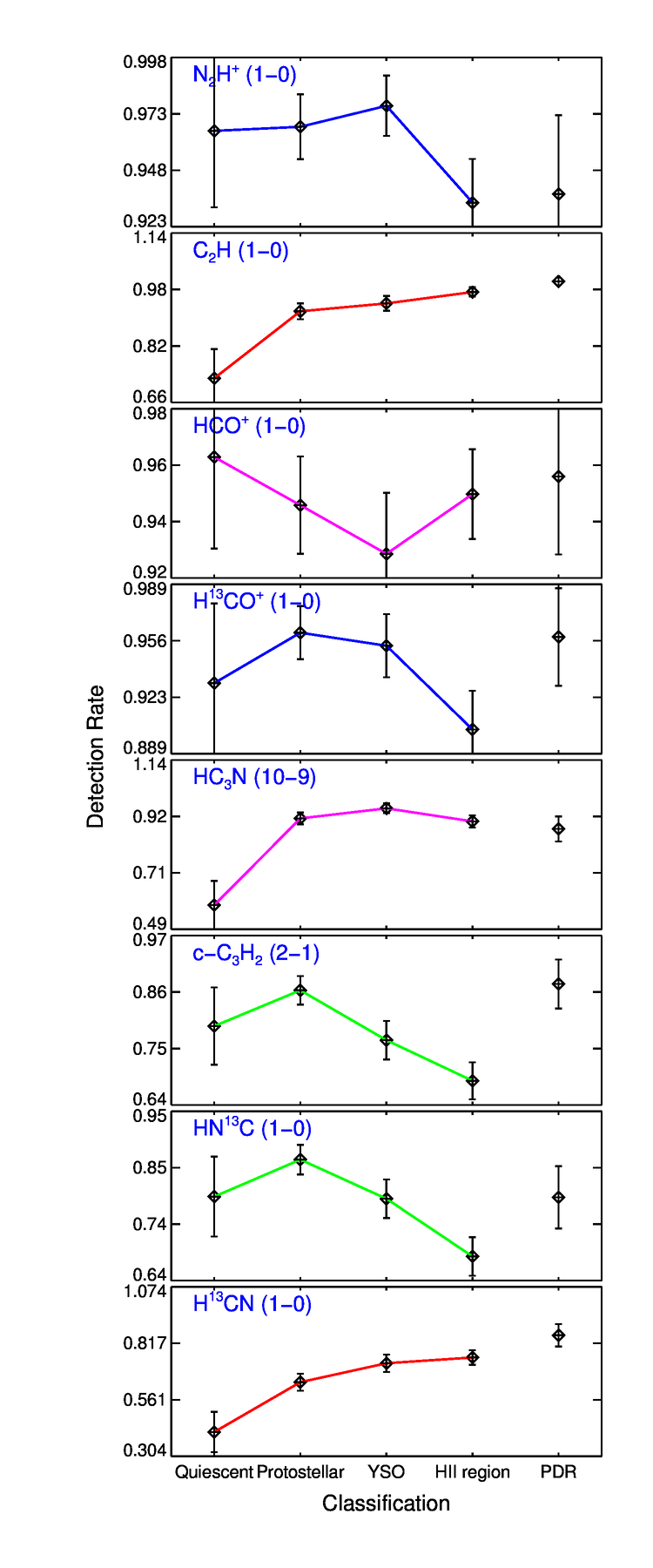}
\includegraphics[width=0.45\textwidth, trim= 45 15 10 0, clip]{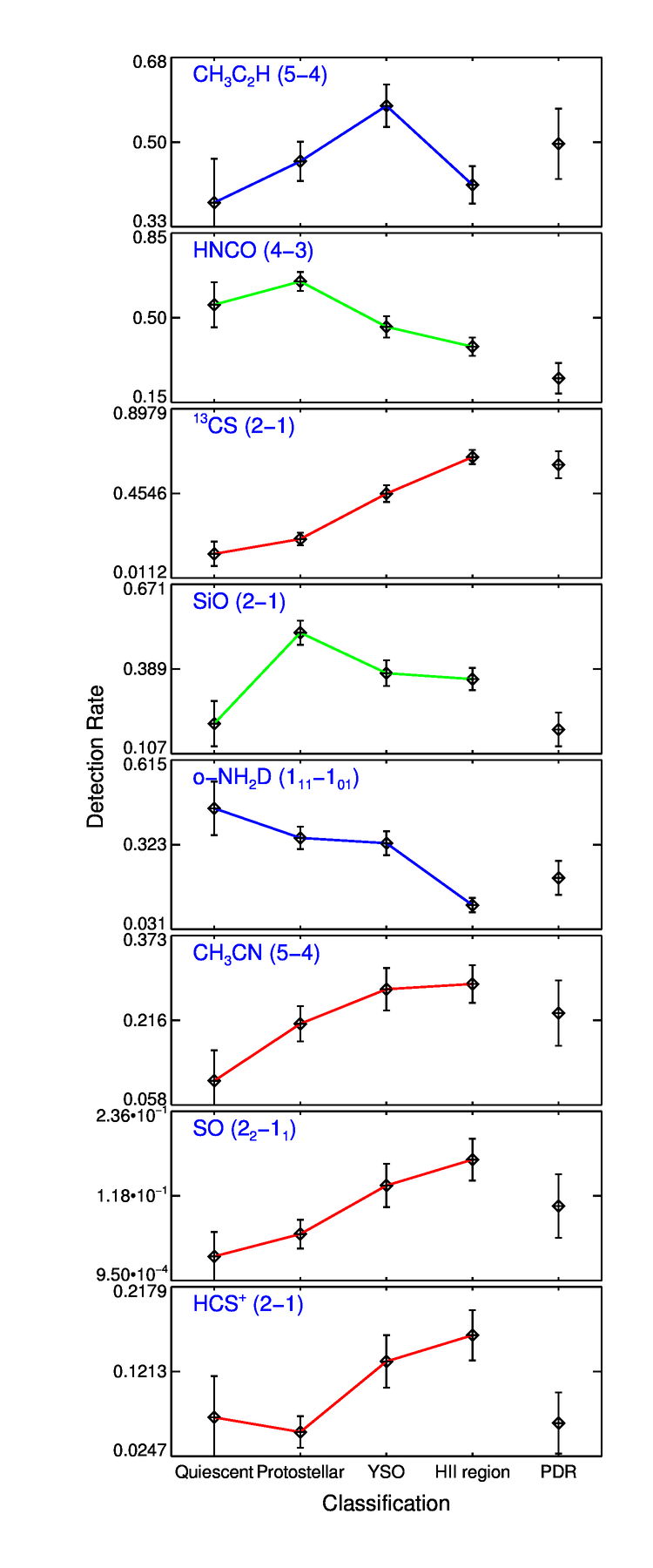}

\caption{Detection rates for the thermal transitions towards ATLASGAL clumps are plotted as a function of source classification. The quiescent, protostellar, YSO and \hii\ region classes are considered to be different stages in the evolution of the clumps and are placed in this sequence (left to right respectively). The coloured lines that join the four evolutionary stages are used to identify broad trends in the profiles:  steadily increasing detection rates (red), detection rates that peak in the protostellar stage (green), detection rates that are relatively constant with  the three early stages of evolution but which show a sharp decline in the final stage (blue), those that increase steadily but decrease in the \hii\ region stage. The error bars are determined using binomial statistics. Note that the y-axis range is dynamically scales for each transition to emphasis the differences between the different stages.}  
  \label{fig:detection_rates_evolution}  
\end{figure*}

\setlength{\tabcolsep}{3pt}
\begin{table*}
\begin{minipage}{\linewidth}


	\begin{center}
	\caption{Detection statistics for all 26 transitions found within the Mopra 8-GHz bandpass used for these observations. The number given in brackets in Column\,5 indicates the number of observations where multiple components are detected. Column\,6 indicates the fraction of sources in which the line was detected.}
	\begin{tabular}{lc.ccccccccl}
	\hline
	Emission {line(s)}  & $\nu$  &  \multicolumn{1}{c}{$E_{\rm u}/k^a$} &  \multicolumn{1}{c}{Log($n_{\rm crit}$)}  & \multicolumn{2}{c}{Number of} &  Detection & RMS noise & $T_a^*$ & Line-width & Intensity &   \multicolumn{1}{l}{Comments}\\

 &  (GHz) & \multicolumn{1}{c}{(K)} & (cm$^{-3}$) &   \multicolumn{2}{c}{detections} & ratio & (mK) & (K) & (\kms) & (\kms\,K) \\
    	\multicolumn{1}{c}{(1)} &  \multicolumn{1}{c}{(2)} & \multicolumn{1}{c}{(3)}  &  \multicolumn{1}{c}{(4)}& \multicolumn{2}{c}{(5)} & \multicolumn{1}{c}{(6)} & \multicolumn{1}{c}{(7)} & \multicolumn{1}{c}{(8)} & \multicolumn{1}{c}{(9)} & \multicolumn{1}{c}{(10)} & \multicolumn{1}{l}{(11)}  \\
	\hline\hline
c-C$_3$H$_2$ (2-1)	&	85.3389	&	6.4	&	6.0	&	446	&	(9)	&	0.78	&	24	&	0.21	&	3.3	&	0.72	&	Cyclic molecules	\\
HCS$^+$ (2-1)	&	85.3479	&	6.1	&	4.5	&	62	&	(0)	&	0.11	&	22	&	0.13	&	3.3	&	0.49	&	Sulphur chemistry	\\
CH$_2$CHCN (9-8)	&	85.4218	&		&		&	3	&	(0)	&	0.01	&	28	&	0.11	&	1.2	&	0.19	&	Temperature	\\
CH$_3$C$_2$H (5-4)	&	85.4573	&		&		&	273	&	(4)	&	0.48	&	26	&	0.19	&	2.8	&	1.34	&	Temperature	\\
HOCO$^+$ (4$_{04}$-4$_{03}$)	&	85.5315	&	10.3	&	5.2	&	3	&	(0)	&	0.01	&	17	&	0.08	&	4.5	&	0.4	&	Indirect gas-phase CO$_2$ tracer	\\
H42$\alpha$	&	85.6884	&		&		&	58	&	(1)	&	0.10	&	11	&	0.16	&	27.6	&	1.03	&	Ionised gas	\\
o-NH$_2$D (1$_{11}$-1$_{01}$)	&	85.9263	&	20.7	&	6.6	&	148	&	(0)	&	0.26	&	24	&	0.15	&	2.1	&	0.89	&	Deuteration, coldest dense gas	\\
SO ($2_2-1_1$)	&	86.0940	&	19.3	&	5.2	&	67	&	(0)	&	0.12	&	26	&	0.31	&	5.2	&	4.64	&	Suphur chemistry, shocks	\\
H$^{13}$CN (1-0)	&	86.3402	&	4.1	&	6.3	&	403	&	(13)	&	0.71	&	26	&	0.25	&	3.5	&	2.03	&	Dense gas	\\
HCO (1-0)	&	86.6708	&		&		&	36	&	(0)	&	0.06	&	22	&	0.11	&	2.8	&	0.35	&	Photon dominated regions	\\
H$^{13}$CO$^+$ (1-0)	&	86.7543	&	4.2	&	4.8	&	536	&	(20)	&	0.94	&	25	&	0.36	&	3.1	&	1.2	&	Dense gas 	\\
SiO (2-1)	&	86.8470	&	6.2	&	5.4	&	218	&	(9)	&	0.38	&	10	&	0.17	&	13.2	&	0.45	&	Shocks, outflows	\\
HN$^{13}$C (1-0)	&	87.0908	&	4.2	&	5.5	&	445	&	(7)	&	0.78	&	26	&	0.26	&	3.1	&	0.82	&	Dense gas	\\
C$_2$H (1-0)	&	87.3169	&	4.2	&	6.0	&	542	&	(131)	&	0.95	&	45	&	0.68	&	3.5	&	6.31	&	Early chemistry/PDR$^b$	\\
HNCO (4-3)	&	87.9253	&	10.6	&	6.0	&	269	&	(1)	&	0.47	&	25	&	0.19	&	3.8	&	0.82	&	Hot core, FIR, chemistry	\\
HCN (1-0)	&	88.6318	&	4.2	&	5.7	&	528	&	(284)	&	0.93	&	34	&	0.98	&	4.5	&	9.15	&	Dense gas	\\
HCO$^+$ (1-0)	&	89.1885	&	4.3	&	4.8	&	542	&	(136)	&	0.95	&	39	&	1.34	&	4.5	&	6.26	&	Kinematics (Infall, outflow)	\\
HNC (1-0)	&	90.6636	&	4.3	&	5.1	&	562	&	(119)	&	0.99	&	33	&	1.24	&	4.2	&	5.54	&	High column density, cold gas tracer	\\
C$_2$S (2-1)	&	90.6864	&		&		&	3	&	(0)	&	0.01	&	21	&	0.09	&	3.8	&	0.31	&	Dense gas, evolutionary phase	\\
$^{13}$C$^{34}$S (2-1)	&	90.9260	&	6.5	&	5.6	&	2	&	(0)	&	0.00	&	25	&	0.11	&	2.8	&	0.34	&	Column density	\\
HC$_3$N (10-9)	&	90.9790	&	24.0	&	5.0	&	514	&	(22)	&	0.90	&	26	&	0.46	&	3.1	&	1.64	&	Hot core	\\
HC$^{13}$C$_2$N (10-9)	&	90.9790	&	23.9	&	5.2	&	3	&	(0)	&	0.01	&	22	&	0.12	&	12.7	&	1.31	&	Hot core	\\
CH$_3$CN (5-4)	&	91.9871	&	13.2	&	5.4	&	142	&	(5)	&	0.25	&	25	&	0.18	&	4.5	&	2.74	&	Hot core temperature	\\
H41$\alpha$	&	92.0345	&		&		&	60	&	(0)	&	0.11	&	10	&	0.14	&	31.8	&	0.92	&	Ionised gas	\\
$^{13}$CS (2-1)	&	92.4943	&	6.0	&	5.5	&	250	&	(1)	&	0.44	&	23	&	0.24	&	3.5	&	0.95	&	Dense gas, infall	\\
N$_2$H$^+$ (1-0)	&	93.1738	&	4.5	&	4.8	&	545	&	(0)	&	0.96	&	27	&	0.66	&	3.1	&	8.62	&	Dense gas, depletion resistant	\\
\hline
	\end{tabular}

	\label{tab:mollines}
	\end{center}
    
\end{minipage}
\begin{flushleft}
$^a$ Excitation energies ($E_{\rm u}/k$), Einstein A coefficients ($A_{\rm u}$), and collisional rates ($\gamma$) were obtained from the Leiden atomic and molecular database (LAMDA; \citealt{Schoeier2005}) assuming a temperature of 20\,K; these have been used to calculate the critical densities using the equation $n_{\rm crit} = A_{\rm u}/\gamma$. For o-NH$_2$D the excitation energy and critical density are take from \citet{machin2006} and correspond to a gas temperature of 25\,K. The excitation temperatures and critical densities for H$^{13}$CN, H$^{13}$CN (1-0) and $^{13}$CS are taken from \citet{sanhueza2012} and for C$^{34}$S and HC$^{13}$C$_2$N they are take from \citet{miettinen2014} and calculated assuming a temperature of 15\,K. The excitation temperature and critical density for HOCO$^+$ are taken from  \citet{hammami2007}. The excitation temperature and critical densities have been determined for C$_3$CN and SO assuming a gas temperature of 30 and 60\,K, respectively. There are currently no collision rates available for HCO, C$_2$S, CH$_2$CHCN and CH$_3$C$_2$H.\\
$^b$ C$_2$H has been found to be enhanced in PDRs (\citealt{fuente1993}), however, \cite{beuther2008_CCH} has detected it towards clumps in early stages of star formation.
\end{flushleft}
\end{table*}

A summary of the 26 molecular transitions and radio recombination lines detected in the observed bandpass is given in Table\,\ref{tab:mollines}. We indicate the detection rates and the physical conditions to which each transition is sensitive. In Fig.\,\ref{fig:transitions_bargraph} we show the detection rates of the various transitions graphically. In Table\,\ref{tab:fitted_parameters}, we present the fitted parameters for all of the transitions observed towards all of the ATLASGAL clumps. We have detected the ten brightest transitions to more than 70\,per\,cent of the sources observed, climbing to over 90\,per\,cent of the sample for the seven brightest transitions. Fig.\,\ref{fig:detection_stats} shows the number of lines detected towards each source, and reveals that on average we have detected 13 transitions or more towards approximately half of the sources observed. Emission is detected to all but 5 ATLASGAL clumps. Inspection of these reveals three spectra that are significantly more noisy  ($\sim$60\,mK) than the rest of the sample (AGAL313.576+00.324 and  AGAL319.399$-$00.012 are detected in MALT90 and AGAL305.321+00.071 is not covered by MALT90), which is likely to have resulted in the non-detection of any molecular emission. One source (AGAL354.711+00.292) has an rms of 26\,mK and is also not detected in MALT90. Inspection of the mid-infrared image would suggest that this is a \hii\ region and the non-detection of any emission may indicate it is quite distant. The remaining source (AGAL303.118$-$00.972) has an rms of 30\,mK and is detected in MALT90 and so it is unclear why no line emission has been detected towards this object.

\setlength{\tabcolsep}{6pt}
\begin{table*}


	\begin{center}
	\caption{Fitted line parameters.}
	\begin{tabular}{ll........}
	\hline
	ATLASGAL  & Transition & \multicolumn{1}{c}{RMS}  &  \multicolumn{1}{c}{\vlsr}& \multicolumn{1}{c}{\vlsr\ error} & \multicolumn{1}{c}{$T_a^*$} &  \multicolumn{1}{c}{$T_a^*$ error} & \multicolumn{1}{c}{$\Delta$\vlsr} & \multicolumn{1}{c}{$\Delta$\vlsr\ error}  & \multicolumn{1}{c}{Intensity} \\

	CSC name  &  & \multicolumn{1}{c}{(mK)}  &  \multicolumn{1}{c}{(\kms)}& \multicolumn{1}{c}{(\kms)} & \multicolumn{1}{c}{(K)} &  \multicolumn{1}{c}{(K)} & \multicolumn{1}{c}{(\kms)} & \multicolumn{1}{c}{(\kms)}  & \multicolumn{1}{c}{(K\,\kms)} \\

	\hline\hline

AGAL300.164$-$00.087	&	N$_2$H$^+$ (1-0)	&	27	&	-39.4	&	0.04	&	0.35	&	0.017	&	0.9	&	0.0	&	2.9	\\
AGAL300.504$-$00.176	&	N$_2$H$^+$ (1-0)	&	21	&	7.9	&	0.14	&	0.09	&	0.008	&	1.5	&	0.1	&	1.4	\\
AGAL300.721+01.201	&	N$_2$H$^+$ (1-0)	&	19	&	-43.3	&	0.05	&	0.22	&	0.011	&	1.0	&	0.1	&	2.1	\\
AGAL300.826+01.152	&	N$_2$H$^+$ (1-0)	&	19	&	-43.1	&	0.05	&	0.48	&	0.017	&	1.3	&	0.0	&	5.9	\\
AGAL300.911+00.881	&	N$_2$H$^+$ (1-0)	&	20	&	-41.7	&	0.04	&	0.53	&	0.016	&	1.2	&	0.0	&	6.2	\\
AGAL301.014+01.114	&	N$_2$H$^+$ (1-0)	&	20	&	-41.9	&	0.03	&	0.54	&	0.016	&	0.9	&	0.0	&	4.6	\\
AGAL301.116+00.959	&	N$_2$H$^+$ (1-0)	&	17	&	-41.1	&	0.03	&	0.57	&	0.015	&	1.0	&	0.0	&	5.7	\\
AGAL301.116+00.977	&	N$_2$H$^+$ (1-0)	&	19	&	-40.5	&	0.02	&	1.01	&	0.026	&	0.9	&	0.0	&	8.7	\\
AGAL301.136$-$00.226	&	N$_2$H$^+$ (1-0)	&	21	&	-39.5	&	0.09	&	0.19	&	0.009	&	1.6	&	0.1	&	2.8	\\
AGAL301.139+01.009	&	N$_2$H$^+$ (1-0)	&	24	&	-40.5	&	0.05	&	0.39	&	0.015	&	1.3	&	0.0	&	5.3	\\    

\hline
	\end{tabular}
	\label{tab:fitted_parameters}
	\end{center}

Notes: Only a small portion of the data is provided here, the full table is available in electronic form at the CDS via anonymous ftp to cdsarc.u-strasbg.fr (130.79.125.5) or via http://cdsweb.u-strasbg.fr/cgi-bin/qcat?J/MNRAS/.\\ 
\end{table*}

Multiple components are detected towards a significant number of sources for some transitions ($\sim$50\,per\,cent for HCN and $\sim$20-25\,per\,cent for C$_2$H, HCO$^+$ and HNC), and so it is necessary to determine those that are likely to be associated with the ATLASGAL clump. To do this, we  use the N$_2$H$^+$ transitions in the first instance, as this is a high density tracer that is resistant to depletion onto dust grains at low temperatures, has been detected towards the vast majority of sources (96\,per\,cent), has a relatively simple line shape (no extended shoulders or self-absorption features), and the velocity can be assigned unambiguously as only one component is detected towards each source (\citealt{vasyunina2011}). For the remaining 25 sources not detected in N$_2$H$^+$, we have inspected the spectra of the other high density tracers and, if one of the components is found to be significantly brighter than the others then that velocity is assigned to the sources. All of the velocity assignments have also been checked against observations made towards these sources using other high-density tracers (e.g., NH$_3$; \citealt{urquhart2018_csc}) and by comparing the morphology of integrated $^{13}$CO\,(2-1) maps produced of each velocity component from the SEDIGISM survey (\citealt{schuller2017}) with the dust emission ({\color{blue} Urquhart et al. in prep.}).

The brightest 10 transitions are able to trace a large range of physical conditions including cold and dense gas (HNC, H$^{13}$CO$^+$, HCN, HN$^{13}$C, H$^{13}$CN), outflows (HCO$^+$), early chemistry (C$_2$H),  gas  associated with protostars and YSOs (HC$_3$N, and cyclic molecules (C$_3$H$_2$). This gives us a significant amount of scope to search for differences in the chemistry as a function of the evolutionary stage of the star formation taking place within this sample of clumps. 

Before we begin to look at the observed parameters to identify evolutionary trends, it is useful to first take a look at the individual detection rates for the different source types identified. In the last four columns of Table\,\ref{tab:mollines} we give the detection statistics for the five different source types outlined in Sect.\,\ref{sect:classification}. Since the detection rates for the brightest lines are very high, we would not expect to find any significant difference in their detection rates as a function of source type, so we exclude them from this analysis (i.e., HNC, N$_2$H$^+$, C$_2$H, HCO$^+$, H$^{13}$CO$^+$, HCN): these can be considered as good universal gas tracers. We can also discard any transitions where the overall detection rates are relatively small ($<$50) as  the uncertainties associated with these are likely to be larger than any differences observed. 

In Fig.\,\ref{fig:detection_rates_evolution} we present plots showing the detection rates of the thermal lines as a function of the different source classifications. The plots can be broadly grouped into a few distinct types. First, those that display a steady increase in the detection rate as a function of evolution (H$^{13}$CN, $^{13}$CS, SO); these are likely to be correlated with the dust temperature of the clump (linked by red lines). Second, there are two transitions that show an increase in the detection rate for the first three evolutionary stages but then either plateau to the \hii\ region stage  (HC$_3$N) or show a significant decrease (CH$_3$C$_2$H); the increase in the detection rate is also likely to be correlated with increasing dust temperature resulting in increased emission from these molecules. The plateau seen in the HC$_3$N detection rate between the YSO and \hii\ region stages might indicate that the release of these molecules from the grains is in equilibrium with the destruction by uv-photons or chemical reactions, while the decrease in the CH$_3$C$_2$H might indicate that the molecule is being destroyed by uv-radiation. In the third type (linked by blue lines), molecules show a detection rate that is relatively constant for the first three stages but show a marked drop for the \hii\ regions (H$^{13}$CO$^+$, NH$_2$D and N$_2$H$^+$); these molecules are likely to be fairly insensitive to the temperature on clump scales but are sensitive to the uv-radiation from the embedded \hii\ region.

Finally, we find four molecules where the detection rate is highest for the protostellar stage (HNCO and SiO, c-C$_3$H$_2$ and HN$^{13}$C --- linked by a green line in Fig.\,\ref{fig:detection_rates_evolution}); these are all good tracers of the earliest stage of star formation in clumps. \citet{sanhueza2012} has also reported an increase in the  detection rate of HNCO and SiO towards IRDCs with signs of star formation. The  SiO transition is often linked to fast shocks driven by molecular outflows (\citealt{schilke97}), and we find that our detection rates for this molecule are consistent with this.  The detection rate for the protostellar stage is almost 4 times larger than for the quiescent stage, and drops slightly for the YSO and \hii\ region stages.  The significantly higher detection rate of SiO towards the protostellar sources indicates that molecular clumps are associated with outflows at very early stage, even before the clumps become mid-infrared bright. This finding is similar to the observations by \citet{beuther+sridharan2007} and \citet{motte+2007}. SiO as an evolutionary indicator was also discussed in \citet{miet06}, \citet{lopez-sepulcre+2011}, and  \citet{csengeri2016_sio}. 

Shocks have also been linked to the enhancement of the HNCO abundance (\citealt{zinchenko2000,rodriguez-fernandez2010,li2013_hnco}) and the sizes (\citealt{yu2017}) and integrated line intensities (\citealt{zinchenko2000, sanhueza2012}) of HNCO have been found to be similar to the SiO transition, suggesting a similar production mechanism. However, \citet{yu2017} found that the abundances are not well correlated, but this may simply indicate that they trace different parts of the shocked gas. This is supported by the findings of \citet{blake1996}, who found the spatial distribution of HNCO and SiO are significantly different from each other and this may be because the HNCO emission may be enhanced by low-velocity shocks (\citealt{flower1995,martin2008}) as the molecules can be ejected into the gas phase via grain sputtering \citep{sanhueza2012}. \citet{yu2017} also found the emission from HN$^{13}$C, HNCO and SiO to be more compact than the ubiquitously detected HCO$^{+}$, HCN, HNC and N$_2$H$^{+}$, consistent with the presence of an embedded thermal heating source.  HNCO can be destroyed by far-uv photons produced in high-velocity shocks (\citealt{viti2002}) and \hii\ regions, which would explain the decrease in the detection rates as the embedded object evolves towards the main sequence.

\begin{figure}

\includegraphics[width=0.45\textwidth, trim = 30 20 20 20]{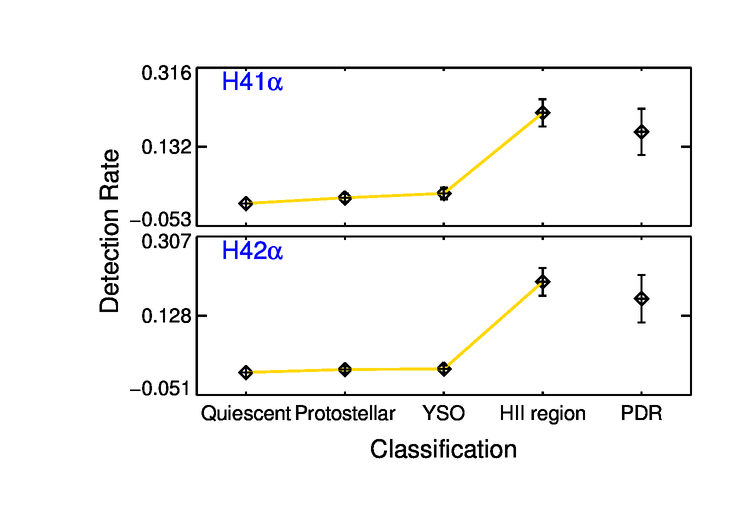}
  \caption{Detection rates for the RRLs towards ATLASGAL clumps are plotted as a function of source classification. The error bars are determined using binomial statistics.}
  \label{fig:rrl_detection_rates_evolution}  
\end{figure}

\subsection{Radio-recombination lines}

The millimetre radio-recombination line (mm-RRL) emission has been the focus of a detailed study presented by \citet{kim2017} and so here we only provide a brief summary of main results. We detected H41$\alpha$ and H42$\alpha$ lines toward 92 (16$\pm$2\,\%) and 84 (15$\pm$1\,\%) clumps targeted in the Mopra data set. They cross-matched the positions of these clumps with radio-continuum catalogues to identify H{\sc ii} regions and determine their properties.

A significant correlation is found between the velocities of the ionized gas traced by mm-RRLs and the molecular gas. This reveals that many of the detected mm-RRLs are emitted from compact H{\sc ii} regions that are still deeply embedded in their natal molecular clumps. H$^{13}$CO$^{+}$ line-widths used for the velocity comparison are additionally found to be significantly broader towards clumps with mm-RRL detections than those towards clumps unassociated with mm-RRLs. They also find that the mm-RRL detections have a good correlation with 6-cm radio continuum emission. This result implies that mm-RRLs and 6-cm continuum emission are both tracing the same ionized regions. 

In Fig.\,\ref{fig:rrl_detection_rates_evolution} we show the detection rates for the two RRLs as a function of evolution. This clearly shows that the detection rates are close to zero for first three evolutionary types but peak sharply for the \hii\ region and PDR classified clumps. Given that the RRLs have already been studied in detail and are only associated with a single evolutionary stage we will not discuss them here and direct the reader to \citet{kim2017} for more details.

\subsection{SiO analysis}

SiO is a commonly used tracer for shocked gas. \citet{csengeri2016_sio} studied a sample of 430 massive clumps selected from the ATLASGAL survey, that are located in the 1\,st Galactic quadrant. The observations have been carried out on the IRAM 30-m telescope in the SiO (2--1) line, and show a high detection rate of 77\,per\,cent in their sample, as well as a $\sim$50\,per\,cent detection rate towards mid-infrared quiet clumps. In total, this adds up to a slightly higher detection rate than found here, however, the IRAM observations are significantly more sensitive. 

In addition, the \citet{csengeri2016_sio} study complemented the SiO\,(2-1) with the higher energy, SiO\,(5-4) transition obtained with the APEX telescope. Excitation analysis of the line-ratios towards a large fraction of sources shows that non-local thermal equilibrium (LTE) effects, as well as varying, distance-dependent beam dilution may bias LTE estimates of SiO column density. Their work finds a significant correlation between the ratio of the SiO\,(5-4) and (2-1) lines and an increasing $L_{\rm bol}/M_{\rm{clump}}$ ratio. These are interpreted as due to a change in excitation conditions with an increasing product of $n$(H$_2$)$T$, i.e., the pressure as a function of the $L_{\rm bol}/M_{\rm{clump}}$ ratio. The $L_{\rm bol}/M_{\rm{clump}}$ ratio is commonly used as a proxy for evolution, although it may also reflect the most massive star forming in the cluster \citep{ma2013}, this suggests that it is the shock condition that changes over time rather than the SiO column density related to the jet activity (c.f.\,\citealp{lopez-sepulcre2011,sanchez-monge2013}). For a proper assessment of the excitation conditions, multiple transitions of this tracer are necessary, and are, therefore, beyond the scope of the current paper.

\begin{figure*}
\includegraphics[width=0.45\textwidth]{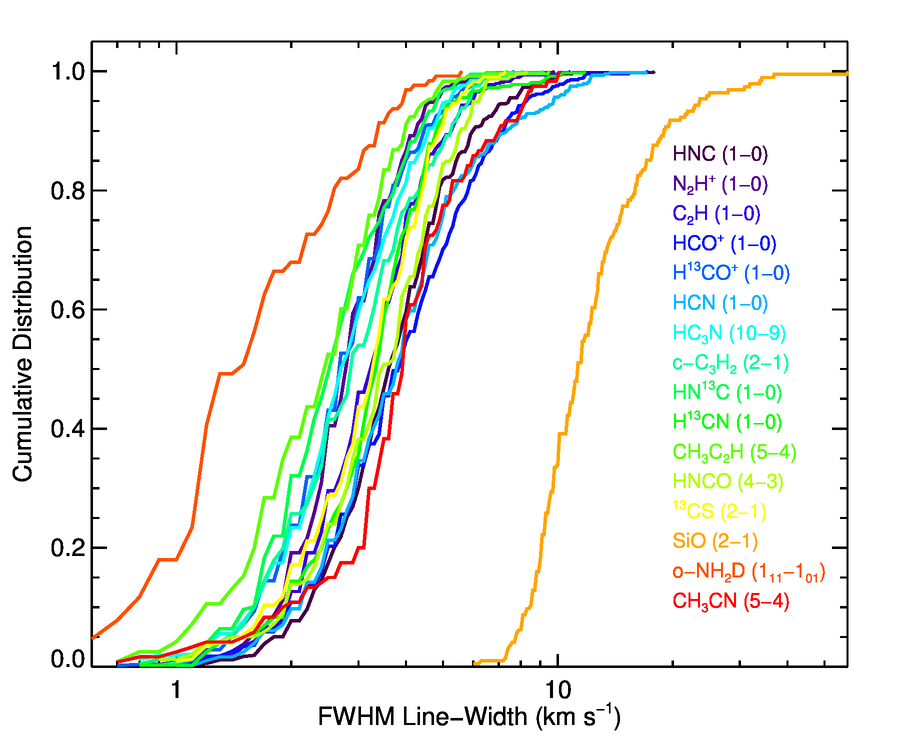}
\includegraphics[width=0.45\textwidth]{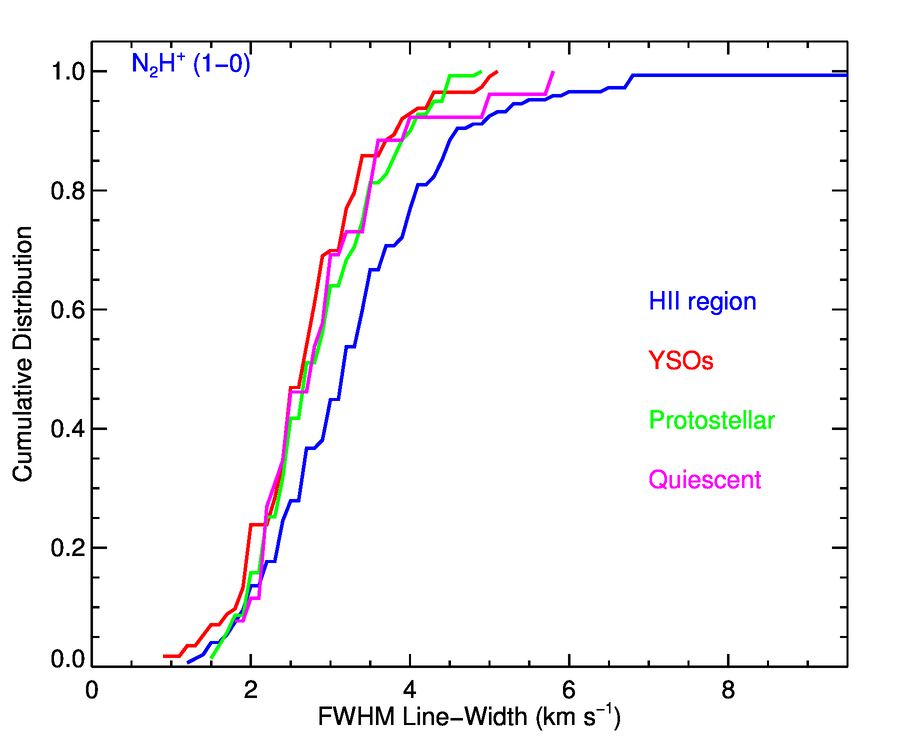}
\includegraphics[width=0.45\textwidth]{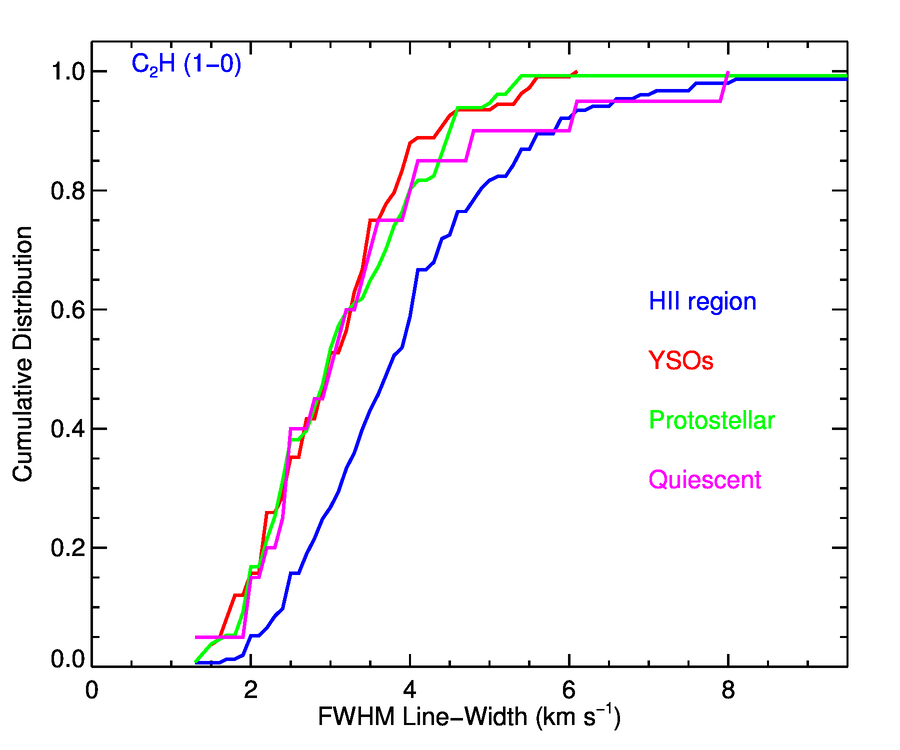}
\includegraphics[width=0.45\textwidth]{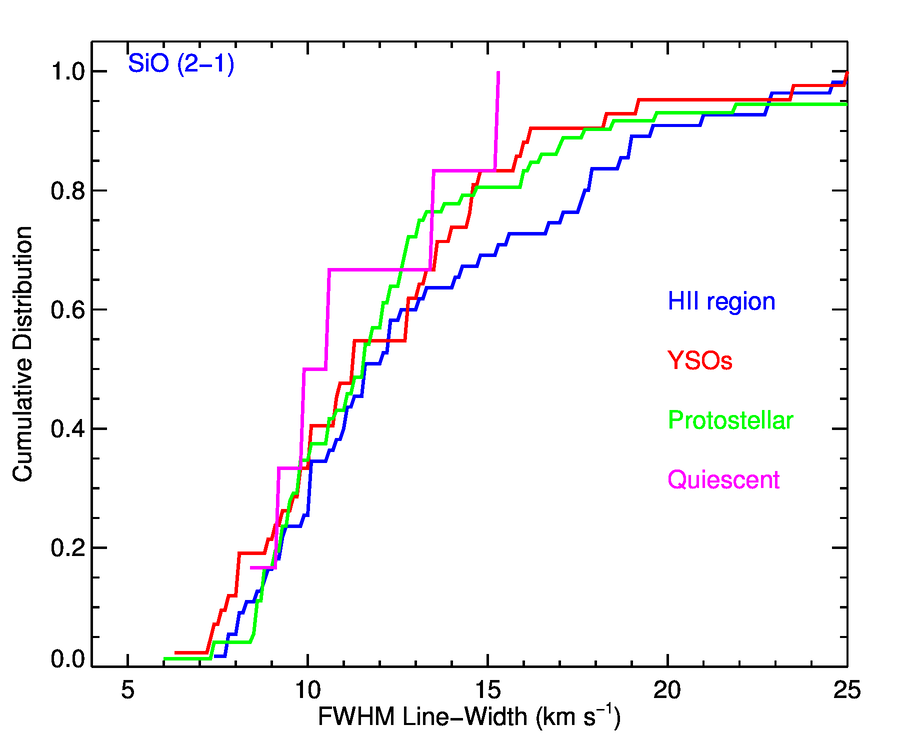}

  \caption{Upper left panel: Cumulative distribution functions of the line-widths for the 16 most detected molecular transitions as determined from Gaussian fits to the emission profiles.
Upper right and lower panels: Cumulative distribution functions of the line-widths and evolutionary stage for three of the six transitions that show a significant difference between the three earliest stages and the \hii\,region stage.}
  \label{fig:show_width_examples}  
\end{figure*}

\subsection{NH$_2$D analysis}

Wienen et al. (2018 subm) has investigated the NH$_3$ deuterium fraction using NH$_2$D lines towards  the same evolutionary sample investigated in this paper. Mopra NH$_2$D observations are combined with NH$_3$ data reported by \citet{wienen2012,wienen2018} to obtain a subsample of 253 sources (36\,per\,cent of our sample) that has been detected with a S/N ratio $>$ 3.  Comparing the distributions of the NH$_2$D and NH$_3$ line-widths, they found a correlation between the two, indicating that the two tracers probe similar regions within a source. They also derived the deuterium fraction of NH$_3$ from the ratio of the NH$_2$D to NH$_3$ column densities. This resulted in a large range in the fractionation ratio (of on order of magnitude) with values up to 1.2, which is the highest NH$_3$ deuteration reported in the literature so far (e.g., \citealt{busquet2010, pillai+2007, crapsi2007}).

Theory predicts a rise in NH$_3$ deuteration with decreasing temperature \citep{caselli2002}. The deuterium fractionation in this sample, however, shows a constant distribution with NH$_3$ line-width and rotational temperature, and we find no significant difference in NH$_3$ deuteration for ATLASGAL subsamples at the various evolutionary stages. We may surmise that high-mass star-forming sites possess a complex temperature structure that prevents us from seeing any trend of deuteration with temperature. A higher fraction of NH$_2$D detections in cold sources, however, still yields a clear difference of deuteration at early evolutionary stages compared to the whole sample (see also \citealt{pillai2011} for a discussion on core scales).

\subsection{Analyses of observed line properties}

\subsubsection{Line-widths}

The observed FWHM line width, $\Delta v_{\rm{obs}}$, is a convolution of the intrinsic line width, $\Delta v_{\rm{int}}$,  the resolution of the spectrometer, $\Delta v_{\rm{channel}}$ (0.9\,\kms) and the instrumental FWHM resolution ($\delta \sim 1.2$; \citealt{longmore2007}). We use the following to obtain the intrinsic line width:

\[
\Delta v_{\rm{int}} = \sqrt{(\Delta v_{\rm{obs}}^2-\left(\delta \Delta v_{\rm{channel}}\right)^2)} 
\]

\noindent The observed molecular lines are only tracing the velocity dispersion of the dense gas. However, to determine the average line-width of gas, $\Delta v_{\rm{avg}}$, we need to take account of the molecular mass of the various tracers. We do this using the following equation taken from \citet{fuller1992}:

\[
\Delta v^2_{\rm{avg}} = \Delta v^2_{\rm{int}} + 8{\rm{ln}}2k_{\rm{B}}T_{\rm{dust}}\left(\frac{1}{\mu_{\rm{mean}}}-\frac{1}{\mu_{\rm{obs}}}\right)
\]

\noindent where $k_{\rm{B}}$ is the Boltzmann constant, $T_{\rm{dust}}$ is the dust temperature, and $\mu_{\rm{mean}}$ and $\mu_{\rm{obs}}$ are the mean molecular mass of the gas (2.33) and the target molecular mass, respectively.

The upper left panel of Fig.\,\ref{fig:show_width_examples} shows the line widths measured from Gaussian fits for the sixteen most highly-detected transitions. The line widths for all but two transitions range between $\sim$2 and 4\,\kms\ (the mean line widths are given in Col.\,9 of Table\,\ref{tab:mollines}). The two transitions that are clearly different are SiO, which is a well known shock/outflow tracer and has the largest measured line widths, and o-NH$_2$D, which has the smallest line widths and also happens to be the most highly detected transition towards quiescent clumps.  o-NH$_2$D is therefore a good tracer of the earliest pre-stellar phase.

We have also compared the line-width distributions for each transition among the four evolutionary stages. We have found no significant differences between the line widths in the quiescent, protostellar and YSO stages for any of the 16 most commonly-detected transitions. We do find, however, significant differences between the YSO and \hii\,region line widths for the following six transitions:  N$_2$H$^+$,  H$^{13}$CO$^+$, c-C$_3$H$_2$,  HN$^{13}$C, HC$_3$N, C$_2$H. For all of these transitions, the \hii\ region line-widths are significantly larger than those of the YSOs. We provide some examples of these in the upper right and lower panels of Fig.\,\ref{fig:show_width_examples}. The first five of these transitions exhibit a sharp fall in the detection rate for \hii\ regions, while the remaining transition indicates a modestly increasing detection rate. The fact that this increase in the line width is only observed for approximately a third of the molecules would suggest this is not due to a bias in the clump sample but is likely to be due to the increased energy output/turbulence injection of the YSOs rather than a change in the dust temperature, as we would expect a change in the temperature to have a gradual effect as a function of evolutionary stage. It is also interesting to note that we observe no significant difference in the SiO line width for the different evolutionary stages.

\begin{figure}
\includegraphics[width=0.45\textwidth]{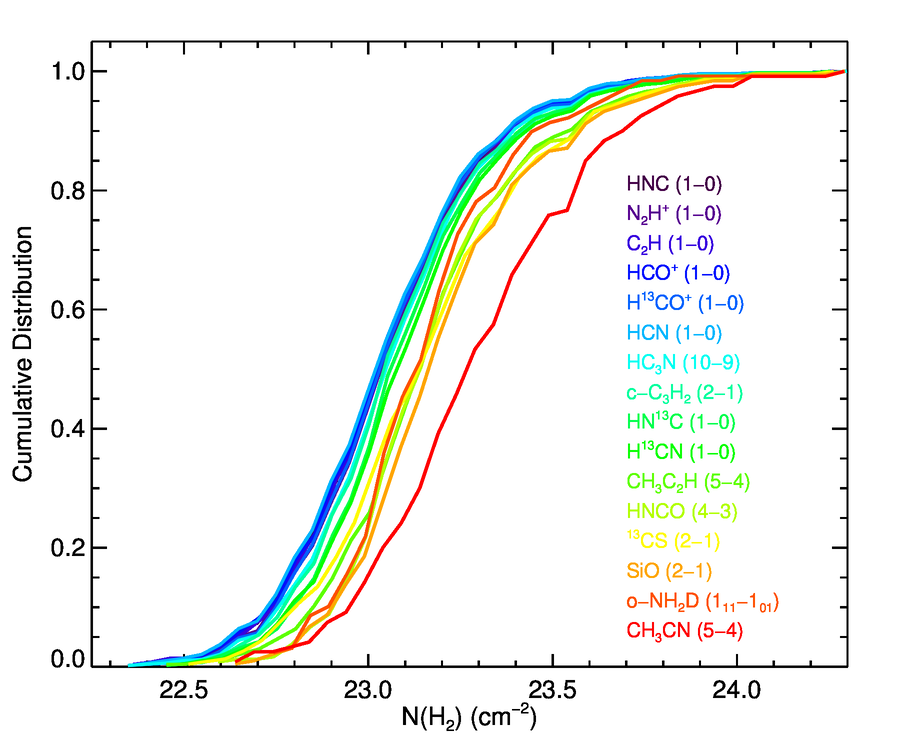}
\includegraphics[width=0.45\textwidth]{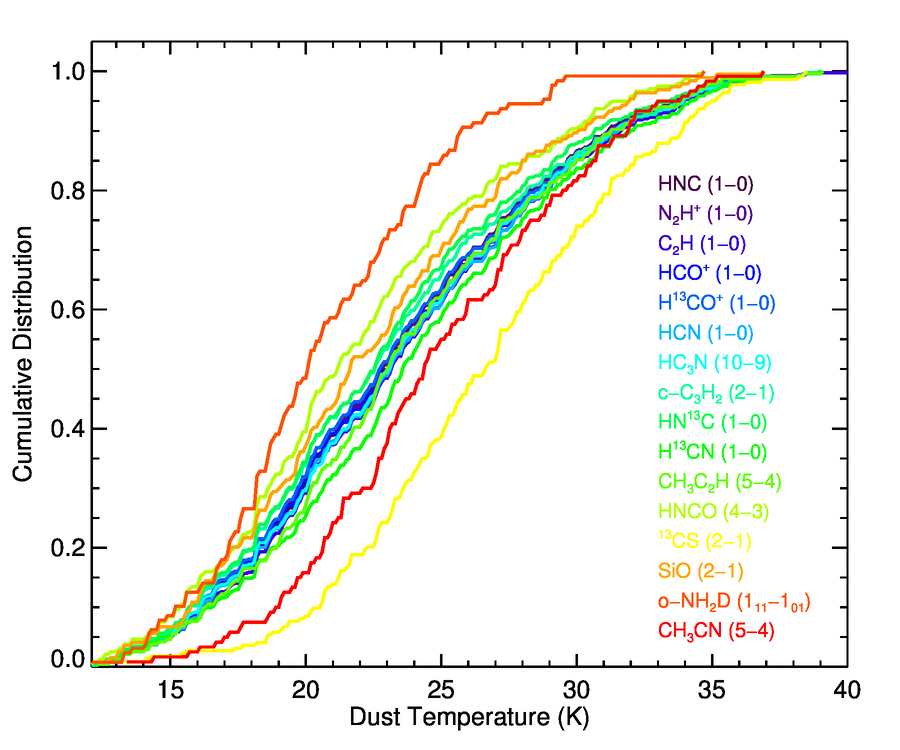}

  \caption{Cumulative distribution functions for the column densities (upper panel) and the dust temperature (lower panel) for the strongest 16 transitions. The names of the various transitions are given to the right of the curves. The H$_2$ column density is calculated from the ATLAGAL dust emission and the dust temperature from the SED fit to the mid-infrared and submillimetre photometry.}
  \label{fig:colden_comparison}  
\end{figure}

\subsubsection{Transitions as a function of H$_2$ column density and dust temperature}

We mentioned previously that the detection rate is highly variable, with some transitions being almost ubiquitously detected while others are only detected towards a handful of clumps. The intensity of the emission from any transition is a function of its abundance and of the excitation conditions. We can assume that the physical conditions are similar for the various transitions, and so comparing the H$_2$ column-density distributions can provide some insight into the abundance of each molecule. We show the H$_2$ column-density distribution, determined from the peak ATLASGAL dust emission, for the 16 strongest detections in the upper panel of Fig.\,\ref{fig:colden_comparison}. All transitions but one (CH$_3$CN) are distributed similarly, with only a $\sim$0.2\,dex spread in the mean H$_2$ column densities. The distribution of the CH$_3$CN transition is the only one that is significantly different from the others and is likely to be the least abundant molecule in the sample.

In the lower panel of Fig.\,\ref{fig:colden_comparison}  we show the cumulative distributions for the dust temperatures determined from the SED fit to the mid-infrared and sub=millimetre photometry as a function of the different transitions. This plot reveals that the dust temperature dependence of all but two transitions are similar: only o-NH$_2$D and $^{13}$CS show any significant deviation from the rest of the transitions. $^{13}$CS has a significantly higher mean dust temperature, and has the steepest increasing detection rate as a function of evolution of all of the transitions observed (a factor of 6). o-NH$_2$D has a lower mean dust temperature and there are no clumps with dust temperatures above 30\,K.  We saw earlier that the detection rate for this transition falls rapidly from the quiescent to the \hii\,region stages (a factor of $\sim$10) and so this molecule is the most closely tied to the earliest stages in our sample. 

We also note that the CH$_3$CN transition deviates from the rest of the transitions at low dust temperatures, but that this deviation becomes less significant as the dust temperature increases. The low dust temperature deviation is likely to result from the fact that this molecule is primarily associated with the highest column density clumps: these tend to be associated with star formation (\citealt{urquhart2014_atlas,giannetti2017}) and so the dust temperature tends to be higher than for the other transitions. CH$_3$CN is thought to be directly desorbed from dust grains, so it is not surprising that you preferentially see it at hotter temperatures (\citealt{thompson2003}). 

The plots presented in Fig.\,\ref{fig:colden_comparison} reveal the range of column densities and dust temperatures to which the various transitions are sensitive, and highlights transitions that stand out in some way (such as o-NH$_2$D predominately tracing colder material and CH$_3$CN tracing high column-densities). This does not provide a very quantitative way to investigate the relationship between the molecular emission and the column density and dust temperature of the gas. In the next subsection we plot the integrated intensity as a function of these two gas properties.

\subsubsection{Integrated intensities}

\begin{figure*}
\includegraphics[width=0.33\textwidth]{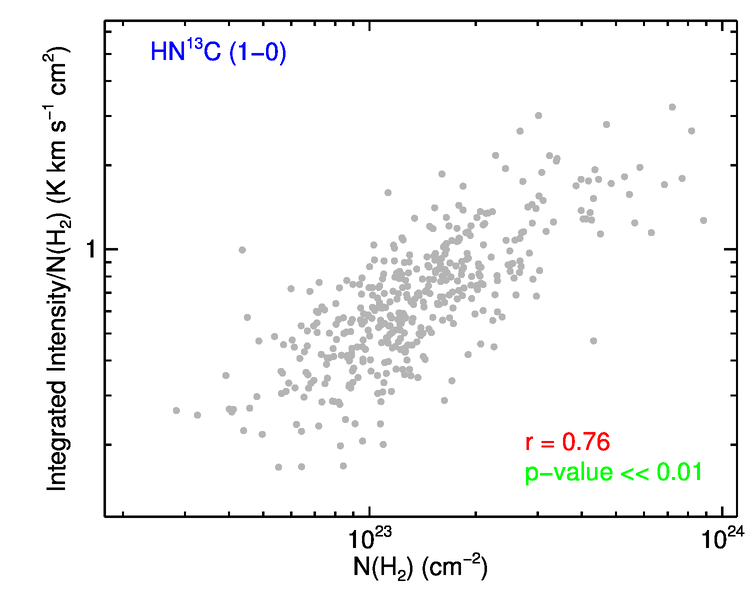}
\includegraphics[width=0.33\textwidth]{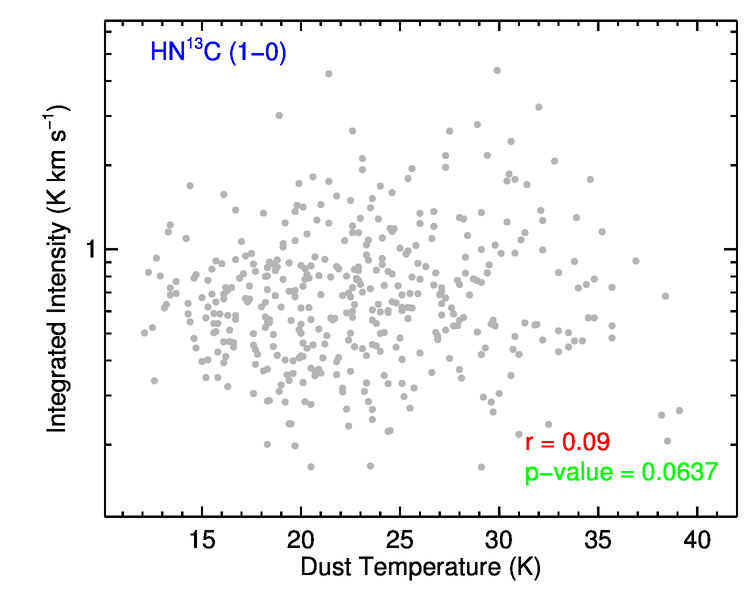}
\includegraphics[width=0.33\textwidth]{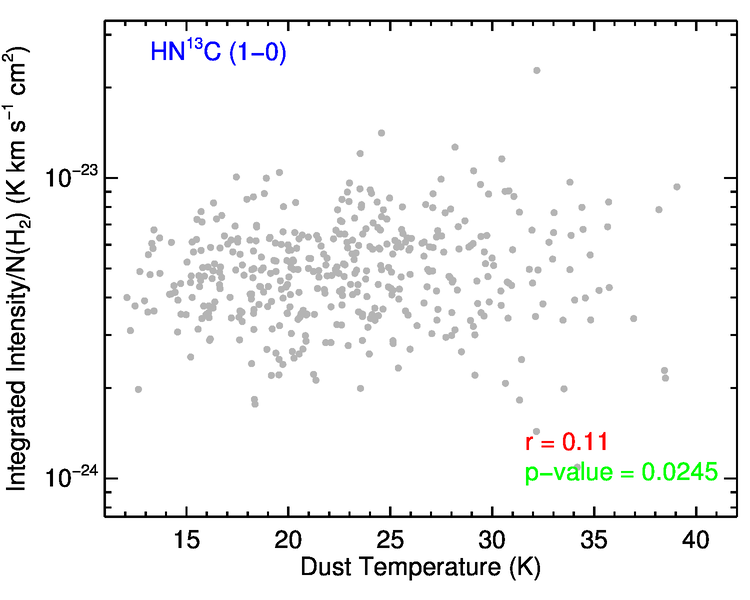}

\includegraphics[width=0.33\textwidth]{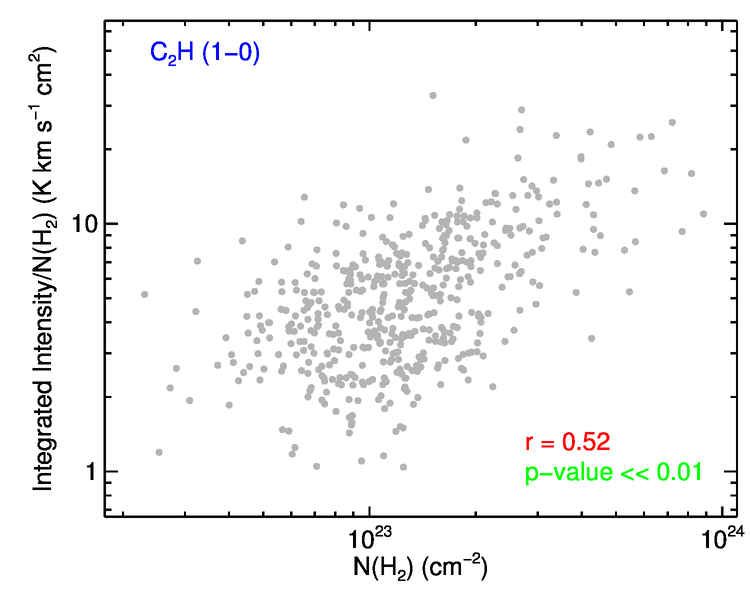}
\includegraphics[width=0.33\textwidth]{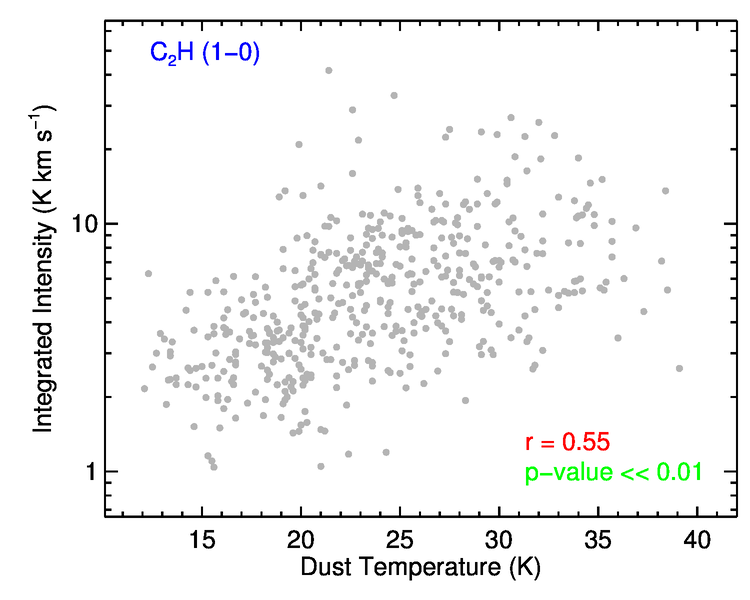}
\includegraphics[width=0.33\textwidth]{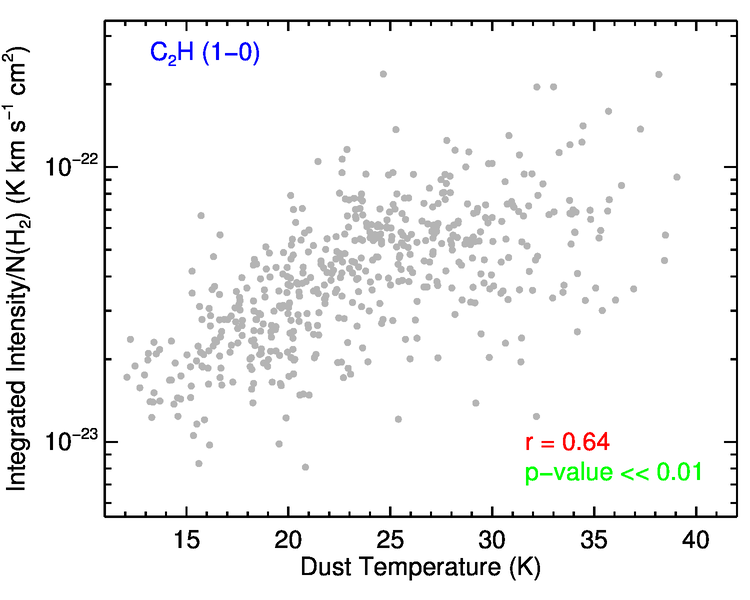}

\includegraphics[width=0.33\textwidth]{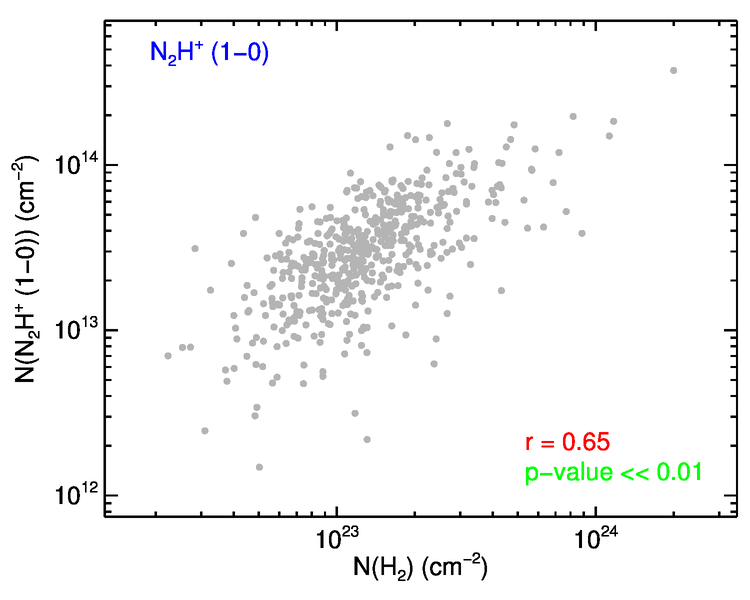}
\includegraphics[width=0.33\textwidth]{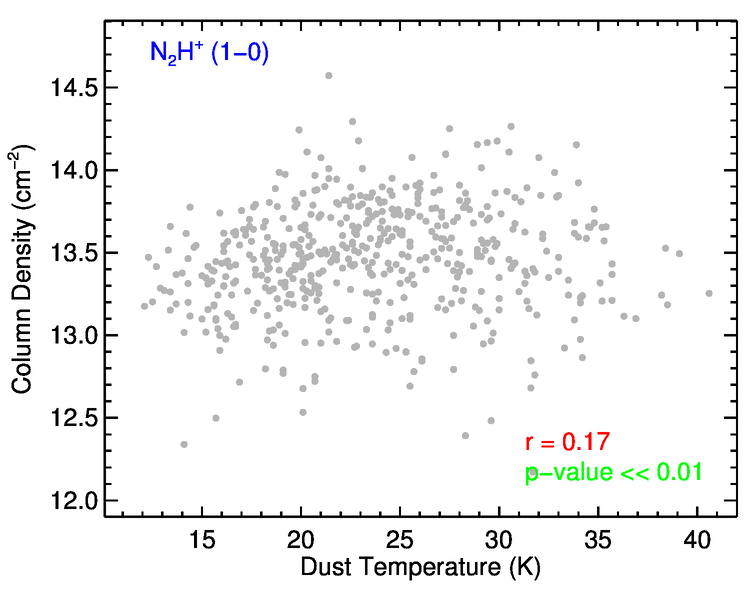}
\includegraphics[width=0.33\textwidth]{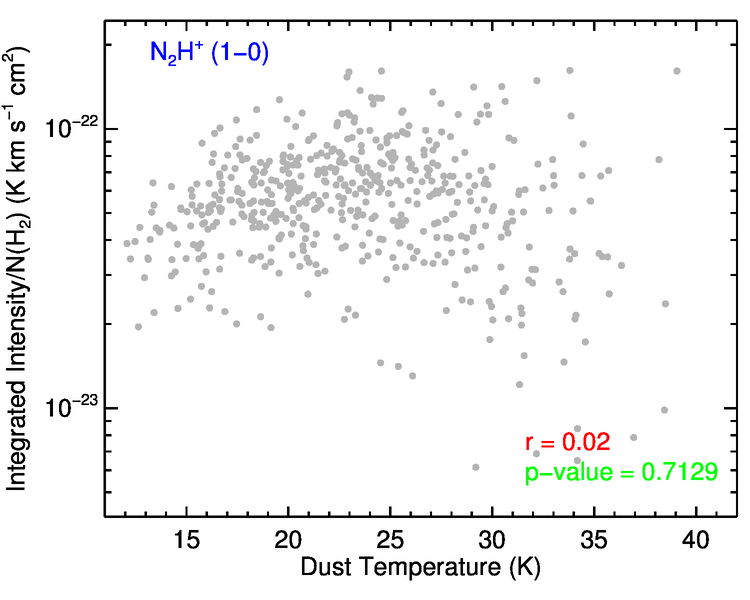}

  \caption{Scatter plots showing the correlation between the observed line intensities and the H$_2$ column density (determined from the dust emission) and dust temperature (determined from a fit to the spectral energy distribution). Here we show the transitions that exhibit strong correlations that are discussed in the text, but we provide these plots for all 16 of the most detected molecules in the Appendix (Fig.\,\ref{fig:appendix_colden_comparison}).}
  \label{fig:colden_temp_comparison}  
\end{figure*}

We have determined the line integrated intensity by summing the emission over the whole velocity range where emission is detected above 3$\sigma$. We present plots of this parameter against the column density and dust temperature for a few transitions in left and middle panels of Fig.\,\ref{fig:colden_temp_comparison}. The dust temperatures  are taken directly from \citet{urquhart2018_csc} but, as mentioned previously, the column densities have been recalculated using the continuum flux within the Mopra beam. In the right-hand panels of this figure we present an estimate of the molecular abundance, which will be explained in the following paragraphs.\footnote{A complete set of plots can be found in Fig.\,\ref{fig:appendix_colden_comparison}.} In all of these plots we give the Spearman correlation coefficients ($r$) and $p$-values in the lower right corners.  Correlations are only considered significant if the $p$-values are lower than 0.0013.

\setlength{\tabcolsep}{6pt}
\begin{table*}

	\caption{Priors used in MCWeeds for CH$_{3}$CCH and CH$_{3}$CN.}\label{tab:priors}

\begin{minipage}{\linewidth}
	
	\begin{tabular}{lcccc}
		\hline
		\hline
		                  & Temperature                            & Column density   & Linewidth            & Velocity     \\
						  & [K]                                    & [log(cm$^{-2}$)] & [km s$^{-1}$]          & [km s$^{-1}$]  \\
		\hline
		Prior             & Truncated normal                       & Normal           & Truncated normal     & Normal       \\
		Parameters (cool) & $\mu = 50$ K                         & $\mu = 14$       & $\mu = 4$            & $\mu=V_{lsr}^{a}$ \\
						  & $\sigma = 15$                          & $\sigma = 1.5$   & $\sigma = 2$         & $\sigma = 1$ \\
						  & $low = 8, high=150$                    &                  & $low = 0.5, high=34$ &              \\
		\hline
	\end{tabular}

	$^{a}$The mean for the radial velocity is obtained from N$_{2}$H$^{+}$.

\end{minipage}
\end{table*}
\setlength{\tabcolsep}{6pt}

We find all of the line integrated intensities of the various transitions are correlated with the  continuum-derived column density at some level ($r=0.41$-0.76 with $p$-values $\ll$0.01), the strongest of which are HN$^{13}$C, CH$_3$CN, H$^{13}$CO$^+$,  N$_2$H$^+$ ($r > 0.7$ for all). In general it is the optically thin isotopologues that have the highest correlation with column density and the optically thick transitions that have the poorest correlation (e.g., HCO$^+$, HNC and HCN),  as expected, although the trends seen in the optically thick and thin transitions of the same species are consistent with each other. 

The correlation between the line integrated intensity and evolution for the HCO$^+$ and HNC transitions ($r=0.25$-0.30, with $p$-values $\ll$0.01) are consistent with the findings of \citet{sanhueza2012}, who found the column densities of these two transitions increased with evolutionary phase. The correlation with dust temperature is significantly weaker ($r=0.18$-0.55 with $p$-values $\ll$0.01) and, surprisingly, the integrated intensities for six transitions show no correlation with temperature at all (NH$_2$D, N$_2$H$^+$, HNCO, SiO, HN$^{13}$C, CH$_3$CN).  We note that five of these are nitrogen-bearing molecules, and all except CH$_3$CN show a decreasing detection rate as a function of increasing evolutionary stage. We have seen that CH$_3$CN has a steeply increasing detection rate and tends to be associated with higher dust-temperature clumps, and so it is a little surprising to find there is not a direct correlation with dust temperature ($r=0.22$ but $p$-values = 0.016). This suggests that the observed line integrated intensities are dominated by the column density, and that the main impact of the feedback from the star formation processes on the local environment is to warm the dust, releasing molecules back into the gas phase and increasing their abundance and, consequently, their detection rate and line integrated intensity; this affects the N-bearing molecules less as they are slow to freeze out onto grain surfaces.

The line integrated intensities divided by the H$_2$ column density should provide a rough estimate of the molecular abundance for each transition. We have plotted this against the dust temperature and find that all but five transitions show a positive correlation, indicating that the abundance of these  molecules is increasing with the  evolutionary stage.

The abundances of HC$_3$N and HCN were also calculated by \citet{vasyunina2011} for a sample of IRDCs and, compared to a sample of high-mass protostellar objects (HMPOs; i.e., \citealt{sridharan2002,beuther+sridharan2007}), were found to be increasing with evolution, which is consistent with our findings. The five transitions that do not show any correlation with increasing dust temperature are o-NH$_2$D, N$_2$H$^+$, HNCO, SiO, HN$^{13}$C: this is effectively the same set with no correlation between line intensity and dust temperature.  CH$_3$CN is the only transition missing from this group; however, the correlation for this transition is weak ($r = 0.29$ with $p$-value = 0.0011).

The transition that has the strongest correlation between abundance and dust temperature is C$_2$H ($r=0.65$  with $p$-values $\ll$0.01), which also happens to be one of the most well-detected molecules. It has a detection rate of 98\,per\,cent for the protostellar through to the \hii\ region stages, but the detection rate lies at $\sim$70\,per\,cent in pre-stellar clumps. The abundance of this transition may provide reliable estimates of the gas temperature, assuming the gas and dust temperatures are correlated, for star-forming clumps, despite the somewhat lower detection rate towards the quiescent clumps.  We note that \citet{vasyunina2011} also calculate the abundance of C$_2$H  and  found it to be 20 times higher than a sample of HMPOs. Since both studies consider C$_2$H to be optically thin,  this disagreement is hard to reconcile, however, their sample size is quite modest (15 IRDCs), and their results have uncertainties of around an order of magnitude and use heterogeneous data sets when comparing the IRDCs and HMPOs, all of which may be factors.

Although N$_2$H$^+$ does not have the strongest correlation with column density ($r = 0.7$), it has a very high overall detection rate (96\,per\,cent), despite a slight drop for \hii\,regions.  As neither its line integrated intensity or abundance is correlated with the dust temperature, it becomes the best tracer of column density. We note that high-resolution studies have reported N$_2$H$^+$ emission being depleted around embedded low-mass protostellar objects in nearby molecular clouds (e.g. \citealt{belloche2004, tobin2013}), however, this is on far smaller physical scales than observed here and any small-scale variations are likely to be averaged out over the clumps.

Our determination that our proxy for the N$_2$H$^+$ abundance is invariant to evolutionary stage is consistent with predictions from the low-mass star-formation models of \citet{bergin1997} and \citet{lee2004}, but disagrees with the findings of \citet{sanhueza2012} who reported a trend for increasing column density with evolutionary stage (a KS-test gives a $p$-value of 0.09\,per\,cent that the quiescent and active clumps are drawn from the same parent population). Given that N$_2$H$^+$ is generally optically thin, the line integrated intensity/$N$(H$_2$) should be proportional to the molecular abundance and so we should expect our results to be consistent with those of \citet{sanhueza2012}; however, we note that the $p$-value calculated by \citet{sanhueza2012} does not meet the 3$\sigma$ threshold we require for a correlation to be considered significant in this paper. \citet{hoq2013} also looked at the fractional abundance of N$_2$H$^+$ compared to HCO$^+$ and found the median abundance ratio increases slightly as a function of evolution; however, they also state that this trend is not statistically significant.

We do find a positive but weak correlation between the HCO$^+$ integrated intensity/$N$(H$_2$) and evolution, in line with  \citet{hoq2013} and \citet{sanhueza2012} but in disagreement with the models of \citet{bergin1997} and \citet{lee2004}.

\subsection{CH$_3$CN and CH$_3$CCH transitions}
\label{sect:k_ladder}

\begin{figure}
\centering
\includegraphics[width=0.45\textwidth]{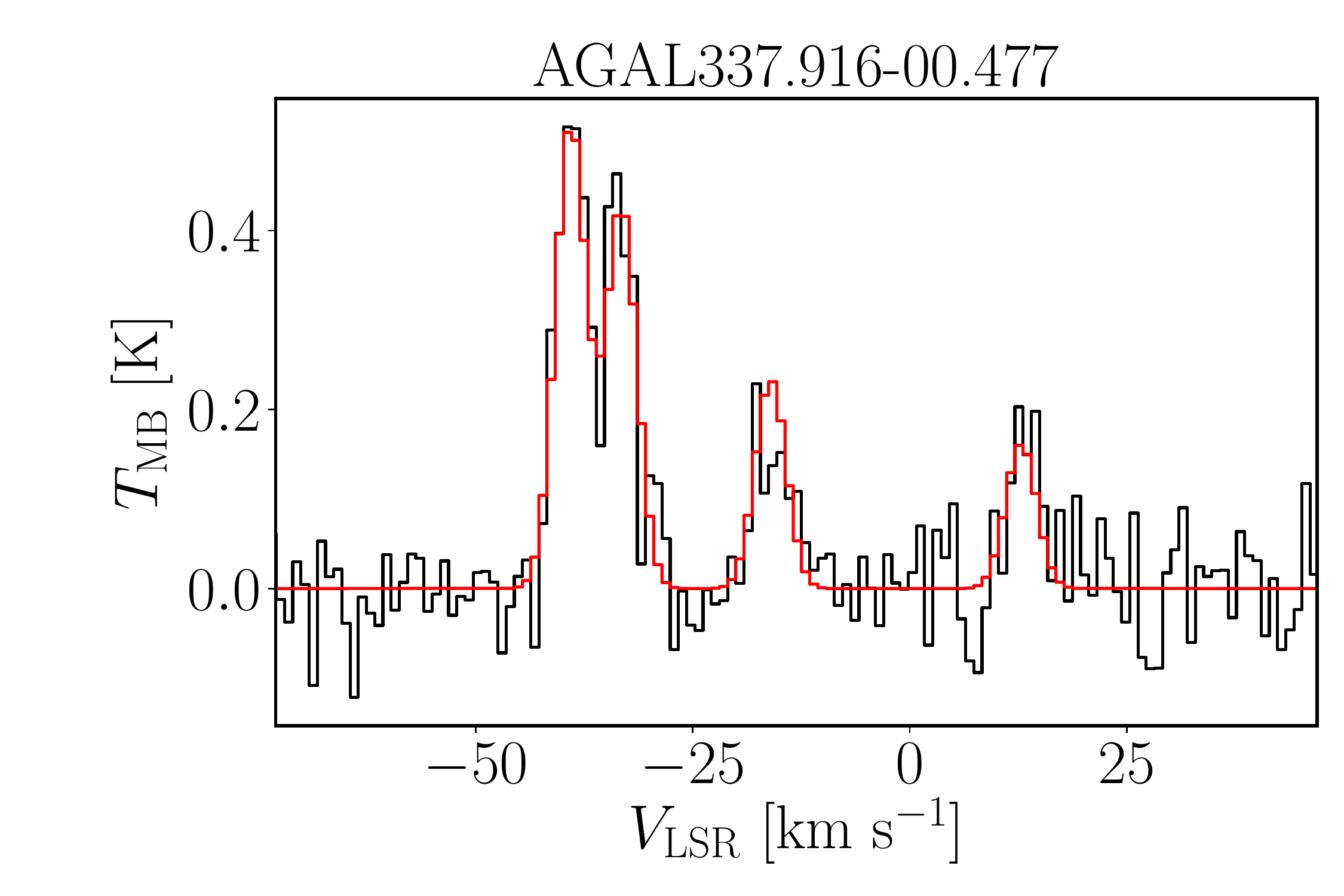}\\
\includegraphics[width=0.45\textwidth]{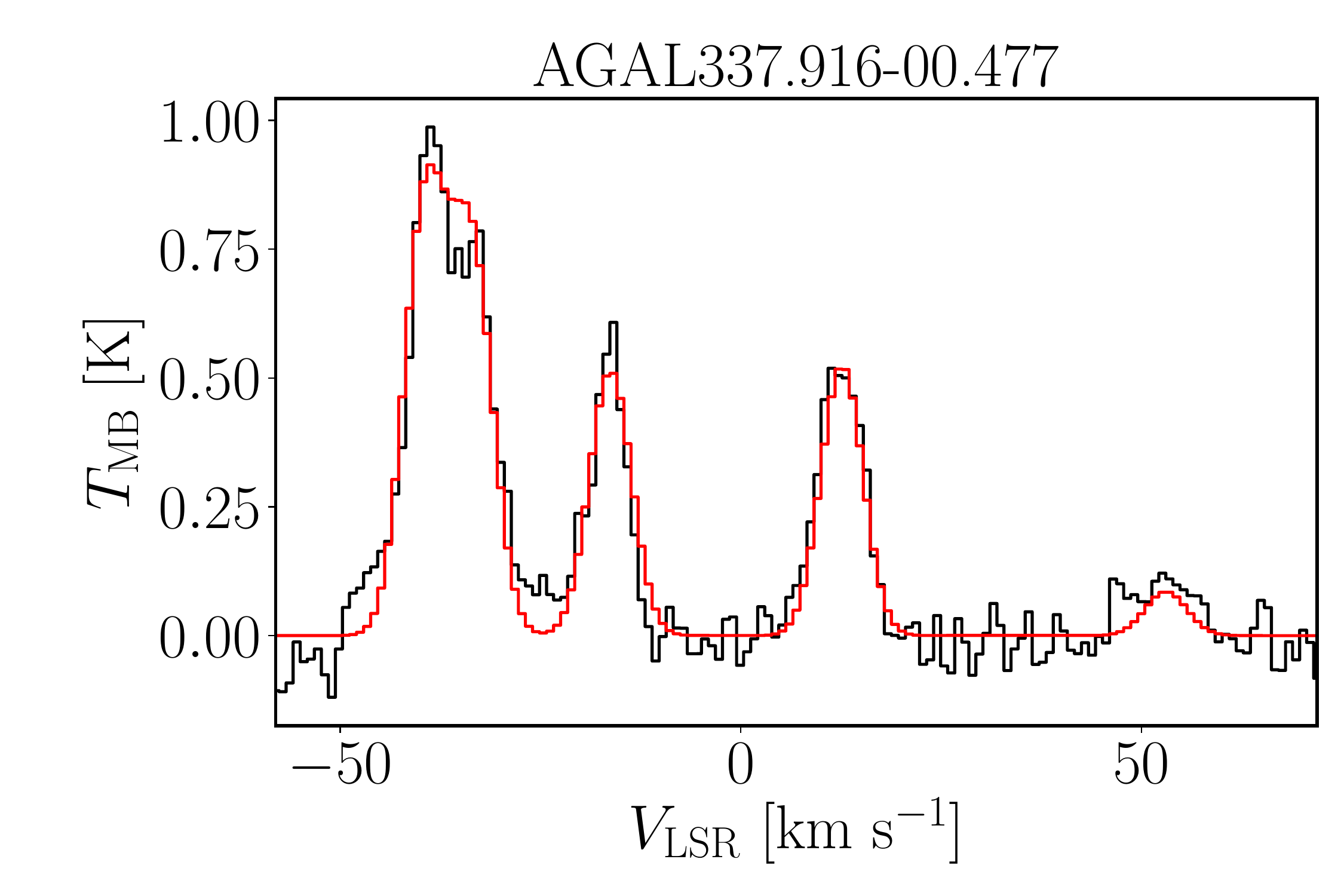}
\caption{Example of the fit performed with \mcweeds\ for one source in the sample. The top panel shows the fit for CH$_3$CCH(5--4), and the bottom one for CH$_3$CN(5--4). The best-fit synthetic spectrum is shown in red.\label{example_mcweeds}}
\end{figure}

\subsubsection{K-ladder fitting procedure}

\begin{figure}
\includegraphics[width=0.45\textwidth]{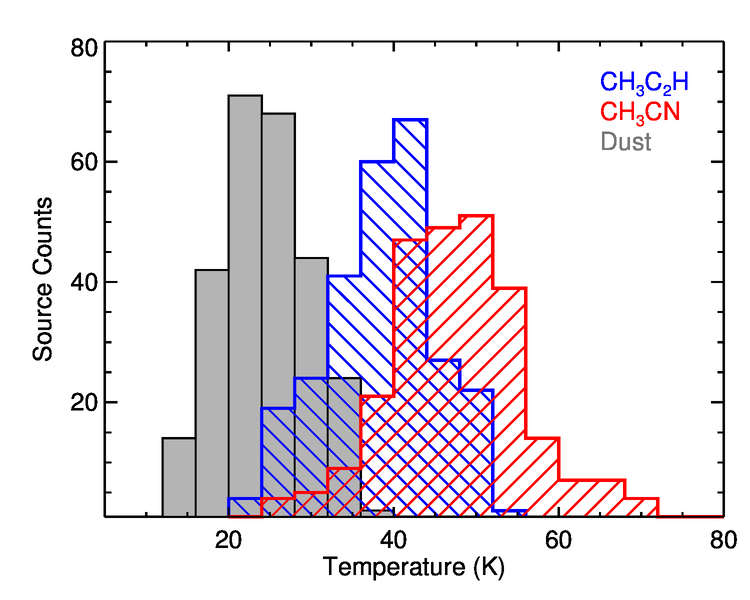}
\includegraphics[width=0.45\textwidth]{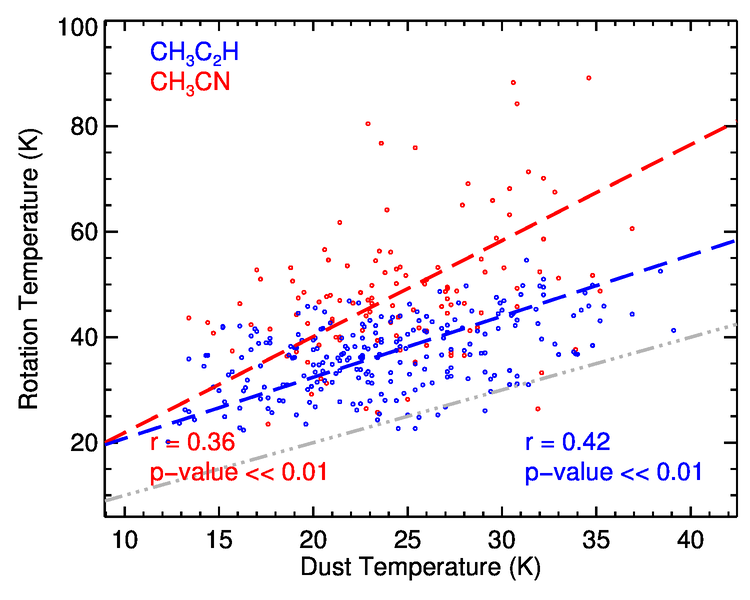}
     \caption{The relationship between the dust temperature and rotation temperatures derived from the CH$_3$CN and CH$_3$CCH transitions. The upper panel shows the distributions of the different temperatures while the lower panel shows the correlation between the two rotation and dust temperatures. The upper panel uses a bin size of 5\,K and different colours are used for each temperature type. The long-dashed lines included in the lower panel show linear fits to the two distributions and the correlation coefficients and $p$-values for the CH$_3$CN and  CH$_3$CCH are given in the lower left and right panels, respectively. The grey dashed-dotted line indicates the line of equality. We have excluded two sources that have rotation temperature greater than 100\,K, as measured from CH$_3$CN. \label{fig:temp-hists} }

\end{figure}  

Several lines from the CH$_3$CN and CH$_3$CCH $K$-ladders are detected in a fraction of sources (excluding weak and cold clumps). To extract rotation temperatures and column densities from the $J = 5\rightarrow4$ multiplets of CH$_3$CCH and CH$_3$CN, we make use of \mcweeds\ \citep{giannetti2017}. \mcweeds\ is partially based on \weeds\ \citep{maret2011}, and combines Bayesian statistical models and fitting algorithms from \texttt{PyMC} \citep{patil2010}, and computes synthetic spectra in an arbitrary number of spectral ranges and lines under the assumption of local thermodynamic equilibrium (LTE).

We fit the 5\,$K$-transitions of CH$_3$CCH and CH$_3$CN following the approach described in \citep{giannetti2017}, i.e., adopting a size for the emitting region that depends on the bolometric luminosity of the source and on the rotation temperature traced \citep[see Eqn.\,5 in][]{giannetti2017}. We use Monte Carlo Markov Chains as the fitting algorithm, with the priors described in Table\,\ref{tab:priors}, and 50000 iterations, a burn-in period of 10000 iterations, a delay for adaptive sampling of 5000 iterations, and a thinning factor of 10. One example for each of the two species is shown in Fig.\,\ref{example_mcweeds}, where we indicate the best fit in red.

Histograms of the derived rotation temperatures from these two transitions and the dust temperature derived from the SED are shown in the upper panel of Fig.\,\ref{fig:temp-hists}. The mean temperatures for the dust, CH$_3$CCH and  CH$_3$CN transitions are 24.6$\pm$5.5\,K, 38.3$\pm$6.7\,K and 48.8$\pm$12.7\,K, respectively, where the uncertainties given are the standard deviations rather than the standard errors on the mean. The difference in the rotation temperatures of these two molecular transitions is likely to be due to different beam filling factors, with the CH$_3$CN emission arising from warmer and denser (CH$_3$CN has a higher critical density than CH$_3$CCH) material  closer to the embedded object and is released at higher temperatures.

\begin{figure}
\includegraphics[width=0.45\textwidth]{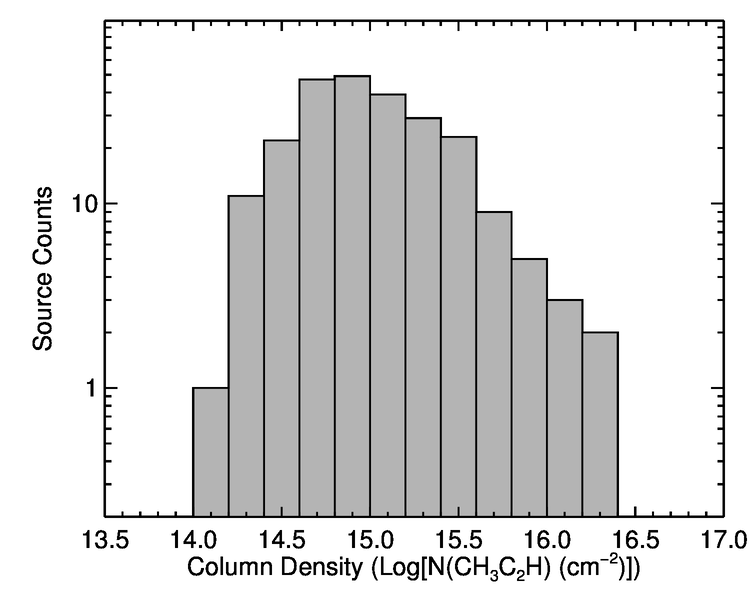}

\includegraphics[width=0.45\textwidth]{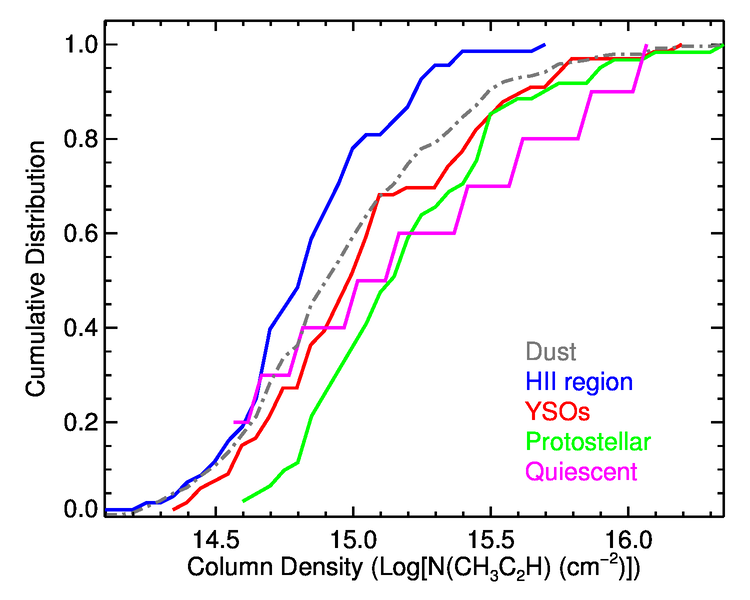}
\includegraphics[width=0.45\textwidth]{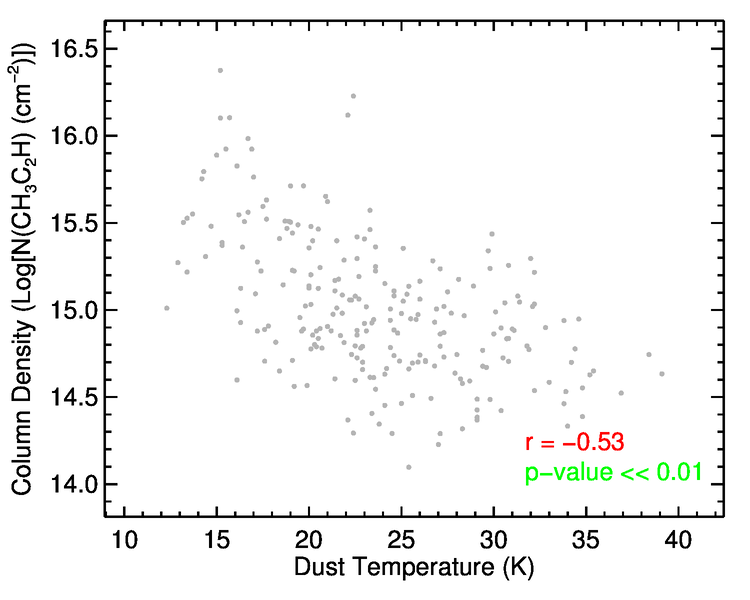}

     \caption{Column density and abundance distributions for CH$_3$CCH. The upper panel shows the column density of the whole sample of 240 detections; the bin size used is 0.2\,dex. The middle shows the column density broken down in to the four evolutionary stages; the colours of the curves match the colours of the sample names given in the lower right corner of the plot. The lower panel is a scatter plot showing the relation between the fractional abundance of CH$_3$CCH relative to H$_2$ and the dust temperature; in the lower right corner we give the Spearman coefficient and associated $p$-value.
       \label{fig:ch3cch_colden} }

\end{figure} 

\subsubsection{Rotation temperatures, column densities and abundances}

The lower panel of Fig.\,\ref{fig:temp-hists} shows the relationship between rotation temperature derived from CH$_3$CCH and CH$_3$CN (5--4) and the dust temperature. This plot reveals that temperatures derived from the dust and the two molecular transitions are correlated. It is clear that there is a better correlation between the CH$_3$CCH transition than the CH$_3$CN, which is significantly more scattered, due to the superposition of different temperature components. The Spearman correlation test reveals a moderate correlation between both sets of rotation temperatures and the dust temperature, and the correlation coefficients are quite similar ($r= 0.36$ for CH$_3$CN and $r= 0.42$ for CH$_3$CCH).  The poor correlation between the rotation and dust temperatures is likely to be due to a mix of different filling factors and the use of too simplistic a model. \citet{giannetti2017} found in their Top100 sample that  the emission from the hot core significantly contaminates the 3-mm transitions and, in general, the low-K transitions start to be fainter for higher  rotation temperatures in the warm gas that dominates at 3-mm. 

In addition to deriving the rotation temperatures, the multiplet fitting can also determine the column densities of the transition, which can be combined with the H$_2$ column densities (derived from the dust emission) to estimate the molecular abundance. We show the column density and abundance distributions for CH$_3$CCH in Fig.\,\ref{fig:ch3cch_colden}. The upper panel of this plot shows the CH$_3$CCH column density distribution of the full sample, while in the middle panel we show the distribution of the three evolutionary subsamples separately as cumulative distributions. Inspection of this plot suggests decreasing CH$_3$CCH column density as a function of evolution; however, the KS test is unable to reject the null hypothesis that the samples are drawn from the same parent population and therefore more data are required before this trend can be substantiated in this way. We show the relation between the column density and dust temperature in the lower panel of Fig.\,\ref{fig:ch3cch_colden},  which reveals a strong negative correlation between these parameters. 

\begin{figure}

\includegraphics[width=0.45\textwidth]{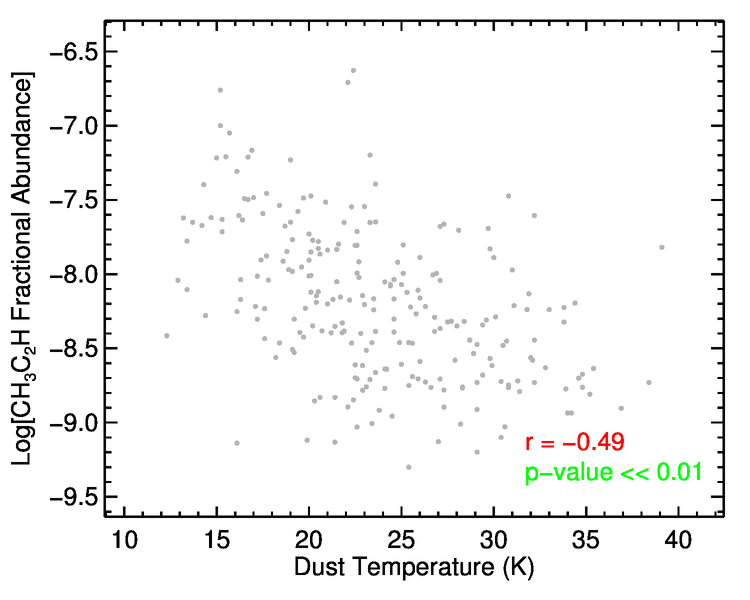}

     \caption{Scatter plot showing the relation between the fractional abundance of CH$_3$C$_2$H and the dust temperature; in the lower right corner we give the Spearman coefficient and associated $p$-value.
       \label{fig:ch3cch_abundance} }

\end{figure}

Since the dust temperature is tightly correlated with the evolution of the embedded sources, it seems clear that the CH$_3$CCH column density decreases as the central source evolves. It is tempting to link this decrease in the CH$_3$CCH column density to a more general trend of decreasing H$_2$ column density in the clumps as a direct consequence of feedback from the evolving protostars. However, we have already found that the H$_2$ column density does not change significantly for the various evolutionary stages, and so this decrease in CH$_3$CCH column density is more likely to be linked to a decrease in abundance. In Fig.\,\ref{fig:ch3cch_abundance} we plot the fractional abundance of the CH$_3$CCH molecule (i.e., $N$(CH$_3$CCH)/$N$(H$_2$)): this plot reveals a trend for decreasing fractional abundance as a function of increasing dust temperature. This would indicate that this molecule is destroyed as the central source evolves and heats up its natal clump. We also note that the detection rate for this transition decreases for the most evolved stage, which is consistent with the trend found in the fractional abundance distribution.

Although we have only shown and discussed the results for the CH$_3$CCH transition, the results are broadly consistent for both transitions.  The correlations are significantly poorer for the CH$_3$CN transition due to the lower detection rate (121 compared to 240 CH$_3$CCH detections). The CH$_3$CN column density has a range between log($N$(CH$_3$CN)) = 14.10-16.62\,cm$^{-2}$ with a mean of 15.01$\pm$0.41, where the uncertainty given is the standard deviation. There is a weak correlation between the column density and dust temperature ($r=-0.31$ with a $p$-value of 0.0006); however, unlike the CH$_3$CCH molecule, the CH$_3$CN fractional abundance is not correlated with the dust temperature ($r=-0.28$ with a $p$-value of 0.0019, although this might  be due to a smaller sample).

CH$_3$CCH and CH$_3$CN are molecules particularly sensitive to the warm-up process \citep{giannetti2017}. Both species trace temperatures considerably warmer than those traced by dust, indicating that their emission comes mainly from dense internal layers of the clump, consistent with the results obtained for the ATLASGAL TOP100 sample \citep{giannetti2014,konig2017}.

\begin{figure}
\centering
\includegraphics[width=0.45\textwidth]{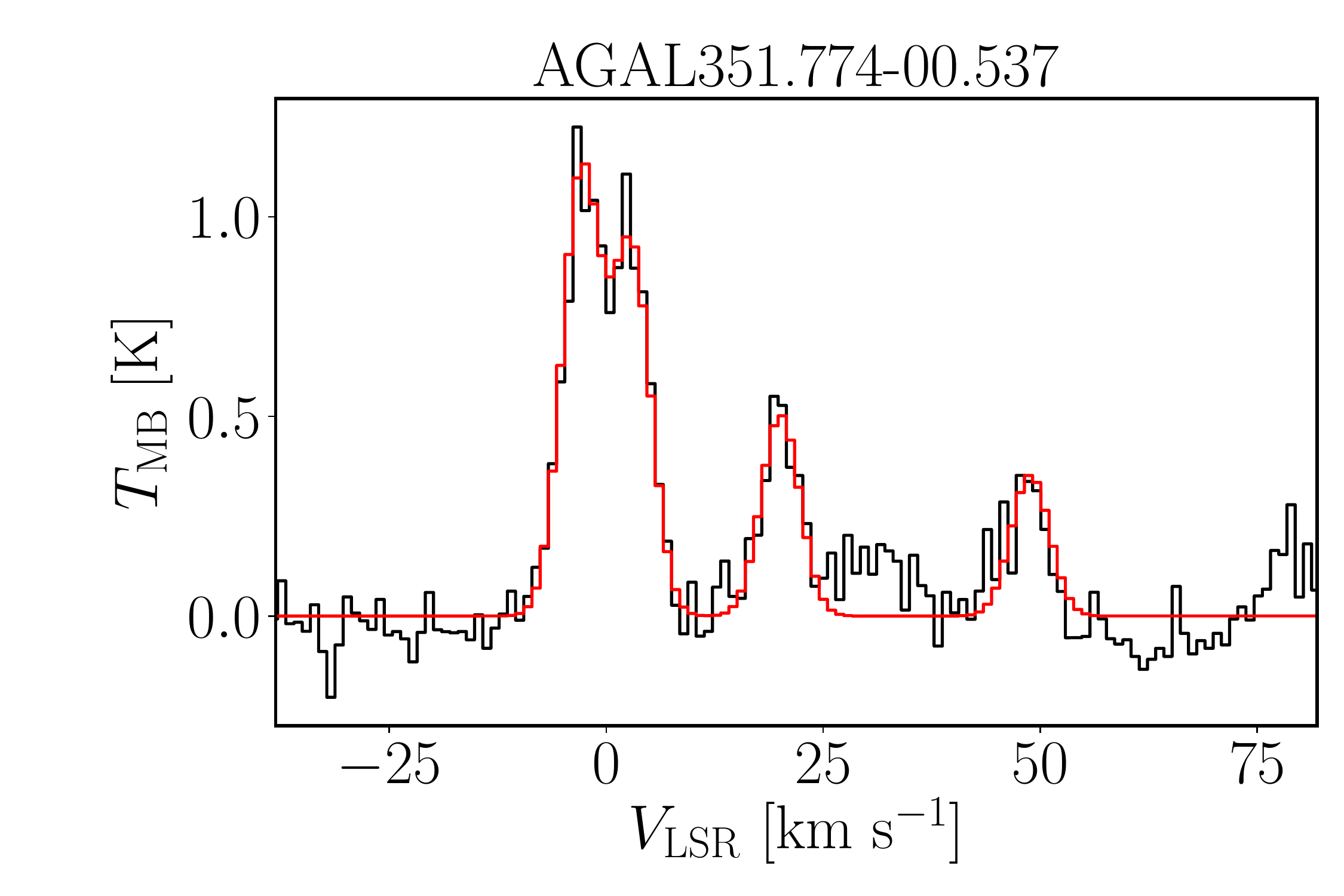}\\
\includegraphics[width=0.45\textwidth]{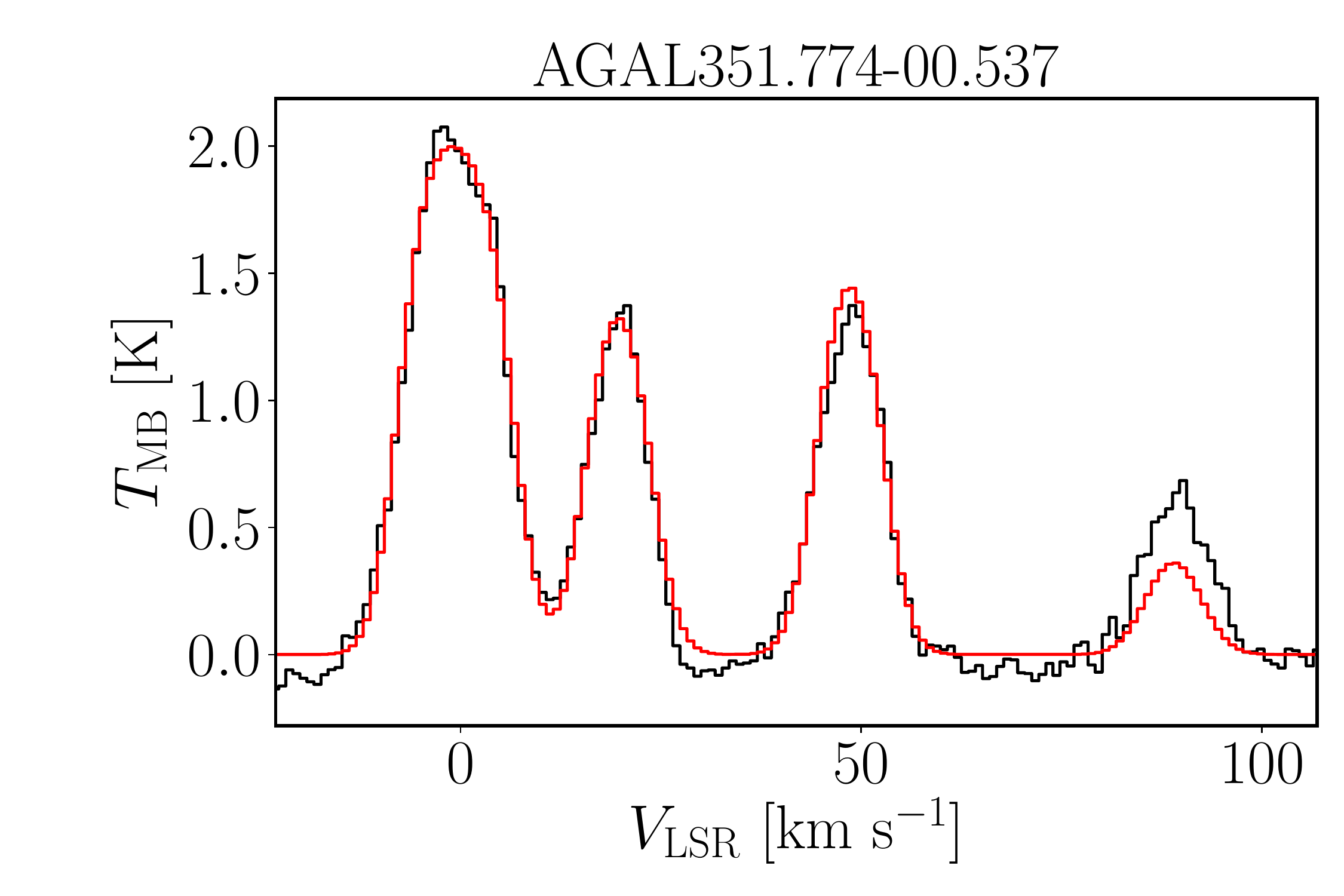}
\caption{Comparison between the fit of CH$_3$CCH(5--4) (top panel) and CH$_3$CN(5--4) (bottom panel) for a known, line-rich hot core. The best-fit synthetic spectrum is shown in red.\label{fig:ch3cn_excess}.}
\end{figure}

The observed range of CH$_3$CCH rotation temperatures is consistent with both the observations of \citet{miet06} towards a sample of clumps harbouring masers and strong SiO emission, as well as that observed for the TOP100 sample. This species is found to show a smooth increase of rotation temperature with evolution, consistent with passive heating from high-mass YSOs, for $L/M\gtrsim2-10\,\,\mathrm{L_\odot \, M_\odot^{-1}}$ \citep{molinari2016,giannetti2017}. CH$_3$CN probes warmer gas in the clumps, also evident from the larger CH$_3$CN line-widths, with a significantly larger scatter. This molecule also has a steeper slope in the dust vs. rotation temperature plot (Fig.\,\ref{fig:temp-hists}). The observed properties of CH$_3$CN can be explained by a much greater abundance in the hot gas surrounding the YSOs caused by the massive release of this species from the icy mantles on the dust grains \citep[e.g.][]{garrod2008}. This may seem in contrast with the findings on CH$_3$CN abundance, but this component is not accounted for in the fitting procedure: it has a small filling factor, because hot cores are very compact. When high-excitation lines are observed and this component can be constrained, the abundance of CH$_3$CN in the hot gas is estimated to be nearly two orders of magnitude higher than in the warm gas \citep{giannetti2017}. Emission from the hot material significantly contributes to the transitions of the 3-mm multiplet, especially those with the highest quantum number $K$, which impacts the temperature determination. Indeed, known hot cores show an excess of emission in the highest excitation lines that are  not reproduced by a single temperature model (Fig.\,\ref{fig:ch3cn_excess}). A similar excess is not present in the spectra of CH$_3$CCH, revealing that the molecule has a different behaviour compared to CH$_3$CN, with no sign of such an important increase in abundance in hot gas. If CH$_3$CCH is predominantly formed onto dust grains, it must therefore be efficiently released already at low temperatures or reprocessed before the sublimation of the ice mantle \citep{giannetti2017}.

\subsubsection{Correlation with quiescent clumps}
\label{sect:protostellar_quiescent_clumps}

\begin{figure*}
\centering
\includegraphics[width=0.45\textwidth]{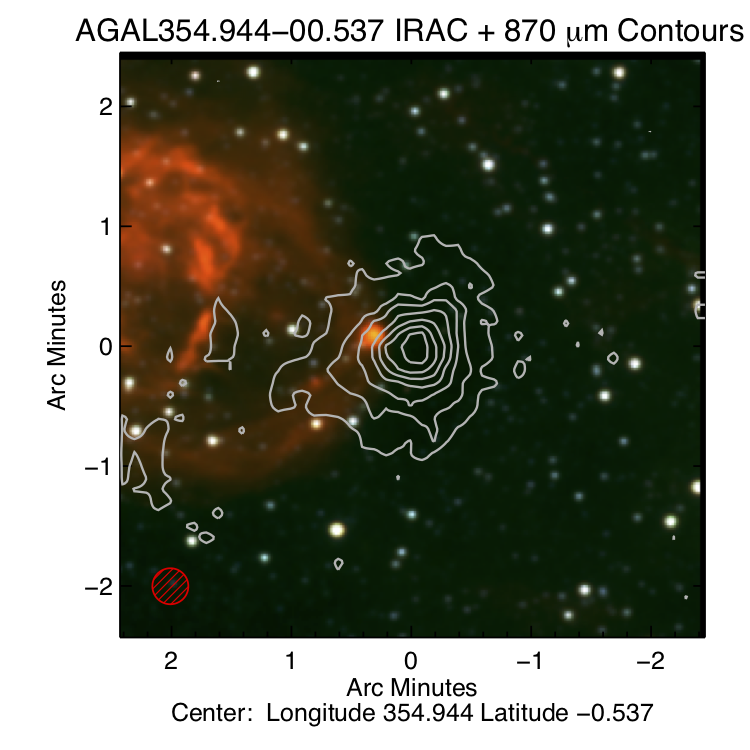}
\includegraphics[width=0.45\textwidth]{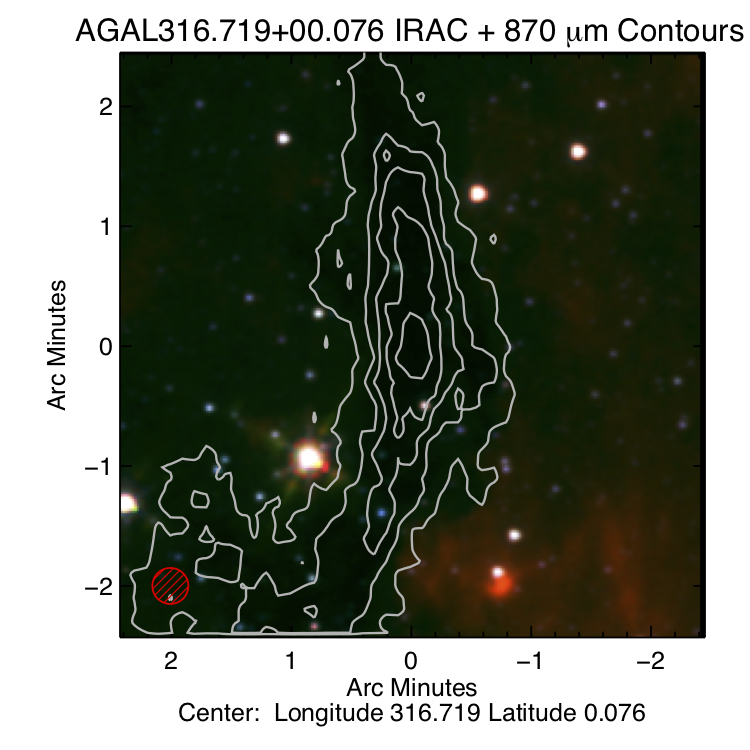}

\caption{Examples of two of the clumps classified as quiescent that are found to be associated with CH$_3$C$_2$H emission, which suggests they may host a young protostellar object. The background image is a three colour composite IRAC image and the contours trace the 870\,\mum\ dust emission mapped by the ATLASGAL survey.  The images are centred on the peak dust emission of each ATLASGAL source and the Mopra beam size is indicated by the hatched red circle shown in the lower left corner of each image.   \label{fig:quiescent}}
\end{figure*}

The rotation temperatures determined for most of the clumps are significantly higher than the dust temperature, suggesting that star formation is taking place in all of the clumps toward which CH$_3$CN and/or CH$_3$CCH is detected. It is, therefore, interesting to note that we find CH$_3$CCH emission towards eleven of the clumps classified as quiescent, three of which are also associated with either CH$_3$CN and/or SiO emission, which are themselves associated with higher rotational temperatures and shocked gas, respectively. The lower detection rate for CH$_3$CN is because it is tracing higher density (see Fig.\,\ref{fig:colden_comparison}) and hotter material (see Fig.\,\ref{fig:temp-hists}) than the CH$_3$CCH transition. The rotation temperatures for these quiescent sources range between 30 and 50\,K, which is much higher than normally found for starless clumps ($\sim 15-20$\,K) and therefore strongly suggest the presence of an internal heating protostellar object. We also find that the quiescent clumps associated with CH$_3$CCH have a significantly higher average $L/M$ than the other quiescent clumps ($33.3\pm9.5$\,\lsun/\msun\ compared to $9\pm1.8$\,\lsun/\msun). We present all of these sources in Table\,\ref{tab:very_young_protostars} and indicate the transitions that are detected towards each of them.

\setlength{\tabcolsep}{4pt}
\begin{table}


	\caption{Sample of very pre-stellar clumps showing emission of CH$_3$C$_2$H.}\label{tab:very_young_protostars}

\begin{minipage}{\linewidth}

	\begin{tabular}{lcccl}
		\hline
		\hline
CSC Name		                  & CH$_3$C$_2$H                            & CH$_3$CN  & SiO & Notes$^1$          \\
\hline
AGAL316.719+00.076	&	X	&		&	& Quiescent	\\
AGAL326.652+00.619	&	X	&	X	&	X & Quiescent/PDR	\\
AGAL331.029$-$00.431	&	X	&		& &Quiescent/PDR		\\
AGAL331.639+00.501	&	X	&		&	&Quiescent	\\
AGAL333.014$-$00.521	&	X	&		&	&Quiescent/PDR	\\
AGAL333.071$-$00.399	&	X	&		&	&Quiescent/PDR	\\
AGAL333.524$-$00.269	&	X	&		&	&Quiescent/PDR	\\
AGAL338.459+00.024	&	X	&		&	X & Quiescent	\\
AGAL351.466+00.682	&	X	&	X	&	X	& Quiescent/PDR\\
AGAL351.571+00.762	&	X	&		&	&Quiescent/PDR	\\
AGAL354.944$-$00.537	&	X	&		&	&Quiescent/PDR	\\
\hline

	\end{tabular}

\end{minipage}
{$^1$ Clumps are classified as Quiescent if there is no compact 70\,$\mu$m emission coincident with the peak submm emission and PDR if there is extended 8\,$\mu$m emission within the 3$\sigma$ submm emission contour (see examples given in Fig.\,19).}

\end{table}
\setlength{\tabcolsep}{6pt}

In Fig.\,\ref{fig:quiescent} we present mid-infrared images of two examples of quiescent clumps that are associated with CH$_3$C$_2$H and CH$_3$CN. Inspecting these images reveals that the majority are located near evolved \hii\ regions and so the observed line emission may include a contribution from the PDR excited by the \hii\ region (an example of these is presented in the left panel Fig.\,\ref{fig:quiescent} and these are identified in the final column of Table\,\ref{tab:very_young_protostars}). However, we also  find three sources that appear to be genuinely quiescent (right panel of Fig.\,\ref{fig:quiescent}) and so, although we cannot rule out the possibility that the emission from the higher temperature gas is due to the presence of the PDR in the sidelobe of the telescope beam, it cannot explain the emission seen in every case.

Although the available evidence is rather circumstantial, it does support the hypothesis that some of the quiescent clumps associated with CH$_3$C$_2$H are actually in a very early protostellar stage. These detections may, therefore, indicate that  10-30\,\,per\,cent of the quiescent clumps are harbouring protostellar objects that are still so extremely young and deeply embedded that they remain dark even at 70\,\mum. This may lower the upper limit for the fraction of quiescent clumps to only a few per\,cent. Further observations are required to confirm the nature of these clumps and investigate their properties.

\setlength{\tabcolsep}{6pt}
\begin{table*}

\begin{center}\caption{Summary of physical properties determined from the $K$-ladder fitting to the four evolutionary subsamples identified. In Col.\,(2) we give the number of clumps in each subsample, in Cols.\,(3-5) we give the mean values, the error in the mean and the standard deviation, in Cols.\,(6-8) we give the median and minimum and maximum values of the samples. }
\label{tbl:derived_para_all}
\begin{minipage}{\linewidth}
\small
\begin{tabular}{lc......}
\hline \hline
  \multicolumn{1}{l}{Parameter}&  \multicolumn{1}{c}{\#}&	\multicolumn{1}{c}{$\bar{x}$}  &	\multicolumn{1}{c}{$\frac{\sigma}{\sqrt(N)}$} &\multicolumn{1}{c}{$\sigma$} &	\multicolumn{1}{c}{$x_{\rm{med}}$} & \multicolumn{1}{c}{$x_{\rm{min}}$}& \multicolumn{1}{c}{$x_{\rm{max}}$}\\
\hline
Log[$N$(CH$_3$C$_2$H) (cm$^{-2}$)] &           240&15.01&0.03 & 0.41 & 14.95 & 14.10 & 16.38\\
\cline{1-1}
\hii\ region &          68&14.87&0.04 & 0.30 & 14.86 & 14.10 & 15.71\\
YSO &           66&15.09&0.05 & 0.41 & 15.02 & 14.34 & 16.23\\
Protostellar &           61&15.23&0.05 & 0.38 & 15.18 & 14.60 & 16.38\\
Quiescent &           10&15.20&0.17 & 0.55 & 15.13 & 14.57 & 16.10\\
\hline
Log[$N$(CH$_3$CN) (cm$^{-2}$)] &          121&14.01&0.07 & 0.72 & 13.98 & 12.82 & 16.62\\

\cline{1-1}
HII &           40&13.77&0.10 & 0.66 & 13.70 & 12.82 & 15.31\\

YSO &           66&14.29&0.08 & 0.64 & 14.73 & 12.82 & 15.31\\
Protostellar &           34&14.30&0.11 & 0.65 & 14.32 & 13.19 & 15.92\\

Quiescent &            3&13.81&0.50 & 0.87 & 13.47 & 13.16 & 14.81\\

\hline\\
\end{tabular}\\

\end{minipage}

\end{center}
\end{table*}

\setlength{\tabcolsep}{6pt}

\section{Correlation of line ratios with evolution}
\label{sect:chemical_clock}

We have detected emission from sixteen of the 27 observed transitions towards more than 25\,per\,cent of the sample, while the remaining eleven lines were detected in at most 12\,per\,cent of the observed sources. Each transition is sensitive to different chemical and excitation conditions, (see Table\,\ref{tab:mollines}), and these tracers can be combined to produce 120 pairings that may be used to investigate the sensitivity of line ratios to changes in the physical properties of clumps that result from the evolution of the embedded protostellar objects.

The ultimate goal of this analysis is to identify line ratios that can be used as chemical clocks that have the ability to reliably determine the evolutionary phase of the embedded object, and to simultaneously provide some insight into the physical processes involved. We will first investigate the trends seen as a function of the evolutionary groups we have identified and then will look more specifically for trends between the line ratios and dust temperature, which is strongly correlated with the $L/M$ ratio (\citealt{urquhart2018_csc}), which itself is a widely used diagnostic of evolution ({\citealt{molinari2008}}). 

\subsection{Group statistics\label{sec:grp_stats}}

We have calculated intensity and line-width ratios for all 120 combinations discussed above as well as their inverses.  Any given pair that exhibits an evolutionary trend may increase with evolutionary stage or decrease: to facilitate comparisons, we have chosen the ratio, that gives a predominantly increasing value with evolutionary stage. The mean values of all ratios were calculated for each of the classification subgroups (quiescent, protostellar, YSO, and \hii). Line ratio detection statistics are given in Table\,\ref{tab:subgroup_detections}.   As each transition traces different physical conditions (and may be more or less likely to be detected at certain evolutionary stages), not all ratios can be calculated for every source.  For nearly all ratios, the protostellar, YSO, and \hii\ stages each represent approximately one-quarter to one-third of the detected sources.  Roughly ten\,per\,cent of detections were classified in the PDR stage, and approximately five\,per\,cent were quiescent sources. As the bulk of detected sources belong to the protostellar, YSO, or \hii\ stages, these three formed the core of our analysis (Sections \ref{sect:QP_class_analysis}).  

\begin{table*}
 
 \caption{$p$-values for ANOVA and Kruskal-Wallis tests for intensity and line-width, with pairwise tests between Protostellar (p), YSO (Y), H{\sc ii} regions, and photon-dominated regions (P) calculated from Tukey Honest Significant Difference or Dunn post-hoc tests, as appropriate.  Group differences that are significant at the two sigma level have values in the post-hoc test columns.  $p$ values that are less than 0.001 are shown as 0.000. The quiescent subsample is too small to test for correlations with the other subsamples. This subset of molecular ratios is ordered from largest number of statistically significant intensity differences to fewest.}
\begin{center}
 \begin{tabular}{l ccccc ccccc}
 \hline
  & \multicolumn{5}{c}{Intensity} & \multicolumn{5}{c}{Line-width}\\
  \cmidrule(lr){2-6}\cmidrule(lr){7-11}
  & & & \multicolumn{3}{c}{post-hoc tests} & & &\multicolumn{3}{c}{post-hoc tests}\\

  Line Ratio 	&	Test	&	$p$	&	$p_{\rm{Q-p}}$ & $p_{\rm{p-Y}}$	& $p_{\rm{Y-H}}$	&	Test	&	$p$	&	$p_{\rm{Q-p}}$ & $p_{\rm{p-Y}}$	& $p_{\rm{Y-H}}$\\
  
   (1) 	&	(2)	&	(3)	&	(4) & (5)	& (6)	&	(7)	&	(8)	&	(9)& (10)	& (11) \\

  \hline
H$^{13}$CN/N$_2$H$^+$ & Kruskal-Wallis & {\color{black}0.000} & {\color{black}1.000} & {\color{black}0.000} & {\color{black}0.000} & Kruskal-Wallis & {\color{black}0.062} & & \\
$^{13}$CS/N$_2$H$^+$ & Kruskal-Wallis & {\color{black}0.000} & {\color{black}1.000} & {\color{black}0.018} & {\color{black}0.000} & ANOVA & {\color{black}0.930} & & & \\
HCN/HNC & ANOVA & {\color{black}0.000} & {\color{black}0.900} & {\color{black}0.036} & {\color{black}0.039} & Kruskal-Wallis & {\color{black}0.017} & {\color{black}0.417} & {\color{black}1.000} & {\color{black}1.000}\\
HCN/N$_2$H$^+$ & Kruskal-Wallis & {\color{black}0.000} & {\color{black}1.000} & {\color{black}0.002} & {\color{black}0.001} & Kruskal-Wallis & {\color{black}0.099} & & \\
HC$_3$N/HN$^{13}$C & ANOVA & {\color{black}0.000} & {\color{black}0.598} & {\color{black}0.004} & {\color{black}0.001} & Kruskal-Wallis & {\color{black}0.002} & {\color{black}0.768} & {\color{black}0.223} & {\color{black}1.000} \\
H$^{13}$CN/HN$^{13}$C & Kruskal-Wallis & {\color{black}0.000} & {\color{black}1.000} & {\color{black}0.001} & {\color{black}0.000} & ANOVA & {\color{black}0.850} & &\\
HC$_3$N/N$_2$H$^+$ & Kruskal-Wallis & {\color{black}0.000} & {\color{black}0.275} & {\color{black}0.001} & {\color{black}0.000} & ANOVA & {\color{black}0.004} & {\color{black}0.003} & {\color{black}0.900} & {\color{black}0.900} \\
CCH/N$_2$H$^+$ & Kruskal-Wallis & {\color{black}0.000} & {\color{black}0.181} & {\color{black}0.000} & {\color{black}0.000} & Kruskal-Wallis & {\color{black}0.000} & {\color{black}0.436} & {\color{black}0.014} & {\color{black}0.838}\\
HCO$^+$/N$_2$H$^+$ & Kruskal-Wallis & {\color{black}0.000} & {\color{black}1.000} & {\color{black}0.009} & {\color{black}0.020} & ANOVA & {\color{black}0.319} & & \\
H$^{13}$CO$^+$/N$_2$H$^+$ & Kruskal-Wallis & {\color{black}0.000} & {\color{black}1.000} & {\color{black}0.000} & {\color{black}0.019} & ANOVA & {\color{black}0.000} & {\color{black}0.444} & {\color{black}0.001} & {\color{black}0.900} \\
H$^{13}$CO$^+$/HN$^{13}$C & ANOVA & {\color{black}0.000} & {\color{black}0.249} & {\color{black}0.008} & {\color{black}0.033} & Kruskal-Wallis & {\color{black}0.221} & &\\
CCH/HN$^{13}$C & ANOVA & {\color{black}0.000} & {\color{black}0.092} & {\color{black}0.012} & {\color{black}0.001} & ANOVA & {\color{black}0.106} & & \\
CCH/c-C$_3$H$_2$ & Kruskal-Wallis & {\color{black}0.000} & {\color{black}1.000} & {\color{black}0.000} & {\color{black}0.011} & ANOVA & {\color{black}0.000} & {\color{black}0.806} & {\color{black}0.122} & {\color{black}0.900} \\

\hline
 \end{tabular}
 \end{center}
 \label{tab:ANOVA}
\end{table*}

Although stellar evolution is a continuous process, the evolutionary classifications that we are using are incremental, based on classification of IR emission.  As such, investigation of difference between subgroups requires categorical statistical analysis.  We have used the ANalysis Of VAriance (ANOVA) technique to investigate significant differences between these stages.  The ANOVA requires subsample variances to be similar in order to produce meaningful results (homoscedasticity): in accordance with typical procedure, we set a cut-off that the maximum evolutionary subgroup variance be no more than twice the minimum subgroup variance. Line ratios that are determined to not be homoscedastic are instead analysed with the non-parametric Kruskal-Wallis (K-W) test.  This test serves the same role as the ANOVA, although its results are not quite as rigorous.

Once the null hypothesis has been rejected with the ANOVA or K-W tests (that is, once the tests have determined that it is 95\,per\,cent likely that at least one of the evolutionary sub-groups is drawn from a different underlying population), a post-hoc test is used to determine distinctions between each of the subgroups.  These tests operate in a pairwise fashion between the different sub-populations.  Line ratios that were analysed with the ANOVA are followed with the Tukey Honest Significant Difference (Tukey HSD) post-test, while the non-parametric Dunn post-hoc test is used in conjunction with the K-W test.  Pairs are rated distinct at the 95\,per\,cent confidence level.  A sample of these results are shown in Table \ref{tab:ANOVA}.  Columns three and eight show the $p$-value for the ANOVA or K-W test for the intensity and line-width ratios, respectively.  If the ANOVA or K-W test failed to reject the null hypothesis, no pairwise post-hoc tests are performed, and these table cells are left blank.

\begin{figure*}
\includegraphics[width=0.45\textwidth,trim= 1cm 0.25cm 0.25cm 0.25cm, clip
]{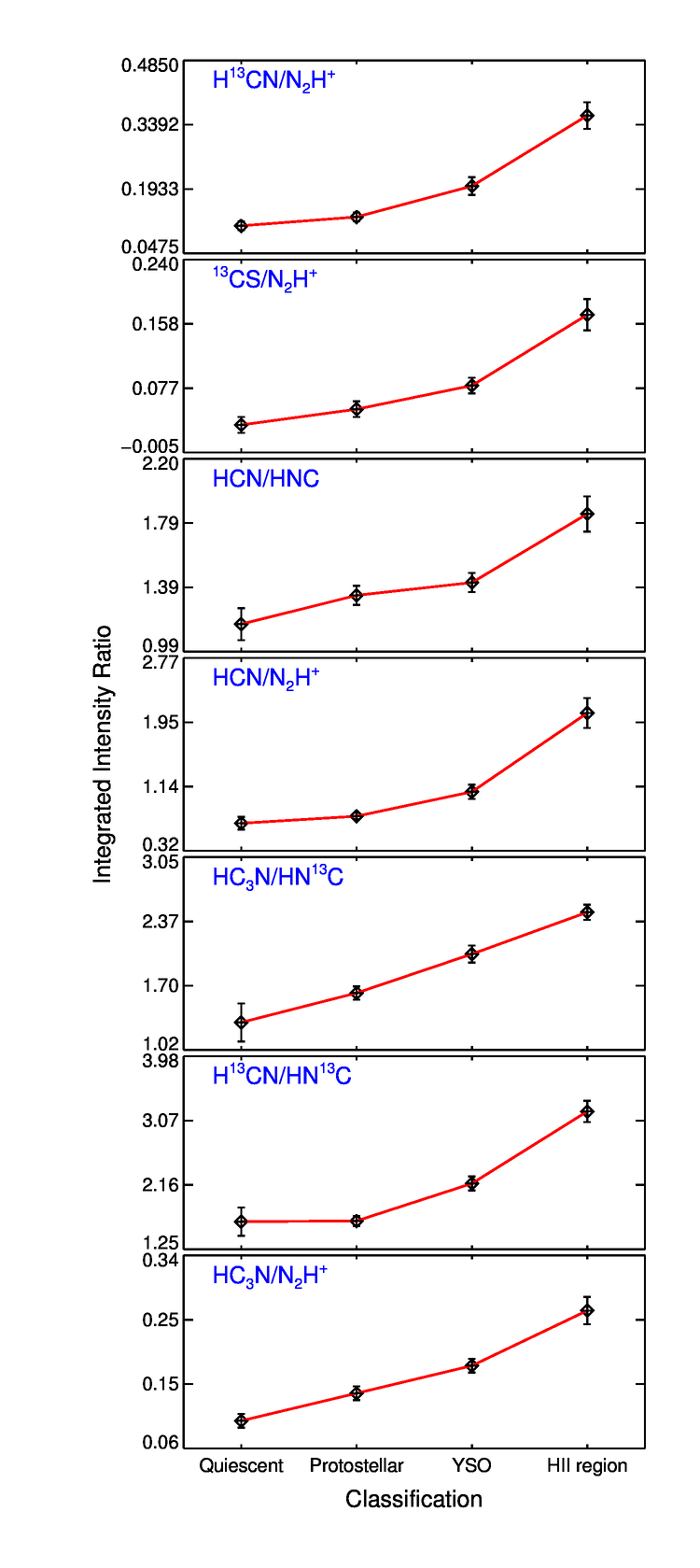}
\includegraphics[width=0.45\textwidth,trim= 1cm 0.25cm 0.25cm 0.25cm, clip]{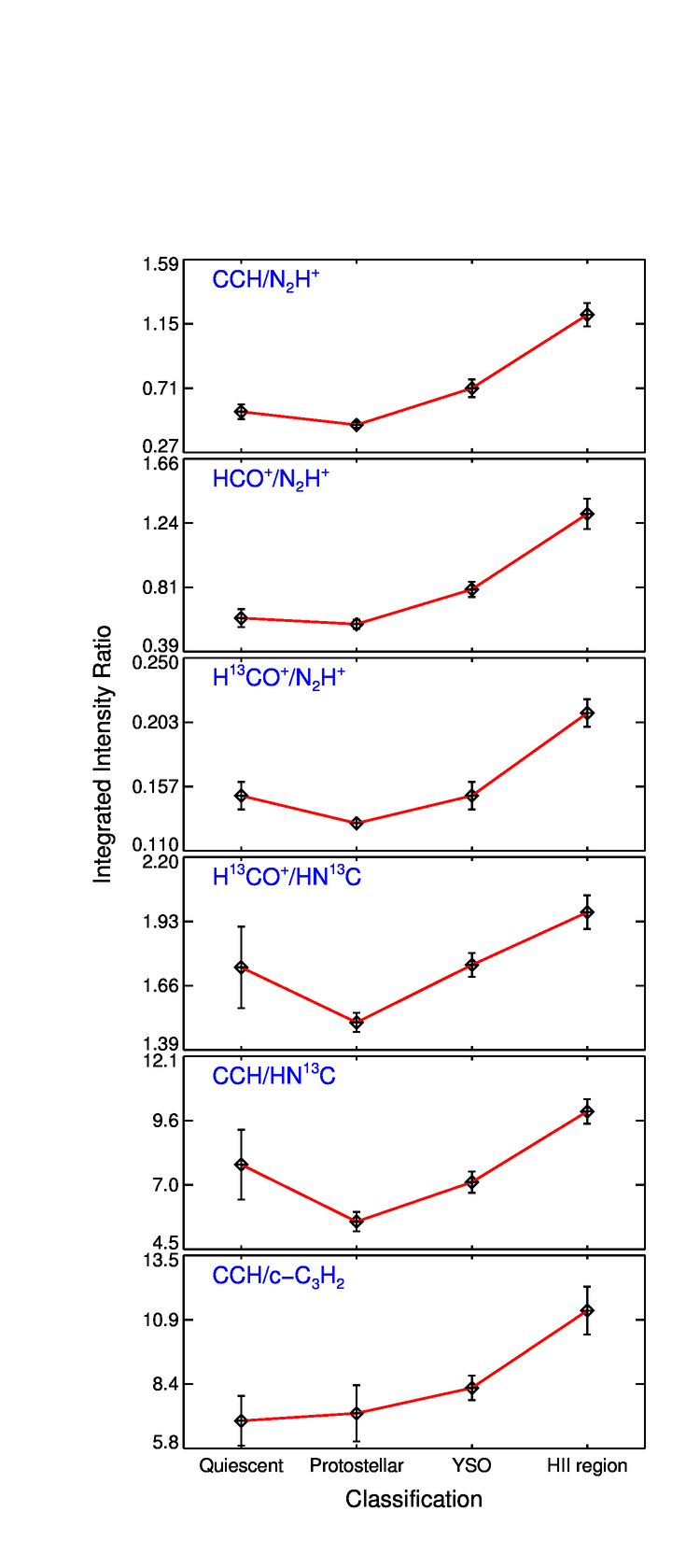}

  \caption{Mean values of integrated-intensity ratios for the thirteen most evolutionarily distinct line ratios identified. These all exhibited statistically significant differences between the protostellar, YSO, and \hii\ classes. These ratios have been ordered to ensure that the ratios increase as a function of evolution and, for many of these ratios, it is clear that the quiescent subsample fits nicely into this sequence. The error bars are standard errors based on classification subgroup sizes. The plots are ordered from top to bottom and left to right in order of decreasing statistically significant differences between evolutionary class subgroups.
}    \label{fig:ratio_plots}  
\end{figure*}

The ratios were then scored by the number of statistically significant pairwise distinctions: ratios that were fully distinct between the three core stages were awarded the highest score, and ratios that were statistically indistinct between the stages scored zero points. Secondary weighting given to those pairs whose Quiescent ratio followed the evolutionary trend. The resulting ordering is used in Table\,\ref{tab:ANOVA} and Fig.\,\ref{fig:ratio_plots}.

\subsubsection{Quiescent sources} \label{sect:QP_class_analysis}

Initial tests showed that in only 13 of the 120 line ratio combinations was the quiescent subsample statistically distinct from the protostellar group.  The overall lack of statistically significant differences between the quiescent and protostellar groups may result from the smaller number of quiescent sources (median of 12 quiescent sources detected per line ratio, less than 20\% of the number of sources in each of the other evolutionary groups; see Table\,\ref{tab:subgroup_detections} for details) or due to the fact that a significant fraction of the clumps classified as being quiescent are actually hosting very young protostellar objects (as discussed in Sect.\,\ref{sect:protostellar_quiescent_clumps}).

None of these thirteen line-ratio combinations showed strong statistical distinction among the remaining groups and, of these thirteen, only one (HC$_3$N/c-C$_3$H$_2$ intensity) implied an evolutionarily-significant trend from the quiescent phase to the \hii\ phase.  Seven showed a trend that was distinctly non-monotonic: i.e, the quiescent ratio was high while the other ratios increased upward with evolutionary stage from a low protostellar value, or vice versa. The remaining five ratios possessed a quiescent ratio that was significantly less than the ratio of the remaining phases (all effectively `constant' at that scale) while the lack of statistically significant differences may suggest little chemical/excitational distinction between the quiescent and protostellar stages, these twelve anomalous values suggest a substantive chemical or excitational shift for some tracers between the quiescent and protostellar stages.

The lack of distinction of the quiescent group results in an analysis that centres on the statistical differences between the protostellar, YSO, and \hii\ evolutionary groups, as discussed below.

\subsubsection{Protostellar, YSO, and \hii\ groups}

A selection of post-hoc-test $p$-values for evolutionarily-sequential pairs is shown in columns 4-6 (for intensity ratios) and 9-11 (for line-width ratios) of Table\,\ref{tab:ANOVA}.  Line pairs are scored by the largest number of significant distinctions between the protostellar-YSO ($p-Y$), YSO-\hii\ ($Y-H$) and quiescent-\hii ($q-H$, not shown), with secondary weighting given to the quiescent-protostellar ($Q-p$) pairing.  Thus, the top row has the largest number of distinct values, with decreasing numbers of pairwise distinctions as one progresses to subsequent rows.   This sorting is also used in arranging the plots shown in Fig.\,\ref{fig:ratio_plots}.

We observe that 13 lines show protostellar, YSO, and \hii\ groups that are fully distinct at the two-sigma confidence level.  We note that none of these show quiescent ratios that are statistically distinct from the protostellar ones, making this group wholly distinct from the set of 13 mentioned in Section\,\ref{sect:QP_class_analysis} above.  Seven of the six (shown on the left side of Fig.\,\ref{fig:ratio_plots}) possess quiescent mean values that are marginally less than the protostellar ones, leading to a suggested small evolutionary trend.

\subsection{Discussion of results}

We have identified thirteen line intensity ratios that are sensitive to the evolutionary stage of the embedded star formation, and so these combinations provide some potential to investigate the chemical evolution of dense clumps.

It is interesting to note that seven out of the ten most significant trends have N$_2$H$^{+}$ in the denominator.  The commonality of this molecule is consistent with our earlier finding that its abundance is not correlated with evolutionary stage: it is more resistant to depletion onto dust grains at low temperature and in dense environments at the size scales probed by these observations (\citealt{stephens2015}). 

The other molecule that features prominently in these ratios is HNC and its isotopologue HN$^{13}$C. \citet{rathborne2016} noted the presence of this molecule in two of the evolutionary trends they reported, and concluded that this is probably due to the fact that HNC is more abundant in colder and less-evolved clumps (\citealt{hoq2013}). Given that  emission from HNC is detected towards every source in our sample this is hard to verify; however, we do see that there is a significant decrease in the detection rate of HN$^{13}$C as a function of evolution (from $\sim$85\,per\,cent for protostellar sources to $\sim$70\,per\,cent for \hii\ regions, see second from bottom left panel of Fig.\,\ref{fig:detection_rates_evolution}) that is consistent with the conclusion that the abundance of this molecule decreases as the star formation proceeds. However, we also found that there is no correlation between the integrated intensity, or our proxies for abundance,  of HN$^{13}$C with evolution (i.e., {integrated intensity}/N(H$_2$) and dust temperature; see top panels of Fig.\,\ref{fig:colden_temp_comparison}). From these two pieces of conflicting evidence we conclude that there is likely to be a modest decrease in the HNC abundance with evolution, but this is not sufficient to have a significant impact on the line ratios. We conclude, therefore, that the observed evolutionary trends seen in the intensity plots presented in Fig.\,\ref{fig:ratio_plots} are primarily due to increases in the abundance of the numerator and the invariance of the abundance of the two denominators since we can assume that the excitation conditions are the same for both transitions. 

Ratios including HNC and those including N$_2$H$^{+}$ exhibit different trends.  We note that HNC ratios tend to increase in even steps with evolutionary age, while N$_2$H$^{+}$-containing ratios show increasing increments for later evolutionary stages. Inspection of the distribution of line intensity and abundance of the N$_2$H$^{+}$ transition as a function of dust temperature (lower panels of Fig.\,\ref{fig:colden_temp_comparison}) reveals a slight decrease at high dust temperatures: this is not significant with respect to the whole sample, but is noticeable and would explain the slightly larger increases in line ratio between the YSO and \hii\ region stages. This slight depletion of the N$_2$H$^{+}$ emission at higher dust temperatures is predicted from chemical models (e.g., \citealt{bergin1997,lee2003,bergin07}) as CO is released from dust grains, which destroys N$_2$H$^{+}$ in the gas phase.

\subsubsection{Comparison with other studies}

\begin{figure}
\includegraphics[width=0.45\textwidth,trim= 1cm 0.25cm 0.25cm 0.25cm, clip]{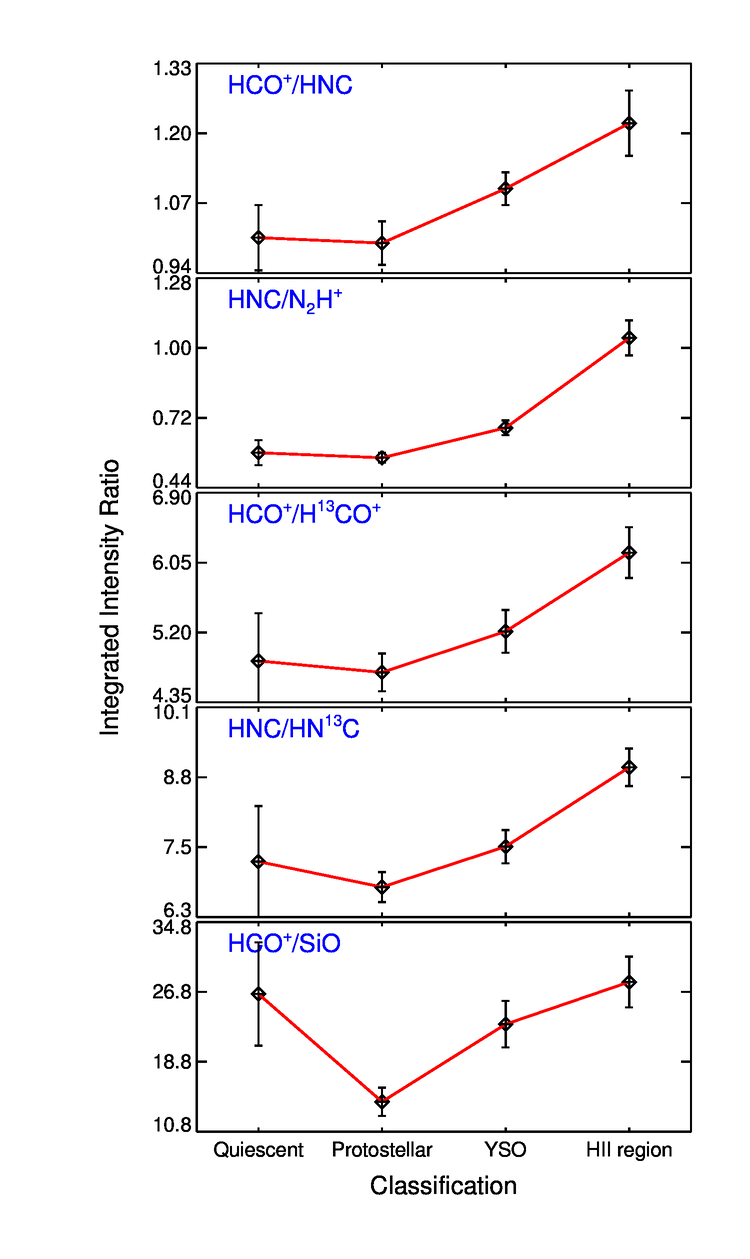}
  \caption{As for Fig.\,\ref{fig:ratio_plots}, but for the 5 line ratios discussed by \citet{rathborne2016} that are not included in our top 13 most significant ratios. }    \label{fig:malt90_plots}  
\end{figure}

\begin{figure*}


\includegraphics[width=0.33\textwidth]{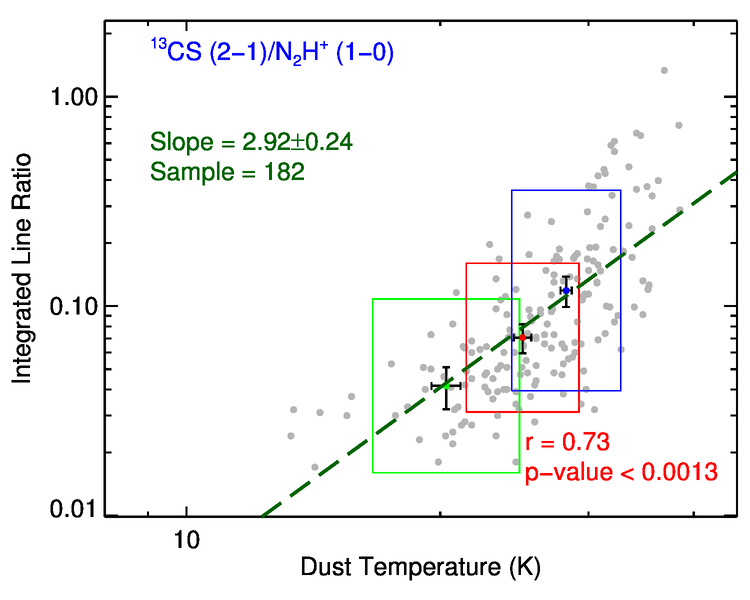}
\includegraphics[width=0.33\textwidth]{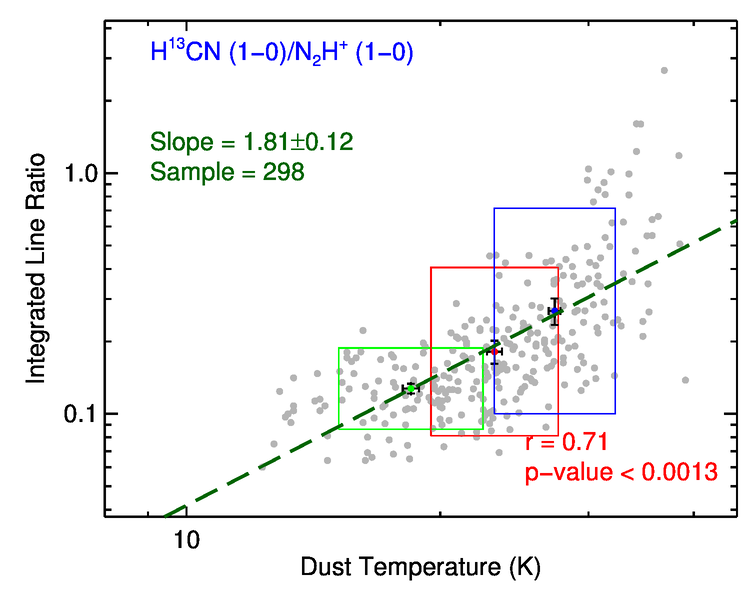}
\includegraphics[width=0.33\textwidth]{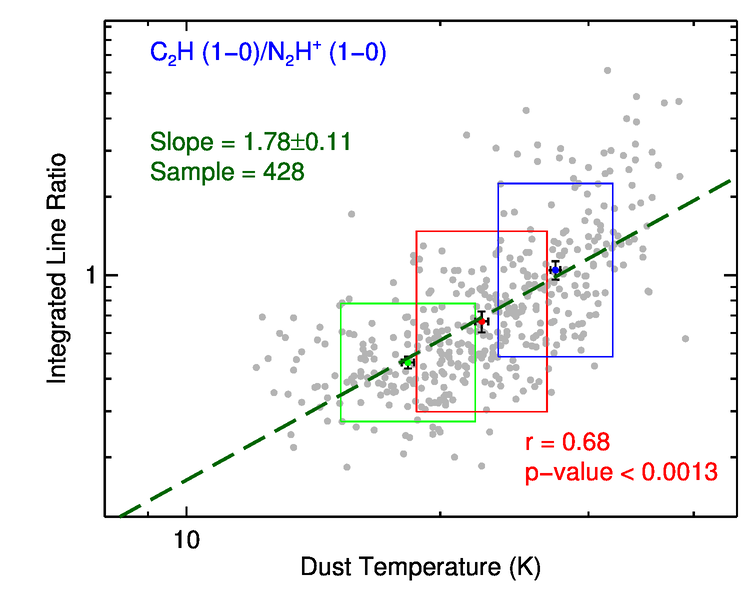}
\includegraphics[width=0.33\textwidth]{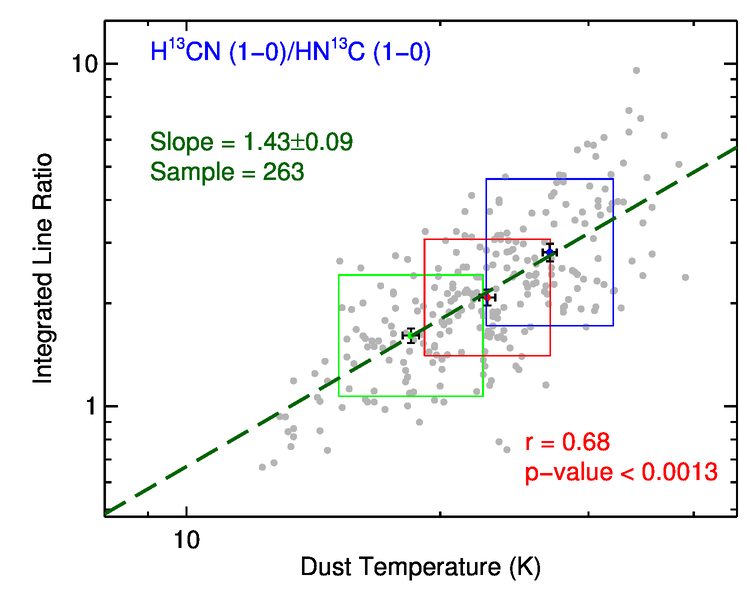}
\includegraphics[width=0.33\textwidth]{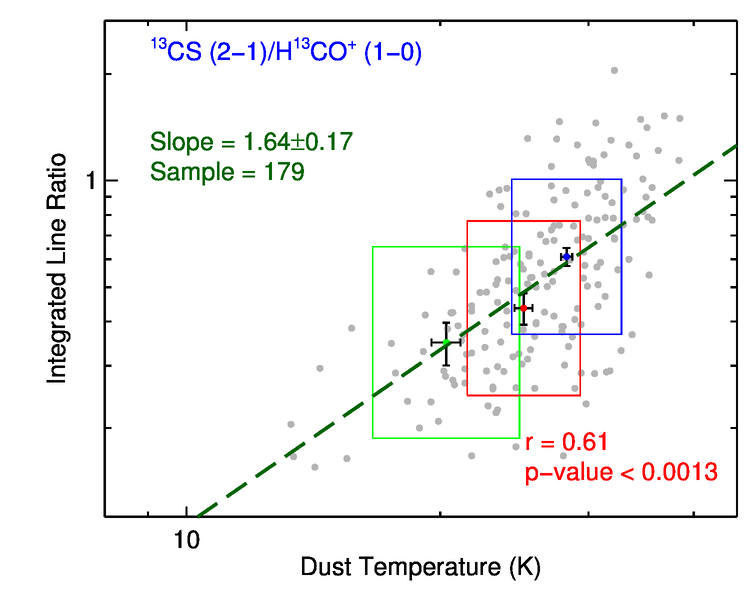}
\includegraphics[width=0.33\textwidth]{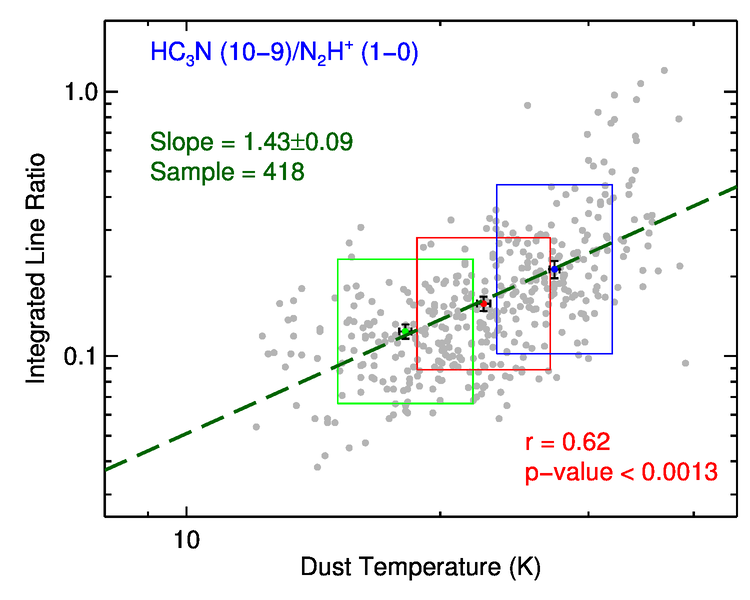}

\caption{The six line-intensity ratios that are most strongly correlated with dust temperature that have sample sizes greater than 200 clumps. The transition ratio is given in the top left corner and the data are shown as filled grey circles. The Spearman correlation coefficient ($r$) and significance ($p$-value) are given in the lower right corner. The dashed green line is the result of a linear fit to the log-log values and the slope and the number of measurements are given in the upper left corner. The mean, the standard error and the standard deviations of these parameters for the protostellar (green), YSO (red) and \hii\ region (blue) subsamples; these are shown as filled circles, error bars and rectangular boxes, respectively.}
\label{fig:james_temp_correlations} 
\end{figure*}

The MALT90 team investigated a similar set of molecular-line observations, and \citet{rathborne2016} highlighted nine line combinations (without statistical analysis) that suggest a trend of increasing line ratio as a function of the evolution. Four of these ratios are identified among the 13 statistically distinct ratios discussed above (i.e., HCN/HNC, CCH/N$_2$H$^+$, HNC/N$_2$H$^+$, HCO$^+$/N$_2$H$^{+}$). Our results for the remaining five line ratios are shown in Fig.\,\ref{fig:malt90_plots}. Visual inspection of these five ratios reveal trends that are consistent with our evolutionary sequence; however, it is also clear that the difference between the protostellar and YSO stages is not large enough to be considered statistically reliable. Taken as a larger group, we find reasonable agreement with the nine line ratios discussed by \citet{rathborne2016}, despite the fact that the samples and the classification schemes are both slightly different.

\citet{hoq2013} investigated the HCN/HNC and N$_2$H$^{+}$/HCO$^+$ ratios (note that the latter is inverted from the \citealt{rathborne2016} study)  in an effort to test their utility as chemical clocks.  They found no significant trend for the N$_2$H$^{+}$/HCO$^+$  ratio, and although they found that the HCN/HNC ratio increased weakly but significantly as function of evolution, they cast doubt on its suitability as a clock due to substantial overlaps between the distributions of the different evolutionary categories used. A similar trend of increasing HCN/HNC abundance as a function of evolution has been found by \citet{vasyunina2011} in a sample of IRDCs. Our investigation also flagged both of these pairs within our 13 most significant evolutionary ratios (HCN/HNC and HCO$^+$/N$_2$H$^{+}$ in the left-middle panel and lower-middle right panels, respectively, of Fig.\,\ref{fig:ratio_plots}). \citet{sanhueza2012} also investigated the HCO$^+$/N$_2$H$^{+}$ abundance ratio and found an increasing ratio with evolution leading them to conclude it could be used as a chemical clock. \footnote{Note that \citet{sanhueza2012} plots the N$_2$H$^{+}$/HCO$^+$ and consequently report a decreasing abundance ratio with evolution.}  This increasing HCO$^+$/N$_2$H$^+$ ratio is consistent with models presented by \citet{bergin1997,lee2003,bergin07}: 
as the clump heats up, CO released from the dust grains not only destroys the N$_2$H$^{+}$, but also acts as a parent molecule of HCO$^{+}$ (\citealt{hoq2013}). 

 The HNC/N$_2$H$^{+}$ abundance ratios were also investigated by \citet{sanhueza2012} who found it decreased with evolution. However, we find the opposite trend. We note that the $p$-value for the KS-test performed by \citet{sanhueza2012} on the quiescent and active clumps had a value of 5\,per\,cent from which they concluded that the chance both samples are drawn from the  same parent population is low, however, this value is above the threshold we use to indicate the samples are significantly different and therefore their result is not conclusive. 

The high detection rates achieved by these observations allow us to identify evolutionary trends for many more line ratios; however, as found by \citet{hoq2013}, the spread in the ratios for the different evolutionary subsamples results in significant overlap between them, which limits their utility as chemical clocks for individual sources. We stress that the analysis presented in this section is based on evolutionary stages  determined by IR, submillimetre, and radio emission characteristics (Section\,\ref{sect:classification}), which are themselves not definitive and, as such, we lack the capability to investigate these possible correlations outside of categorical methods. In the next section we will investigate the correlations between these line ratios and their dust temperature, which has also been shown to be a good tracer of evolution (\citealt{urquhart2018_csc,elia2017}).

\subsection{Viability of line ratios as chemical clocks}

\subsubsection{Line ratio as a function of dust temperature}

We have already found that the dust temperature increases with the evolutionary state, and this has also been shown to be tightly correlated with the luminosity/mass ratio (\citealt{urquhart2018_csc,elia2017}), which is a commonly used diagnostic of evolution (e.g., \citealt{molinari2008,urquhart2014_csc}). We have therefore plotted the line-intensity ratios as a function of their dust temperatures and calculated the Spearman correlation coefficient ($r$) and significance ($p$-value) in an effort to identify ratios that are correlated with evolution. Only correlations that have a $p$-value $<$0.0013 are considered to be significant and the line ratios of any negative correlations found were inverted to ensure that all correlations are positive. 

In total, we have identified 80 line-intensity ratios that are positively correlated with the dust temperature. The  correlation coefficients range from $r$ = 0.16-0.71 and the sample sizes range from 34 to 487 clumps. Since all of these 80 line ratios are significant (i.e., $p$-value $<$0.0013) we consider line ratios with $r > 0.5$ to be strongly correlated and those with $r < $0.5 to be weakly correlated i.e., a measure of the intrinsic scatter; these two groups have 30 and 50 ratios, respectively. The majority (19) of the strongly correlated ratios consist of at least one optically thin transition (e.g., $^{13}$CS/N$_2$H$^+$, C$_2$H/N$_2$H$^+$ and $^{13}$CS/HN$^{13}$C). The strong correlation between optically thick transitions and N$_2$H$^+$, which are almost universally detected, have the advantage of being readily applied to both Galactic and extragalactic studies.

By comparison of these results with the analysis presented, in the previous section where we used pair-wise statistical tests to identify correlations, we find that all 13 of the ratios shown in Fig.\,\ref{fig:ratio_plots} are also picked up in this analysis. The correlation coefficients of those identified in the previous analysis range from $r = 0.31$-0.71 with eight found in the strongly correlated group and therefore there is good agreement between the two analysis techniques.

For a line ratio to be useful as a chemical clock it needs to have three qualities: there should be 1) a strong functional correlation between the line ratio and clump dust temperature; 2) a strong dependence between the line ratio and dust temperature to provide powerful diagnostic ability (i.e., steep slopes determined from linear fits to the data); and 3) high detection rates over all of the evolutionary stages. In Fig.\,\ref{fig:james_temp_correlations} we present plots of the six line ratios that are most strongly correlated with the dust temperature and detection rates of $>$35\,per\,cent ($>$ 200 clumps). All of these ratios are strongly correlated with dust temperature ($0.6 < r < 0.71$). It is interesting to note that 4 of the 6 most strongly correlated ratios have N$_2$H$^+$ in the denominator; we have already noted that the intensity and abundance of this molecule appears to be invariant to increases in the dust temperature, which is probably what makes it such a good molecule to identify change in the abundances of other molecules.

It is clear from the correlations shown in Fig.\,\ref{fig:james_temp_correlations} that feedback from the evolving protostars is having a significant effect on the chemistry of their natal clumps. On these plots we also show the mean, the standard error and the standard deviations of these parameters for the protostellar, YSO and \hii\ region subsamples; these are shown as filled circles, error bars and rectangular boxes, respectively. These plots reveal that, despite the strong correlation between these line ratios and the dust temperatures, there is a significant overlap in the line ratios for the different evolutionary categories confounding their utility to accurately determine the evolutionary stage of a single embedded object; however, it is possible to use these results in a probabilistic manner to estimate the evolutionary stage. It is also worth noting that the overlap in the evolutionary stages is lower for the dust temperature than  the line ratio, indicating that $T_{\rm dust}$ is a better evolution parameter.

We also note that the line ratios display a smooth distribution with no obvious jumps as a function of dust temperature that would indicate significant changes  linked to specific stages or processes. This is in agreement with many observational works that show that star formation  is a continuous process and the observational classification scheme is imposing an artificial set of stages on the data. Classifying sources into distinct types is therefore likely to be a somewhat crude approximation, and this has limited the number of evolutionary correlations identified. Using the dust temperature as our probe of evolution allows us study the star formation as a continuous process. However, we also note that there is  significant scatter in the distributions and so using these line ratios has a diagnostic for evolution is likely to be somewhat imprecise, particularly given that there may be several evolutionary stages present within the beam (see figure\,5 of \citealt{urquhart2014_csc}).

\subsubsection{Summary}

We have identified a number of line ratios that are strongly correlated with the dust temperature and with an observationally defined set of evolutionary stages. We have found that there is a significant amount of overlap between adjacent evolutionary stages and scatter in the data that make using these correlations unreliable for using either the dust temperature or line ratio to definitively determine the evolutionary stage of an embedded object. We have found that the overlap in the parameters for the different evolutionary stages is lower for the dust temperature and so this may be more definitive than using the line ratios. 

It is possible that our inability to clearly separate the different stages may be due to the relatively poor spatial resolution of these observations ($\sim$0.5\,pc). These scales are typical of entire cluster-forming regions, where YSOs of different mass and in different evolutionary stages, as well as hot and cold material coexist, effectively washing out physical and chemical differences in close proximity of the (proto-)stars.
A clear example in this sense is represented by hot cores: we are not able to observe the high temperatures ($\gtrsim100$~K) and the sharp jumps in the abundance of molecular species released from the ice mantles of dust grains in these conditions, that are clearly visible at small scales \citep[e.g.][]{oberg2013, fayolle2015} or with selective tracers \citep[e.g.][]{bisschop2007, giannetti2017}.
Most likely, the observationally-derived evolutionary stages are only able to capture snapshots of the physical and chemical properties, that evolve in a continuous manner, especially when averaging on large spatial scales. Therefore, despite their statistical validity, they will probably always show some degree of superposition.

Although we are unable to use any of the relationships discussed to discrimination of the evolutionary stage of any single object these results can be used to determine the evolutionary state in a probabilistic way.

\section{Summary and Conclusions}
\label{sect:summary}

We present the result of a 3-mm molecular-line survey conducted towards a sample of 570 high-mass star-forming clumps. These clumps have been selected from the ATLASGAL Compact Source Catalogue (CSC; \citealt{contreras2013, urquhart2014_csc}) and care has been taken to ensure the sample is representative of the full range of observationally defined evolutionary stages associated with high-mass star formation. The sample has been classified using the  evolutionary sequence described by \citet{konig2017} and consists of 29 quiescent clumps, 153 protostellar clumps (70-\mum\ bright), 128 YSOs (mid-infrared bright) and 166 \hii\ regions (mid-infrared bright + compact radio continuum emission). We also classify 48 as PDRs while the emission seen towards the remaining 46 clumps is too complicated to classify reliably. The distances, luminosities and clump masses for the vast majority of the clumps have been reported by \citet{urquhart2018_csc}.

These observations have detected $\sim$30 molecular and radio recombination-line transitions with high detection rates for large numbers of these ($>90$\,per\,cent for 7 transitions). We have compared the integrated line intensities, the relative abundances, line ratios and dust temperatures to identify correlations with the evolutionary stage of the clumps. \\

\noindent Our main findings are:

\begin{enumerate}

\item The detection rates of many of the transitions reveals that they increase steadily with evolution (e.g., CCH, H$^{13}$CN). However, we also identify four molecular lines that show a peak in the detection rates for the protostellar stage. This group also includes the HNCO and SiO transitions, which are enhanced by slow and fast shocks, respectively,  and are often associated with molecular outflows. We also note that several molecules show a significant decline in the detection rates for the \hii\ region stage (N$_2$H$^+$ and o-NH$_2$D), indicating their abundances are sensitive to the far-uv radiation field.\\

\item We find good correlation between the integrated intensities and the H$_2$ column densities derived from the dust emission for the majority of molecules. However, it is interesting to note that there is no correlation between the integrated line intensity, or the relative abundances, with the dust temperature for either the HN$^{13}$C and the N$_2$H$^+$ transitions. These two molecules are invariant to evolution and therefore provide an excellent baseline with which to compare changes in the abundance of molecules that are more sensitive to dust temperature, excitation or far-uv induced  changes, such as C$_2$H. The main reason for the invariance of HN$^{13}$C and N$_2$H$^+$ is likely to be due to the fact that they efficiently trace cold gas on large scales, and because they are still very bright in the submillimetre, there is still plenty of it even in the clumps hosting \hii\ regions.\\

\item We use lines from different K-ladders of the CH$_3$C$_2$H and CH$_3$CN transitions to determine the  mean rotational temperatures for approximately half of the clumps; these are 38.3 and 48.8\,K, respectively. We find that these rotational temperatures are significantly larger than determined from the dust emission (24.6\,K). This is likely to be because the emission from these two tracers is emitted from dense compact regions located around the embedded objects. 

\item We find CH$_3$C$_2$H emission towards a third of the quiescent sample suggesting that a significant fraction of these clumps may be harbouring a population of very young protostellar objects that are so deeply embedded that they are not yet detected at 70\,\mum. This sample has larger dust temperatures and $L/M$ ratios and a small fraction are also associated with SiO emission, all of which is consistent with these clumps hosting very young protostars. The number of quiescent clumps is very small ($\sim$5\,per\,cent), and given that 11 of the 28 quiescent clumps may host very young protostellar objects, this is likely to be an upper limit to the true fraction of high-mass starless clumps in the Galaxy.\\

\item We have used categorical statistical analysis methods to identify thirteen line-intensity ratios that are distinct between the protostellar, YSO, and \hii\ region stages, showing an increasing value with advancing evolutionary stage.  These lines are consistent with models of chemical evolution within star-forming clumps. We do not find any consistent trend for the quiescent clumps; however, the sample is very small ($\sim$5\,per\,cent) and is likely to be contaminated with clumps hosting very young protostars, as discussed in the previous bullet point. \\

\item We have elaborated on the molecular-line intensity ratios presented by the MALT90 team, and identify the same general evolutionary trends identified by \citet{rathborne2016}. We find four of the nine ratios therein to be fully distinct between protostellar, YSO, and \hii\ stages, with the other five showing lesser degrees of distinction, suggesting some utility of these ratios as chemical clocks. We have investigated the HCO$^{+}$/N$_2$H$^{+}$ ratio discussed by \citet{hoq2013} and find that it is both statistically distinct and increases with evolutionary stage, in agreement with chemical evolution models.\\

\item We have investigated correlations between line ratios and dust temperatures as a continuous measure of star formation.  We have identified 30 ratios that correlate strongly with dust temperature, with a mix of both optically thin and optically thick line ratios.  We have found that the evolutionary classes overlap, suggesting that the static description of evolutionary stages are artifacts of a much more continuous evolutionary process that are imposed by the observationally-defined classification scheme. These stages can provide useful snapshots of the high-mass star formation process, but may possess limited utility for deeper sequence investigation.\\

\item We have robustly identified a set of six line ratios in our large observational sample, which show clear evolutionary trends in both infrared classification and dust temperature. The lines that form these ratios are strongly detected across the sample and possess strong utility as evolutionary tracers.

\end{enumerate}

\noindent We have identified strong trends for a number of line ratios that can be used to predict the evolutionary stage of the embedded object.  However, due to a significant amount of scatter and overlap between the physical parameters derived for the different evolutionary subsamples, these may be too imprecise for estimating the parameters for an individual source. Although the current results are not definitive in terms of providing a reliable chemical clock, higher-resolution observations will significantly increase the range of temperatures and line ratios and, together with accounting for the appropriate excitation for each species, decrease the scatter in the data, providing greater diagnostic power.

\section*{Acknowledgments}

We would like to extend our thanks to an anonymous referee for their comments and suggestions that have helped clarify many aspects of this work. This work was partly carried out within the Collaborative Research Council 956, sub-project A6, funded by the Deutsche Forschungsgemeinschaft (DFG). This work as made extensive use of the astropy and pyspeckit PYTHON libraries. This document was produced using the Overleaf web application, which can be found at www.overleaf.com.

\bibliographystyle{maras}
\bibliography{atlasgal_mopra,urquhart2016}

    \begin{landscape}
    
   	\begin{table}
    	\begin{center}
        \caption{Integrated intensities $W\,($K\,km\,s$^{-1})$ of the ten most commonly-detected molecular transitions in the sample. The first 35 sources are shown: all transitions and sources tabulated in on-line materials. Three-sigma upper limits with an assumed 1\,\kms\ line-width are shown in parentheses for undetected lines.}
        \begin{tabular}{l c c c c c c c c c c}
\hline
Source  & \rot{HNC (1-0)} & \rot{N$_2$H$^+$ (1-0)} & \rot{HCO$^+$ (1-0)} & \rot{CCH (1-0)} & \rot{H$^{13}$CO$^+$ (1-0)} & \rot{HCN (1-0)} & \rot{HC$_3$N (10-9)} & \rot{c-C$_3$H$_2$ (2-1)} & \rot{HN$^{13}$C (1-0)} & \rot{H$^{13}$CN (1-0)} \\
\hline
\hline
AGAL300.164-00.087 & $2.57\pm0.08$ & $2.91\pm0.05$ & $4.70\pm0.08$ & $3.35\pm0.10$ & $0.52\pm0.07$ & $5.21\pm0.08$ & $0.22\pm0.05$ & $(0.09)$ & $0.16\pm0.05$ & $(0.09)$ \\
AGAL300.504-00.176 & $2.40\pm0.07$ & $1.40\pm0.06$ & $4.27\pm0.19$ & $6.34\pm0.05$ & $0.38\pm0.06$ & $5.76\pm0.12$ & $0.31\pm0.08$ & $(0.08)$ & $(0.07)$ & $0.30\pm0.03$ \\
AGAL300.721+01.201 & $2.00\pm0.05$ & $2.09\pm0.04$ & $2.93\pm0.07$ & $2.04\pm0.10$ & $0.36\pm0.03$ & $3.98\pm0.06$ & $0.46\pm0.04$ & $(0.07)$ & $(0.08)$ & $0.36\pm0.04$ \\
AGAL300.826+01.152 & $4.59\pm0.08$ & $5.94\pm0.05$ & $6.00\pm0.10$ & $3.59\pm0.17$ & $0.87\pm0.06$ & $6.41\pm0.10$ & $0.54\pm0.03$ & $0.76\pm0.06$ & $0.62\pm0.04$ & $0.57\pm0.06$ \\
AGAL300.911+00.881 & $2.70\pm0.09$ & $6.18\pm0.05$ & $2.15\pm0.05$ & $2.32\pm0.07$ & $0.92\pm0.06$ & $1.88\pm0.04$ & $0.58\pm0.06$ & $0.60\pm0.04$ & $0.61\pm0.05$ & $0.97\pm0.06$ \\
AGAL301.014+01.114 & $3.00\pm0.04$ & $4.57\pm0.04$ & $3.21\pm0.05$ & $2.61\pm0.06$ & $0.52\pm0.04$ & $4.01\pm0.06$ & $0.43\pm0.03$ & $0.31\pm0.03$ & $0.26\pm0.02$ & $0.63\pm0.04$ \\
AGAL301.116+00.959 & $4.34\pm0.07$ & $5.69\pm0.04$ & $5.50\pm0.19$ & $5.32\pm0.14$ & $0.79\pm0.05$ & $8.38\pm0.12$ & $0.58\pm0.03$ & $0.41\pm0.05$ & $0.37\pm0.05$ & $0.65\pm0.04$ \\
AGAL301.116+00.977 & $4.77\pm0.08$ & $8.70\pm0.03$ & $5.51\pm0.11$ & $6.96\pm0.08$ & $0.88\pm0.04$ & $8.64\pm0.08$ & $0.87\pm0.03$ & $0.50\pm0.05$ & $0.45\pm0.04$ & $0.81\pm0.04$ \\
AGAL301.139+01.009 & $3.40\pm0.06$ & $5.30\pm0.06$ & $4.51\pm0.12$ & $1.89\pm0.08$ & $0.73\pm0.06$ & $3.40\pm0.10$ & $0.40\pm0.05$ & $0.69\pm0.05$ & $0.65\pm0.05$ & $(0.07)$ \\
AGAL301.136-00.226 & $10.11\pm0.10$ & $2.80\pm0.07$ & $18.21\pm0.20$ & $10.87\pm0.12$ & $1.60\pm0.08$ & $34.16\pm0.18$ & $2.62\pm0.08$ & $0.37\pm0.05$ & $0.47\pm0.08$ & $4.50\pm0.09$ \\
AGAL301.679+00.246 & $2.12\pm0.07$ & $5.52\pm0.05$ & $3.79\pm0.12$ & $3.26\pm0.07$ & $0.60\pm0.04$ & $3.61\pm0.07$ & $0.56\pm0.04$ & $0.74\pm0.04$ & $0.52\pm0.04$ & $0.31\pm0.03$ \\
AGAL301.739+01.102 & $4.47\pm0.11$ & $5.08\pm0.07$ & $5.31\pm0.10$ & $4.32\pm0.16$ & $1.07\pm0.09$ & $8.90\pm0.14$ & $1.02\pm0.05$ & $0.81\pm0.06$ & $0.52\pm0.07$ & $1.46\pm0.09$ \\
AGAL301.814+00.781 & $2.20\pm0.06$ & $1.83\pm0.04$ & $2.52\pm0.05$ & $2.96\pm0.10$ & $0.29\pm0.03$ & $5.46\pm0.04$ & $0.27\pm0.06$ & $0.20\pm0.02$ & $(0.08)$ & $(0.07)$ \\
AGAL302.021+00.251 & $5.13\pm0.12$ & $3.86\pm0.06$ & $5.01\pm0.15$ & $4.68\pm0.15$ & $0.81\pm0.05$ & $10.61\pm0.48$ & $1.48\pm0.15$ & $0.49\pm0.05$ & $0.42\pm0.05$ & $1.89\pm0.15$ \\
AGAL302.032+00.626 & $1.20\pm0.04$ & $2.76\pm0.06$ & $(0.31)$ & $0.69\pm0.15$ & $0.34\pm0.03$ & $2.35\pm0.15$ & $0.62\pm0.11$ & $0.26\pm0.04$ & $0.12\pm0.03$ & $(0.07)$ \\
AGAL302.032-00.061 & $3.53\pm0.10$ & $2.38\pm0.11$ & $5.85\pm0.10$ & $5.28\pm0.18$ & $0.68\pm0.08$ & $7.93\pm0.12$ & $0.53\pm0.12$ & $(0.10)$ & $(0.10)$ & $1.04\pm0.12$ \\
AGAL302.391+00.281 & $5.03\pm0.05$ & $7.92\pm0.05$ & $8.48\pm0.08$ & $4.81\pm0.08$ & $0.98\pm0.06$ & $8.23\pm0.05$ & $0.55\pm0.04$ & $0.56\pm0.05$ & $0.31\pm0.04$ & $0.66\pm0.04$ \\
AGAL302.486-00.031 & $2.42\pm0.05$ & $2.73\pm0.05$ & $2.43\pm0.05$ & $3.69\pm0.09$ & $0.60\pm0.07$ & $3.56\pm0.05$ & $0.50\pm0.05$ & $(0.08)$ & $(0.06)$ & $0.98\pm0.07$ \\
AGAL303.118-00.972 & $(0.08)$ & $(0.08)$ & $(0.06)$ & $(0.13)$ & $(0.07)$ & $(0.07)$ & $(0.07)$ & $(0.07)$ & $(0.08)$ & $(0.08)$ \\
AGAL303.931-00.687 & $3.69\pm0.08$ & $3.07\pm0.05$ & $5.67\pm0.09$ & $13.45\pm0.32$ & $0.49\pm0.08$ & $9.17\pm0.08$ & $0.51\pm0.05$ & $(0.07)$ & $(0.06)$ & $0.61\pm0.05$ \\
AGAL304.021+00.292 & $(0.16)$ & $2.40\pm0.07$ & $(0.08)$ & $(0.13)$ & $0.40\pm0.09$ & $(0.08)$ & $(0.09)$ & $0.16\pm0.03$ & $(0.07)$ & $0.33\pm0.07$ \\
AGAL304.556+00.327 & $3.25\pm0.07$ & $4.78\pm0.03$ & $3.47\pm0.07$ & $2.53\pm0.05$ & $0.49\pm0.05$ & $2.19\pm0.07$ & $0.22\pm0.04$ & $0.21\pm0.06$ & $(0.10)$ & $(0.08)$ \\
AGAL304.761+01.339 & $2.48\pm0.07$ & $4.44\pm0.04$ & $1.66\pm0.04$ & $3.60\pm0.08$ & $0.90\pm0.04$ & $2.44\pm0.03$ & $0.61\pm0.06$ & $1.01\pm0.05$ & $0.50\pm0.05$ & $0.57\pm0.04$ \\
AGAL304.886+00.636 & $1.75\pm0.03$ & $2.94\pm0.03$ & $2.04\pm0.11$ & $2.67\pm0.03$ & $0.50\pm0.02$ & $3.26\pm0.11$ & $0.29\pm0.02$ & $0.34\pm0.03$ & $0.23\pm0.03$ & $0.27\pm0.03$ \\
AGAL304.932+00.546 & $0.88\pm0.05$ & $(0.07)$ & $1.70\pm0.06$ & $2.59\pm0.08$ & $(0.07)$ & $2.47\pm0.06$ & $(0.08)$ & $0.43\pm0.08$ & $(0.06)$ & $(0.06)$ \\
AGAL305.137+00.069 & $6.56\pm0.10$ & $14.30\pm0.09$ & $7.48\pm0.10$ & $5.57\pm0.15$ & $1.00\pm0.09$ & $10.06\pm0.11$ & $1.89\pm0.08$ & $(0.12)$ & $0.71\pm0.12$ & $1.58\pm0.10$ \\
AGAL305.196+00.034 & $7.94\pm0.27$ & $3.01\pm0.11$ & $6.49\pm0.24$ & $8.32\pm0.21$ & $1.64\pm0.10$ & $11.06\pm0.12$ & $1.52\pm0.14$ & $0.91\pm0.14$ & $0.39\pm0.09$ & $1.43\pm0.12$ \\
AGAL305.192-00.006 & $6.26\pm0.09$ & $13.42\pm0.10$ & $4.13\pm0.07$ & $7.07\pm0.06$ & $1.16\pm0.09$ & $8.18\pm0.09$ & $1.45\pm0.08$ & $0.44\pm0.10$ & $0.70\pm0.08$ & $1.28\pm0.10$ \\
AGAL305.197+00.007 & $5.03\pm0.08$ & $6.76\pm0.05$ & $2.91\pm0.09$ & $10.52\pm0.10$ & $0.64\pm0.05$ & $10.81\pm0.12$ & $1.33\pm0.05$ & $0.79\pm0.08$ & $0.40\pm0.04$ & $1.01\pm0.05$ \\
AGAL305.201+00.227 & $7.07\pm0.09$ & $7.89\pm0.08$ & $8.56\pm0.12$ & $9.15\pm0.12$ & $0.61\pm0.06$ & $15.81\pm0.13$ & $1.19\pm0.07$ & $0.66\pm0.06$ & $0.40\pm0.06$ & $0.94\pm0.05$ \\
AGAL305.209+00.206 & $11.53\pm0.23$ & $11.40\pm0.06$ & $13.96\pm0.36$ & $11.97\pm0.16$ & $1.42\pm0.08$ & $31.29\pm0.20$ & $4.66\pm0.11$ & $0.98\pm0.09$ & $1.28\pm0.09$ & $4.99\pm0.09$ \\
AGAL305.226+00.274 & $10.07\pm0.10$ & $17.82\pm0.06$ & $10.49\pm0.10$ & $9.78\pm0.12$ & $1.38\pm0.06$ & $13.78\pm0.10$ & $1.73\pm0.06$ & $0.70\pm0.11$ & $0.92\pm0.07$ & $1.76\pm0.07$ \\
AGAL305.236-00.022 & $5.98\pm0.13$ & $11.82\pm0.09$ & $6.20\pm0.18$ & $7.61\pm0.12$ & $0.94\pm0.09$ & $11.61\pm0.15$ & $0.85\pm0.08$ & $0.51\pm0.08$ & $0.70\pm0.09$ & $1.11\pm0.09$ \\
AGAL305.234+00.262 & $8.30\pm0.10$ & $11.93\pm0.05$ & $10.30\pm0.17$ & $13.76\pm0.11$ & $0.97\pm0.06$ & $13.59\pm0.11$ & $1.84\pm0.04$ & $1.07\pm0.08$ & $0.74\pm0.06$ & $1.47\pm0.06$ \\
AGAL305.242-00.041 & $6.01\pm0.16$ & $9.11\pm0.10$ & $6.00\pm0.21$ & $6.29\pm0.27$ & $0.93\pm0.09$ & $9.27\pm0.17$ & $1.08\pm0.09$ & $0.62\pm0.14$ & $0.83\pm0.10$ & $0.55\pm0.07$ \\
AGAL305.246+00.246 & $4.51\pm0.10$ & $5.95\pm0.07$ & $4.75\pm0.11$ & $8.22\pm0.32$ & $0.44\pm0.09$ & $11.27\pm0.11$ & $1.38\pm0.07$ & $(0.11)$ & $0.44\pm0.06$ & $1.11\pm0.10$ \\
AGAL305.272+00.296 & $5.65\pm0.08$ & $7.87\pm0.04$ & $6.81\pm0.09$ & $5.86\pm0.07$ & $0.54\pm0.02$ & $10.40\pm0.06$ & $0.99\pm0.03$ & $0.45\pm0.06$ & $0.47\pm0.04$ & $1.08\pm0.04$ \\
AGAL305.271-00.009 & $10.57\pm0.15$ & $14.11\pm0.10$ & $11.66\pm0.23$ & $13.00\pm0.18$ & $1.64\pm0.13$ & $21.53\pm0.13$ & $2.54\pm0.10$ & $0.81\pm0.11$ & $0.89\pm0.11$ & $2.41\pm0.11$ \\
\hline
\end{tabular}

	    \label{table:integrated_intensities}
        \end{center}

Notes: Only a small portion of the data is provided here, the full table is available in electronic form at the CDS via anonymous ftp to cdsarc.u-strasbg.fr (130.79.125.5) or via http://cdsweb.u-strasbg.fr/cgi-bin/qcat?J/MNRAS/.\\ 
        \end{table}
 
    \end{landscape}
    \clearpage

\end{document}